\documentclass[review]{elsarticle}

\usepackage{lineno,hyperref}

\usepackage{amsmath,amssymb,amsfonts}
\usepackage{algorithm}
\usepackage{algpseudocode}
\usepackage{epsfig}
\usepackage{hhline}
\usepackage{times}
\usepackage{etoolbox}
\usepackage{multirow}
\usepackage{ctable}
\usepackage{soul}
\usepackage{rotating}
\usepackage{setspace}
\usepackage{pslatex}
\usepackage[T1]{fontenc}
\usepackage{type1ec}
\usepackage{aecompl}
\usepackage{makecell}
\usepackage{float}
\usepackage{graphicx}
\usepackage{textcomp}
\usepackage{xcolor}
\usepackage{lscape,pdflscape}

\newcommand{\figscale}{0.24}

\begin{document}

\begin{frontmatter}

\title{Ambiguity of Objective Image Quality Metrics: A New Methodology for Performance Evaluation}
\tnotetext[mytitlenote]{A preliminary study \cite{cheon2016ambiguity} was presented at the International Conference on Quality of Multimedia Experience, Lisbon, Portugal, 2016.}

\author[address1]{Manri Cheon}
\author[address2]{Toinon Vigier}
\author[address2]{Luk\'{a}\v{s}~Krasula}
\author[address1]{Junghyuk Lee}
\author[address2]{Patrick Le Callet}
\author[address1]{Jong-Seok Lee\corref{mycorrespondingauthor}}
\cortext[mycorrespondingauthor]{Corresponding author}
\ead{jong-seok.lee@yonsei.ac.kr}

\address[address1]{School of Integrated Technology, Yonsei University, 21983 Incheon, Korea}
\address[address2]{LS2N UMR CNRS 6004, Universit\'{e} de Nantes, 44306 Nantes, France}

\begin{abstract}
Objective image quality metrics try to estimate the perceptual quality of the given image by considering the characteristics of the human visual system.
However, it is possible that the metrics produce different quality scores even for two images that are perceptually indistinguishable by human viewers, which have not been considered in the existing studies related to objective quality assessment.
In this paper, we address the issue of ambiguity of objective image quality assessment.
We propose an approach to obtain an ambiguity interval of an objective metric, within which the quality score difference is not perceptually significant.
In particular, we use the visual difference predictor, which can consider viewing conditions that are important for visual quality perception.
In order to demonstrate the usefulness of the proposed approach, we conduct experiments with 33 state-of-the-art image quality metrics in the viewpoint of their accuracy and ambiguity for three image quality databases.
The results show that the ambiguity intervals can be applied as an additional figure of merit when conventional performance measurement does not determine superiority between the metrics.
The effect of the viewing distance on the ambiguity interval is also shown.
\end{abstract}

\begin{keyword}
Quality of experience \sep objective quality assessment \sep ambiguity interval \sep viewing distance
\end{keyword}

\end{frontmatter}


\section{Introduction}
\label{sec:introduction}
Multimedia systems operating in resource-constrained environments usually strive to achieve two conflicting objectives: achieving efficiency and providing high quality content.
For instance, compression, e.g., JPEG \cite{wallace1991jpeg} and JPEG2000 \cite{skodras2001jpeg} for images and H.264/AVC \cite{wiegand2003overview} and HEVC \cite{sullivan2012overview} for videos, is a representative way to deal with this issue; it can reduce the amount of data to enhance storage and transmission efficiency at the cost of degradation of perceptual quality.
Quality degradation introduced through enhanced efficiency tends to lower the quality of experience (QoE) of the consumers.
Therefore, it is important to carefully consider the trade-off relationship between the two objectives in designing the target multimedia systems and services.

The first step toward this goal is to accurately measure the perceptual quality of the content as perceived by human viewers, which can be performed via subjective quality assessment or objective quality assessment \cite{sheikh2006statistical,chikkerur2011objective,cheon2017subjective}.
The former is the most accurate way of assessing the QoE, where human subjects are asked to rate the given content in terms of perceptual quality.
However, it is time-consuming and expensive, and cannot be used in real time applications for controlling or optimizing the quality of the delivered content.
Thus, objective quality assessment performed by objective metrics is widely used to replace subjective quality assessment, which tries to automatically predict perceived quality.
A number of objective quality metrics have been developed and used for various applications including compression, transmission, enhancement, etc. \cite{wang2011applications}.

It has been considered that the primary goal of an objective metric is to predict subjective quality scores, usually denoted as mean opinion scores (MOS), as accurately as possible.
The ITU-T P.1401 standard \cite{itu_t_p1401} specifies recommended procedures to evaluate the accuracy of an objective quality metric.
For instance, the Pearson's linear correlation coefficient (PLCC) and Spearman's rank ordered correlation coefficient (SROCC) are computed to evaluate linearity and monotonicity of metrics with respect to subjective data, respectively.
In addition, the prediction error and consistency are also measured using the root-mean-square error (RMSE) and outlier ratio (OR), respectively.
Additional statistical measures of performance have also been proposed in \cite{krasula2016accuracy}.

In this paper, however, we argue that the accuracy is not the only perspective in which objective quality metrics should be judged, and propose that considering an additional figure of merit provides much more informative insight into the performance and behavior of the metrics, which is their \textit{ambiguity} or, conversely, \textit{reliability}.
In general, the output of an objective metric for a given visual stimulus is expressed as a single value on a continuous scale.
This means that when the predicted quality scores for a pair of stimuli by a metric are obtained, the quality superiority between the stimuli is always formed, no matter how small the difference is.
However, a nonzero quality score difference between two similar stimuli may cause misleading conclusions when the quality difference is not perceivable by human viewers.
In fact, the visual sensitivity of humans is limited in the sense that a small amount of pixel value difference is sometimes visually indistinguishable depending on several factors such as overall luminance and neighboring pixel values \cite{jayant1993signal}.

\begin{figure} %
	\small
	\centering
	\begin{tabular}{cc}
		\includegraphics[trim=2.5cm 1cm 2cm 0cm,scale=0.25]{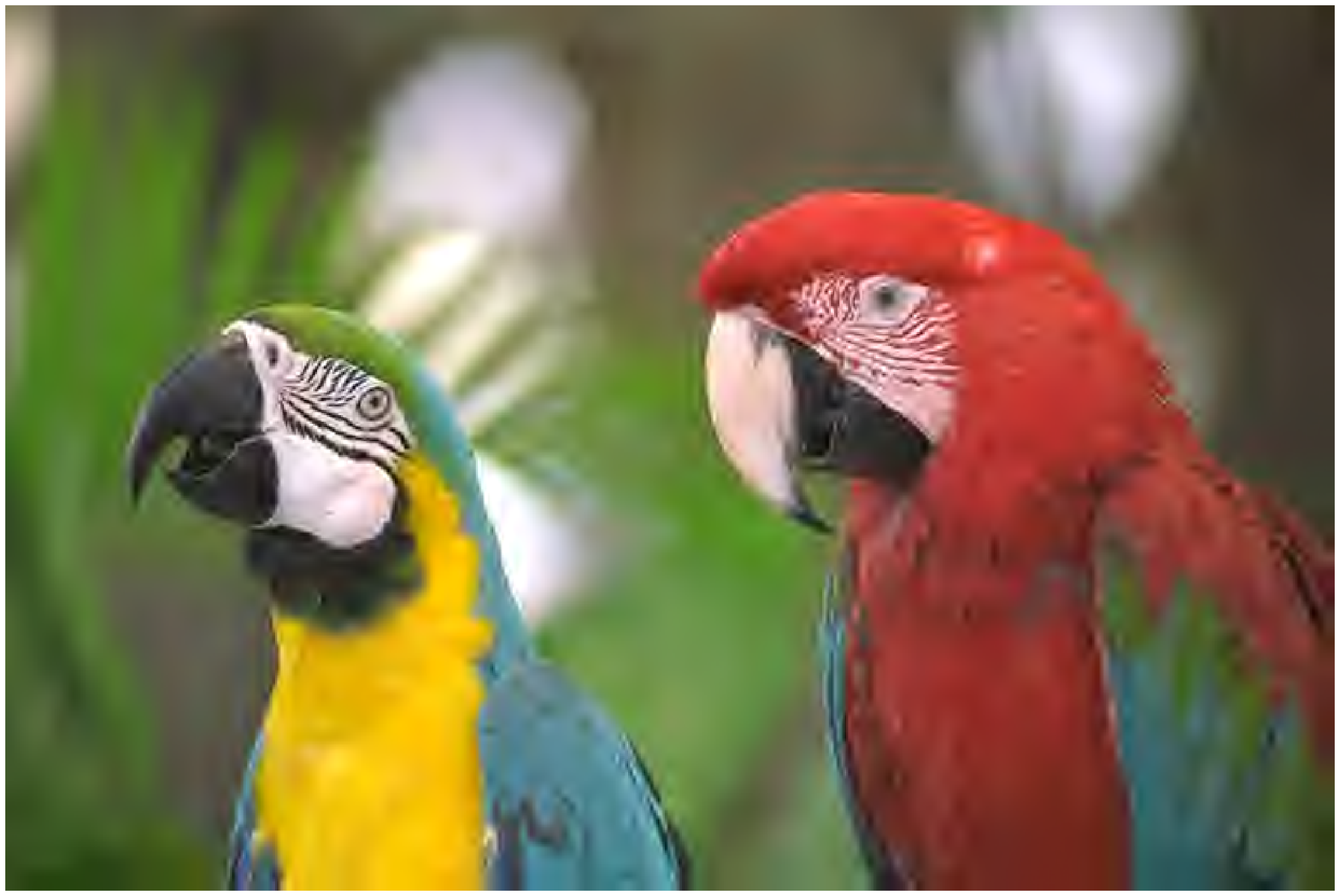}&
		\includegraphics[trim=2cm 1cm 2.5cm 0cm,scale=0.25]{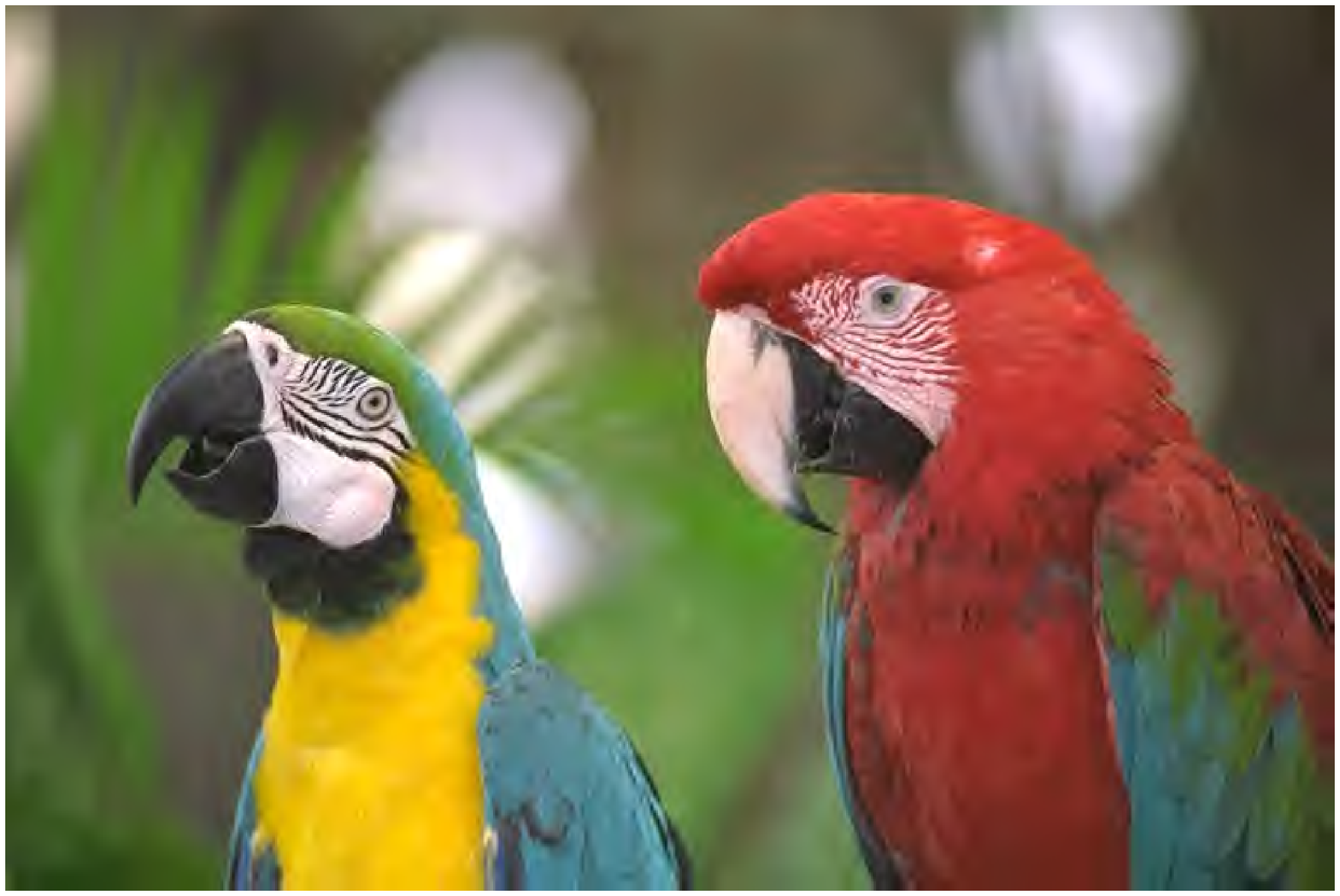}\\
		(a) 30.46 dB &(b) 32.95 dB\\
		\includegraphics[trim=2.5cm 1cm 2cm 0cm,scale=0.25]{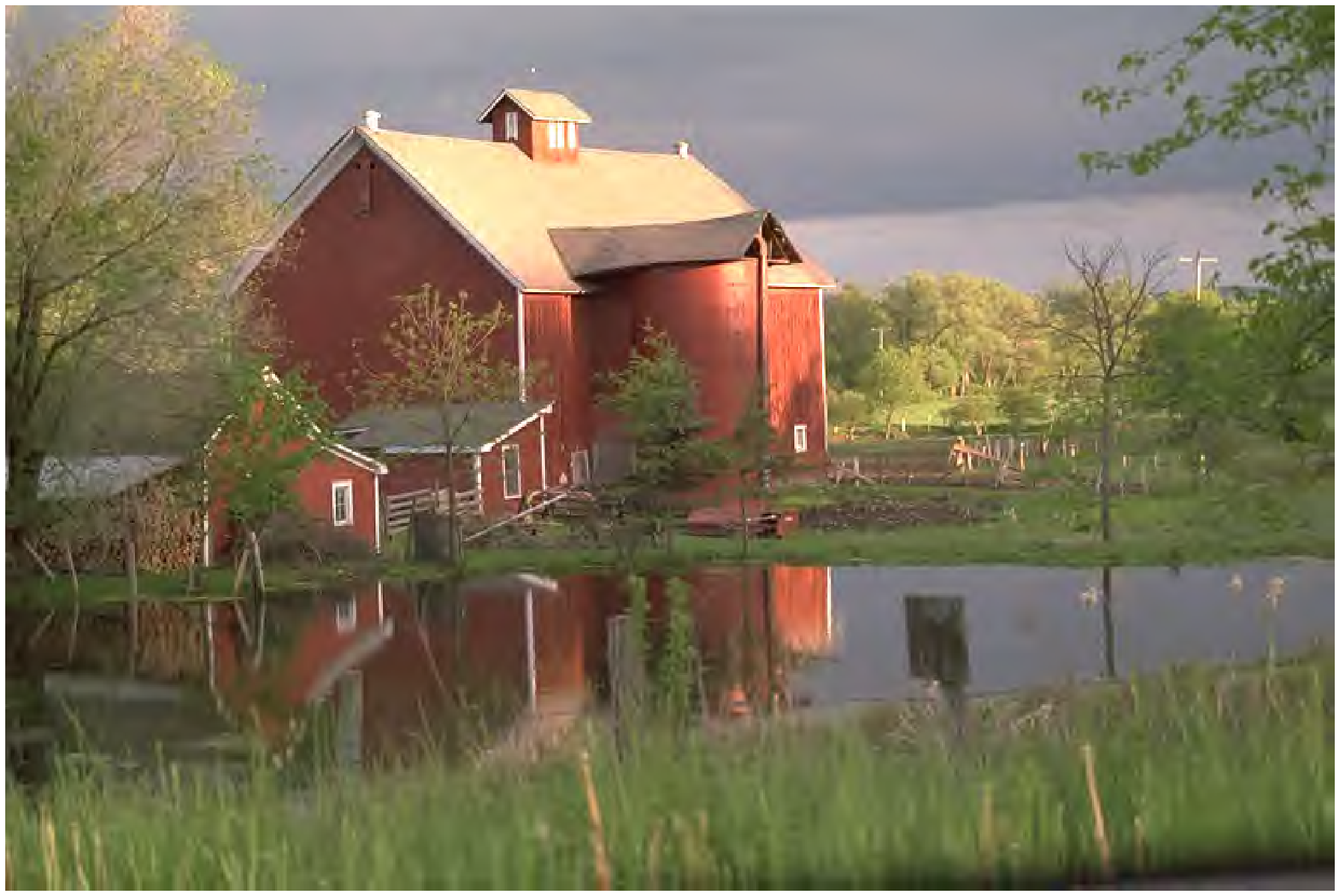}&
		\includegraphics[trim=2cm 1cm 2.5cm 0cm,scale=0.25]{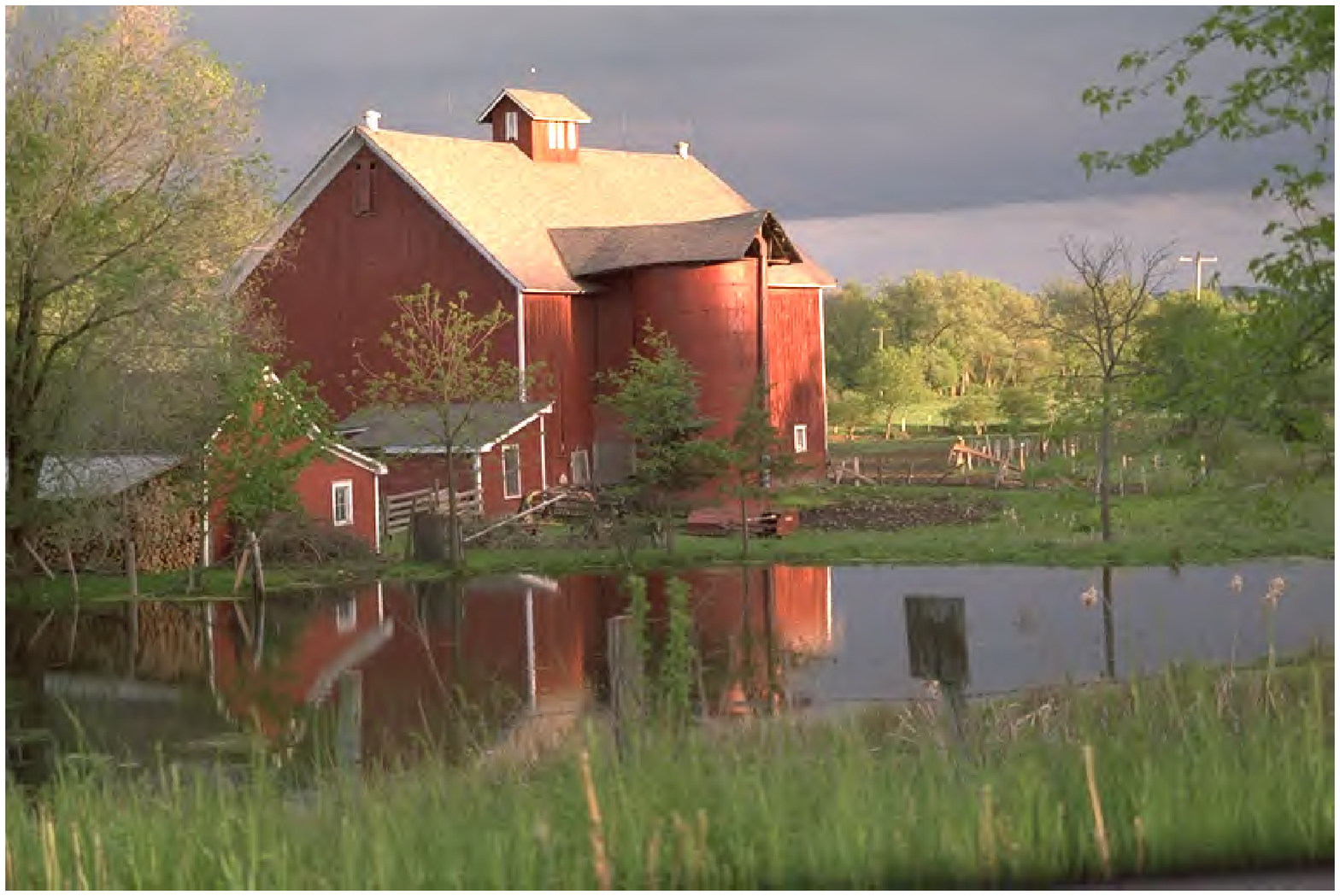}\\
		(c) 30.39 dB &(d) 32.93 dB\\
	\end{tabular}
	\caption{\label{fig:ex_ambiguity} Example images from the LIVE Image Quality Assessment Database \cite{sheikh2006statistical}, demonstrating the ambiguity of objective quality metrics (in this case, PSNR).}
\end{figure}

Figure \ref{fig:ex_ambiguity} shows example images demonstrating the existence of ambiguity of objective metrics \cite{cheon2015ambiguity}.
For two reference images (\textit{parrots} and \textit{house}) from the LIVE Image Quality Assessment Database \cite{sheikh2006statistical}, JPEG2000 compression is applied to corrupt them with different bitrates.
When Figures \ref{fig:ex_ambiguity}(a) and \ref{fig:ex_ambiguity}(b) are visually compared, their quality difference can be easily perceived.
We conducted a subjective quality assessment experiment using the paired comparison scheme \cite{itu_r_bt500, lee2014on}, where most of the hired subjects (14 out of 15) chose Figure \ref{fig:ex_ambiguity}(b) as the one having better quality.
An objective metric, peak signal-to-noise ratio (PSNR), also rates Figure \ref{fig:ex_ambiguity}(b) as having better quality (with a difference of 2.49 dB), which is consistent with the quality superiority perceived by humans.
On the other hand, the difference between Figures \ref{fig:ex_ambiguity}(c) and \ref{fig:ex_ambiguity}(d) is hardly noticeable; nearly a half of the subjects (6 out of 15) chose Figure \ref{fig:ex_ambiguity}(c).
However, the quality measured by PSNR still determines that Figure \ref{fig:ex_ambiguity}(d) is better, showing a difference of 2.54 dB, which is even larger than the difference between Figures \ref{fig:ex_ambiguity}(a) and \ref{fig:ex_ambiguity}(b).
Such inconsistent results between subjective and objective quality measurements are undesirable for quality-optimized multimedia systems.
For instance, a system relying on PSNR may try to deliver Figure \ref{fig:ex_ambiguity}(d) instead of Figure \ref{fig:ex_ambiguity}(c) to improve QoE at the cost of an increased bitrate (20 to 35 kbytes), which is actually not so worthy for users.
An additional observation in this example is the content-dependence of the ambiguity of objective metrics.
In other words, the perceptual insignificance of the PSNR difference is observed only for \textit{house}.

Even the state-of-the-art objective quality metrics showing good performance on predicting perceived quality (e.g.,~\cite{narwaria2015hdr,sheikh2006image,wang2004image}) have the issue of indistinguishable quality ranges because all the existing metrics produce single numerical values representing the perceptual quality of given stimuli, which is highly related to the reliability of the metrics. In this paper, therefore, we address the issue of ambiguity of objective quality assessment and propose an approach to measure the ambiguity as an interval defining the indistinguishable quality score range, which can be applied to any quality metrics to supplement their usefulness in a new direction. Furthermore, we present use cases where the proposed approach can be useful, i.e., one for performance comparison of quality metrics' and the other for analysis of metrics performance in terms of reliability with respect to the viewing distance. 

The main contributions of this paper can be summarized as follows:
\begin{enumerate}

\item \textbf{We propose an approach to measure the ambiguity of objective quality metrics.} \par
The ambiguity is expressed as an interval on the scale of a metric’s score, called \textit{ambiguity interval}, within which the quality difference is perceptually indistinguishable.
In obtaining ambiguity intervals, we incorporate the viewing conditions, in particular, viewing distance, because it is one of the most important factors that significantly influence the visual sensitivity of human viewers.
Our approach employs the visual difference predictor (VDP) \cite{daly1992visible}, which automatically estimates a threshold for perceptually indistinguishable pixel value difference at each pixel location.
Using VDP also eliminates the necessity to conduct subjective experiments to obtain the ambiguity intervals, which maximizes the applicability of the proposed approach.

\item \textbf{We provide a practical use case, i.e., objective metric benchmarking, to demonstrate the effectiveness of the proposed approach.} \par %
We use the ambiguity characteristics of metrics for performance comparison of metrics in addition to the accuracy measure. 
It is shown that the ambiguity can play an important role to determine the superiority among the metrics.
In the research community of multimedia quality assessment, systematic evaluation of objective metrics has been considered important to analyze their advantages and disadvantages \cite{sheikh2006statistical, chikkerur2011objective, cheon2017subjective,mohammadi2015subjective, lin2011perceptual, lee2012comparison,cheon2016evaluation}. 
The Video Quality Experts Group (VQEG), an international forum for perceptual quality assessment towards standardization, also puts a significant amount of efforts for this.
Thus, this use case proposes a novel framework for benchmarking of objective quality metrics, which enables performance analysis of the metrics in multidimensional perspectives.

\item \textbf{As another practical use case, we evaluate state-of-the-art metrics in terms of viewing distance.} \par %
We show that the behavior of a metric depending on the viewing distance also provides valuable information in analyzing the metric’s performance.
Such information can be exploited as a part of benchmarking of objective metrics.
In addition, it can be used to identify proper viewing conditions where the metrics are reliable.

\end{enumerate}

The rest of this paper is organized as follows.
The following section presents the proposed approach in detail.
Section \ref{experiments} describes the experimental setup.
The two use cases, where the ambiguity intervals are exploited, are given in Sections \ref{use1} and \ref{use2}, respectively.
Finally, conclusions are given in Section \ref{conclusion}.

\section{Proposed Method}

\subsection{Approach}

As mentioned in the introduction, the goal of the proposed approach is to obtain an interval for a given objective quality metric, so that a score difference within the interval at that particular quality level is considered as being perceptually insignificant. The core idea to obtain such an interval is to change the amount of distortion (e.g., noise, compression artifacts, etc.) in an image and check using a perceptual model if the change of the distortion would be detected by human observers.

\begin{figure*} %
\centering
\hspace*{-1.5cm}
\includegraphics[trim=0cm 0cm 0cm 0cm,scale=0.3]{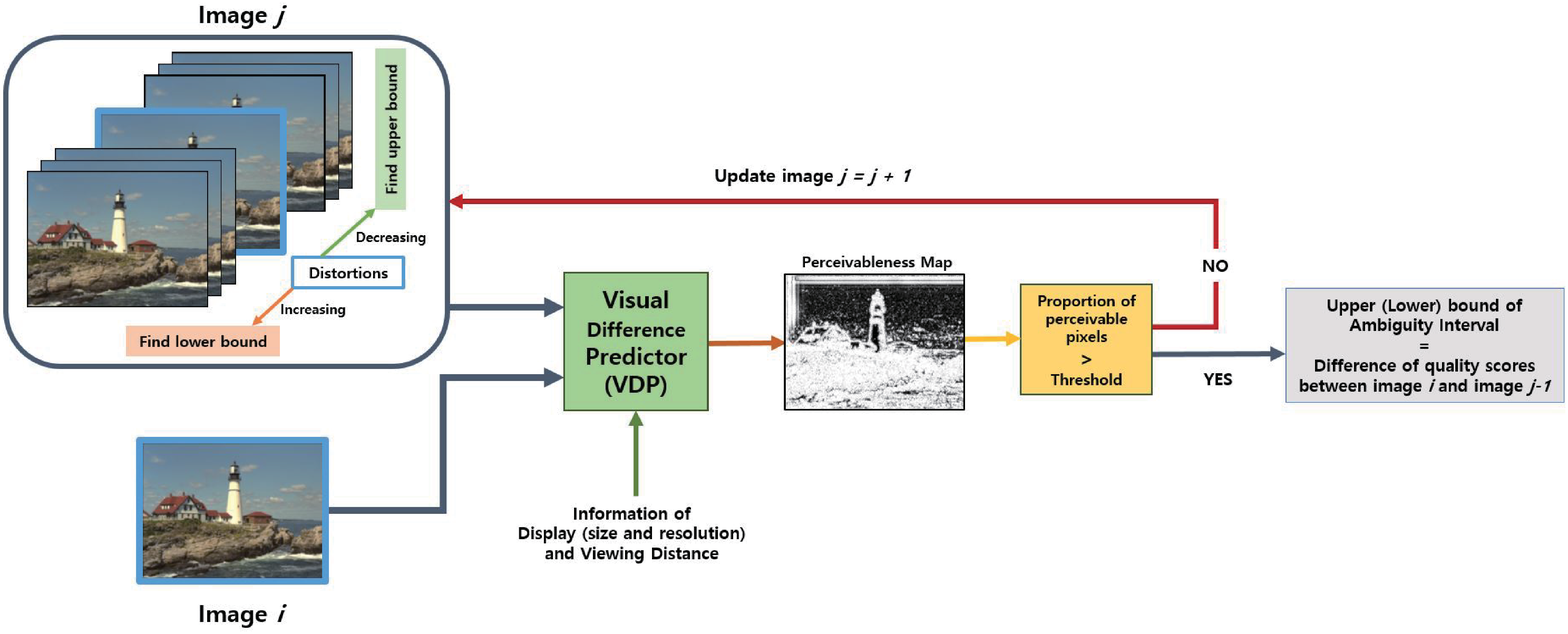}
\caption{\label{fig:procedure} Procedure to obtain an ambiguity interval based on a perceivableness map, which judges whether the two images are perceptually distinguishable or not. Note that the white pixels of the perceivableness map mean distinguishable pixels determined by VDP.}
\end{figure*}

Algorithm \ref{algo} summarizes the procedure of the proposed approach to obtain the ambiguity interval (i.e., the upper and lower bounds of the interval) over the whole quality range for a source image and a type of distortion. Figure \ref{fig:procedure} illustrates the process to obtain the ambiguity interval for a particular quality level corresponding to a degraded image, which corresponds to lines 6 to 19 in Algorithm \ref{algo}.

\begin{algorithm}
\caption{Computing the ambiguity interval}\label{algo}
\begin{algorithmic}[1]
\Require Source image $I_0$ having $M$ pixels
\Ensure Upper bound width $U\in \mathbb{R}^{N}$ and lower bound width $L\in \mathbb{R}^{N}$ of the ambiguity interval

\For {$i \gets 1, N$} \Comment $N$: number of considered quality levels
	\State $I_i\gets$degrade\_image$(I_0,i)$ \Comment Apply quality degradation (compression, blurring, etc.) to $I_0$ ($I_i$ is more degraded than $I_{i-1}$)
	\State $Q_i\gets$measure\_quality$(I_i)$ \Comment Measure the objective quality (assume that a higher $Q_i$ indicates higher quality)
\EndFor

\For {$i \gets 1, N$}
	\For {$j \gets i+1, N$}
		\State PMap $\gets$ vdp($I_i$, $I_j$) \Comment Obtain the perceivableness map
		\If {count(PMap > 0.5)$/M$ > $k$}
			\State $L_i \gets Q_i-Q_{j-1}$ \Comment Obtain the width of the lower bound
			\State break
		\EndIf
	\EndFor
	\For {$j \gets i-1, 1$}
		\State PMap $\gets$ vdp($I_i$, $I_j$) \Comment Obtain the perceivableness map
		\If {count(PMap > 0.5)$/M$ > $k$}
			\State $U_i\gets Q_{j-1}-Q_i$ \Comment Obtain the width of the upper bound
			\State break
		\EndIf
	\EndFor
\EndFor
\State\Return $U$ and $L$

\end{algorithmic}
\end{algorithm}

First, a quality degradation for the distortion type is applied to the source image ($I_0$) with various amounts of distortion and the objective quality levels of the resulting images are measured. 
Then, we determine perceptual distinguishability between two images having different amounts of artifacts. 
For a given image $I_i$ containing a certain type of artifacts, we obtain the level of ambiguity at the corresponding objective quality score ($Q_i$) as an interval around the score. 
We assess perceivable difference of the given image ($I_i$) compared to an image from the same source image but with different amounts of artifacts ($I_j$). 
We gradually increase (or decrease) the amounts of artifacts in $I_j$, until the images that are perceptually distinguishable from the given image are found. 
Among the images that are perceptually indistinguishable from $I_i$, the one with the highest (or lowest) quality level is identified, and the difference between the corresponding quality score and the quality score of $I_i$ is recorded as the width of the upper (or lower) bound of the interval, $U_i$ (or $L_i$).

A visual just-noticeable difference (JND) model is used to determine whether two images having different amounts of distortion are perceptually distinguishable. 
The JND model compares the two images and produces a map having the same size to that of the input images, called \textit{perceivableness} map. 
Each pixel of the map represents the probability that the pixel value difference of the two images at the corresponding location is perceptually distinguishable.
A probability of 0.5 (i.e., random chance) is considered as the threshold of distinguishability.
Therefore, if at most a certain proportion (denoted as $k$) of the pixels of the perceivableness map have values above 0.5, the two images are considered to be perceptually indistinguishable.%

The JND model considered in this study is VDP, originally proposed by Daly \cite{daly1992visible}.
It enables to specify the viewing conditions including the type, resolution, and parameters of the display, together with the viewing distance \cite{mantiuk2011hdr}.
In particular, we use the latest version, known as HDR-VDP 2.2 \cite{narwaria2015hdr}\footnote{An implementation is publicly available at http://hdrvdp.sourceforge.net/wiki/}.
The model quantifies the visible difference between two input images under specific viewing conditions.
The images are firstly passed through a model of the optical retinal pathway, including a simulation of intra-ocular light scatter, photoreceptor spectral sensitivity, luminance masking, and achromatic response.
Further on, they are compared on multiple scales considering the model of neural noise, neural contrast sensitivity, and contrast masking.
Note that when producing a perceivableness map, VDP takes into account the contextual information for each pixel (i.e., its relationship with neighboring pixels).

\begin{figure} %
\small
\centering
\begin{tabular}{cc}
\includegraphics[trim=1.5cm 0cm 1.5cm 0cm, scale = 0.5]{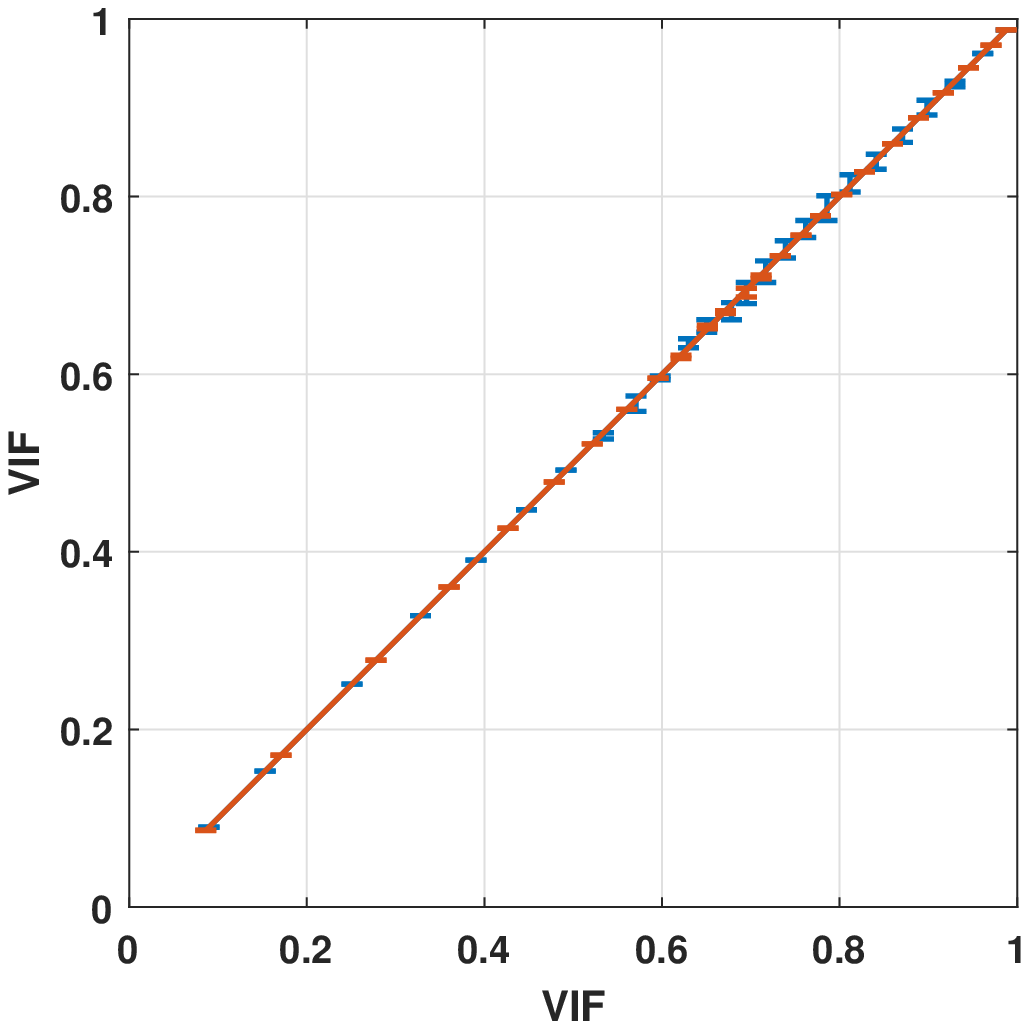}&
\includegraphics[trim=1.5cm 0cm 1.5cm 0cm, scale = 0.5]{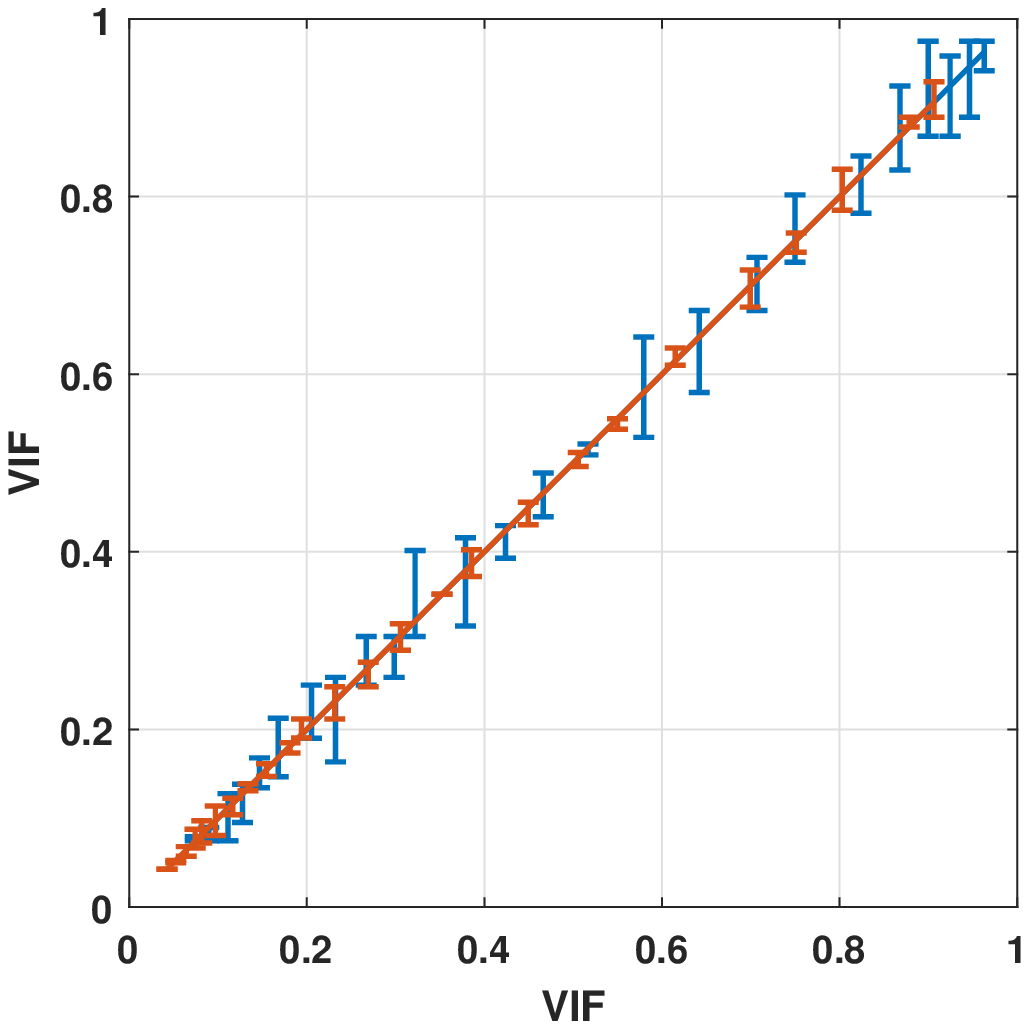}\\
(a) JPEG & (b) JPEG2K\\
\includegraphics[trim=1.5cm 0cm 1.5cm 0cm, scale = 0.5]{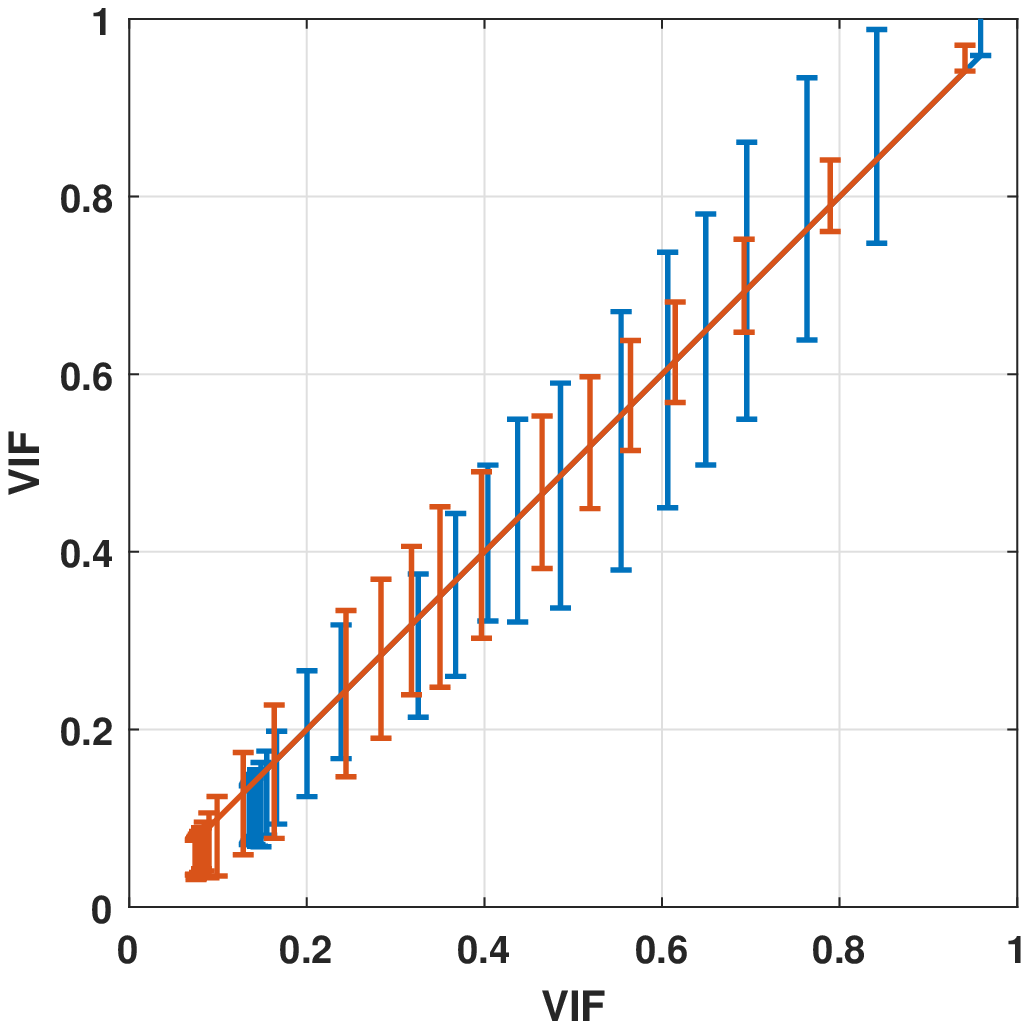}&
\includegraphics[trim=1.5cm 0cm 1.5cm 0cm, scale = 0.5]{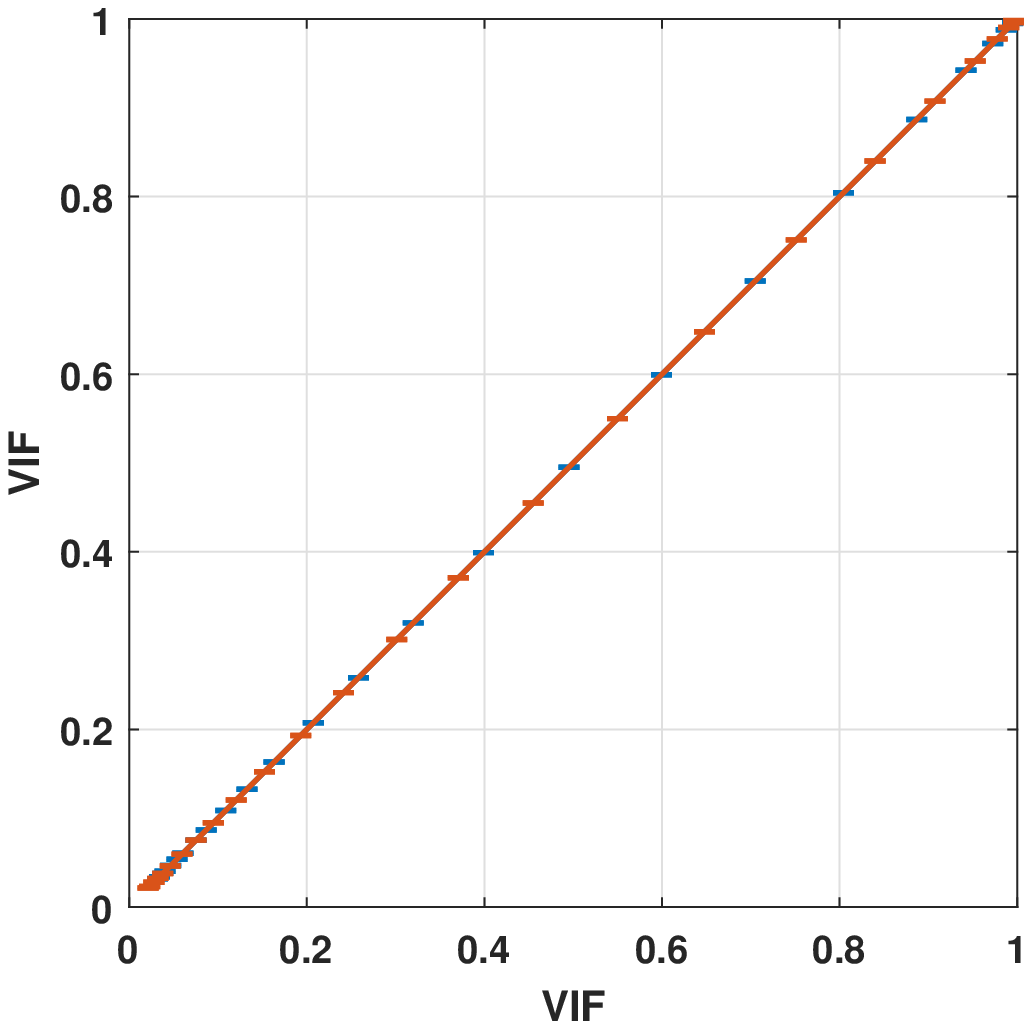}\\
 (c) GB & (d) WN \\
\end{tabular}
\caption{\label{fig:ex_interval} Examples of obtained ambiguity intervals of VIF for the LIVE database. The upper and lower bounds for two different reference images are expressed in different colors. (a) JPEG (b) JPEG2000 (JPEG2K) (c) Gaussian blur (GB) (d) white Gaussian noise (WN)}
\end{figure}

Figure \ref{fig:ex_interval} shows examples of the ambiguity intervals, which are obtained for the visual information fidelity (VIF) metric \cite{sheikh2006image}.
To determine the intervals, we generate $N$=100 images having different amounts of distortion (spanning the whole quality range) for each distortion type and each reference image in the LIVE Image Quality Assessment Database \cite{sheikh2006statistical}, and apply Algorithm \ref{algo} to them.
In the figure, a higher score means a higher quality level, i.e., less artifacts.
Three types of dependency of the interval are observed.
First, the width of the interval is not necessarily uniform over the quality range.
In Figure \ref{fig:ex_interval}(c), for instance, the width of the interval is large for the intermediate quality range and small for low quality (near zero).
This implies that the perceptual scale of the metric is not perfectly linear. 
Second, the interval width is dependent on the content, which is in line with the observation made from Figure \ref{fig:ex_ambiguity}.
This is related to the fact that the visibility of quality degradation is dependent on the image content due to perceptual mechanisms such as frequency-dependent contrast sensitivity, spatial masking, etc.
Third, the type of distortion also influences the interval because the detectability of quality difference depends on the type of artifacts.
Detailed analysis is given in Section \ref{use1}.
In summary, the interval is dependent on the visual components included in the image, which are affected differently by the quality level, the distortion type, and the content itself.

\subsection{Measures for Ambiguity Intervals}

\begin{figure}
\small
\centering
\begin{tabular}{cc}
\includegraphics[trim=1.5cm 0cm 1.5cm 0cm, scale = 0.5]{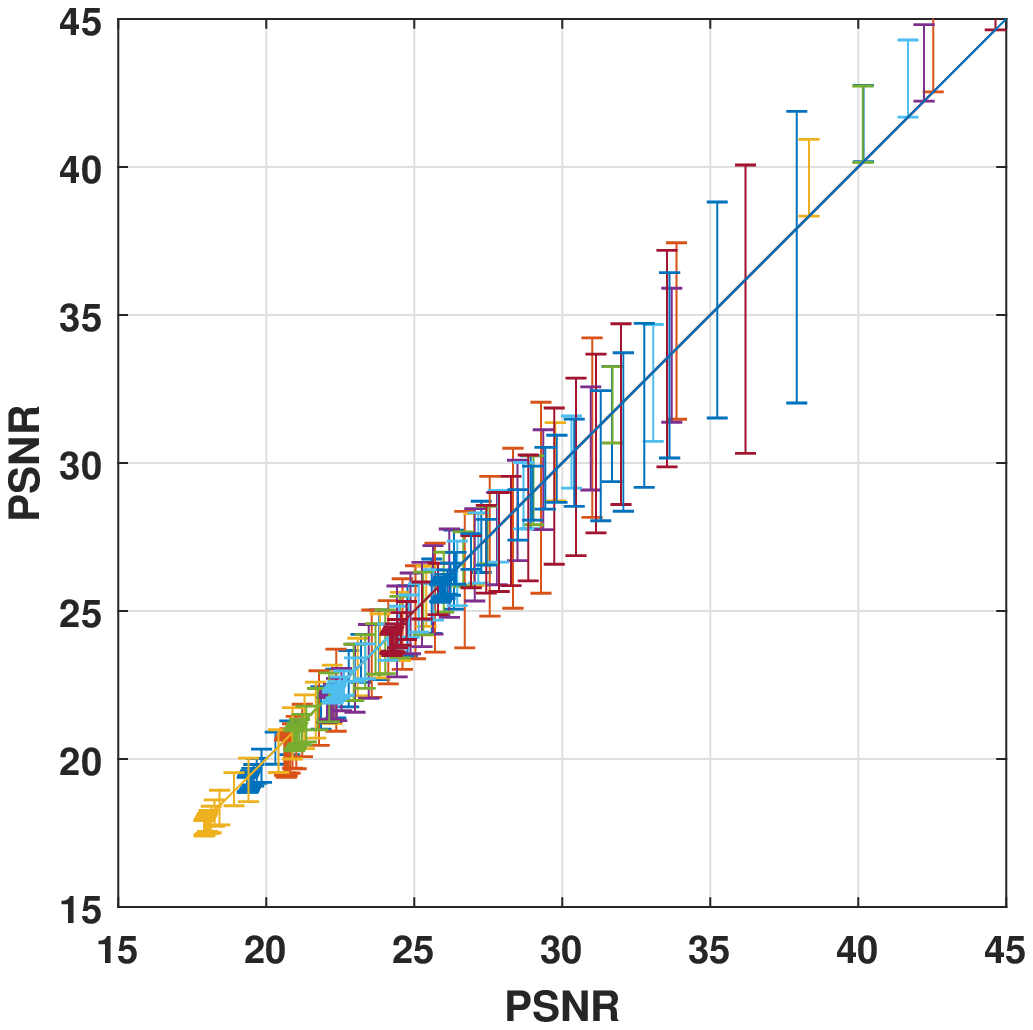}&
\includegraphics[trim=1.5cm 0cm 1.5cm 0cm, scale = 0.5]{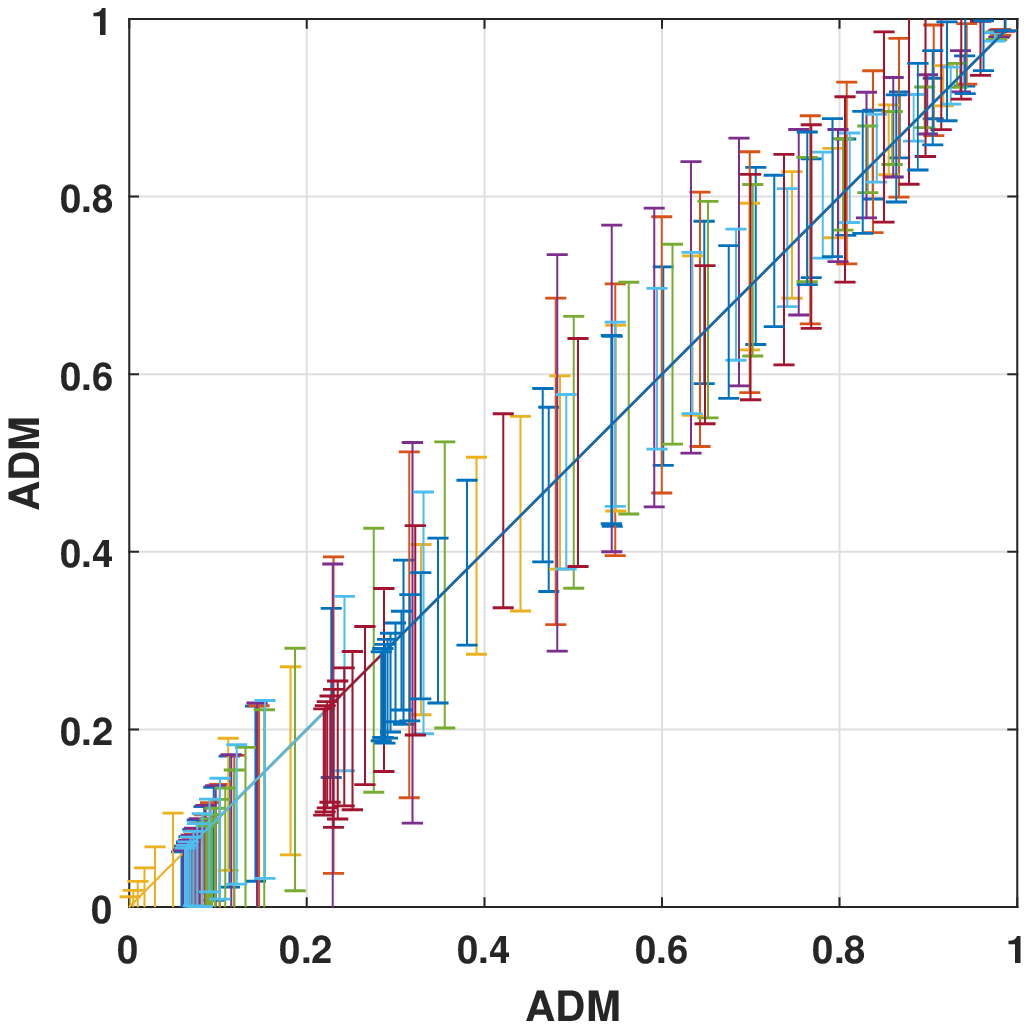}\\
(a) PSNR & (b) ADM\\
\end{tabular}
\caption{\label{fig:ex_ambiguity_metric} Examples of obtained ambiguity intervals for GB of the LIVE database. Different colors mean different reference images. (a) PSNR (b) ADM}
\end{figure}

The ambiguity intervals of an objective metric can be used to measure the performance of the metric in terms of quality resolution.
Figure \ref{fig:ex_ambiguity_metric} shows examples for two different metrics, i.e., PSNR and additive impairment and detail loss measure (ADM) \cite{li2011image}, which have different output ranges and ambiguity interval widths.
Overall, for instance, the intervals of ADM are larger than those of PSNR; the intervals of PSNR are relatively small for the low quality range and get larger as the quality increases, whereas the intervals of ADM are more uniform over the whole range.
To enable easy comparison between the intervals of different metrics, we compute measures that summarize the ambiguity intervals of a metric.
As the first step, the ambiguity intervals of a metric are normalized with the obtained output range of the metric, since different metrics may have different ranges and units.
Note that in our preliminary work \cite{cheon2016ambiguity}, nonlinear regression using subjective rating data was employed for normalization, which limits the applicability of the method only to the cases where subjective data are available.
In addition, only the quality levels associated with subjective ratings were used, which permitted ambiguity evaluation only at a coarse level.

We propose to compute three statistics of the ambiguity intervals, namely, the mean, maximum, and standard deviation of the widths of the ambiguity intervals over the whole quality range in order to measure the performance of a metric in multiple aspects of ambiguity.
They are measures of the sensitivity of a metric in an average sense, the coarsest quality resolution, and the uniformity of the quality resolution, respectively.
The smaller each of these measures is, the better the performance of the metric is.

\section{Experimental setup}
\label{experiments}

We conduct experiments in order to demonstrate applications where the proposed approach can be exploited effectively, which are shown in the following sections.
This section explains the employed databases and the objective metrics considered in the experiments.

\subsection{Databases}

\begin{table}
\small
\centering
\caption{\label{tb:dbinfo} Characteristics of the three databases used for the experiments. A viewing distance is expressed as a multiple of the height of the display.}
\scalebox{0.85}{
\begin{tabular}{c|c|c|c|c|c|c}  \specialrule{.2em}{.1em}{.1em} 
Database & \# Contents & \makecell{Image \\ resolutions} & \makecell{Distortion \\ types} & Screen & \makecell{Viewing \\ distance} & \makecell{Subjective \\ ratings} \\ \hline
\multirow{2}{*}{LIVE \cite{sheikh2006statistical}} & \multirow{2}{*}{29} & 768$\times$512 & JPEG, JPEG2K&sRGB, CRT &\multirow{2}{*}{2H} & \multirow{2}{*}{DMOS} \\
& & 512$\times$512 & GB, WN & 21-inch, 1024$\times$768 & & \\ \hline
\multirow{2}{*}{VDID \cite{gu2015quality}} & \multirow{2}{*}{8} & 768$\times$512 & JPEG, JPEG2K&sRGB, LCD &4H & \multirow{2}{*}{DMOS} \\
& & 512$\times$512 & GB, WN & 23-inch, 1920$\times$1080 & 6H & \\ \hline
\multirow{2}{*}{CIDIQ \cite{liu2014cid}} & \multirow{2}{*}{23} & \multirow{2}{*}{800$\times$800} & JPEG, JPEG2K&sRGB, LCD &1.5H & \multirow{2}{*}{MOS} \\
& & & GB, PN & 24-inch, 1920$\times$1080 & 3H & \\ \specialrule{.2em}{.1em}{.1em} 
\end{tabular}}
\end{table}

We employ three databases that are popularly used in the research of perceptual quality assessment, i.e., the LIVE Image Quality Assessment Database (LIVE) \cite{sheikh2006statistical}, which is one of the most popular databases for benchmarking objective metrics, the Viewing Distance-changed Image Database (VDID) \cite{gu2015quality}, which is the first image quality assessment database specifically established for varying viewing distances, and the Colourlab Image Database: Image Quality (CIDIQ) \cite{liu2014cid}, which also contains subjective data for multiple viewing distances.
The databases were produced based on different experimental setups such as reference images, distortion types, screens, viewing distances, etc.
We select them to ensure reproducibility of distortion types and availability of information regarding viewing environments, e.g., information of the screen and viewing distance.
Table \ref{tb:dbinfo} summarizes the characteristics of the databases.
Four common distortion types are selected, i.e., JPEG compression, JPEG2000 (JPEG2K) compression, Gaussian blur (GB), and white Gaussian noise (WN).
For the CIDIQ database, Poisson noise (PN) is considered instead of WN.
JPEG and JPEG2K are well known compression schemes for images, and GB and WN (or PN) are distortions that can easily occur in pre- or post-processing of images.
VDID and CIDIQ have subjective results from two different viewing distances.

\subsection{Objective metrics}

We consider 33 state-of-the-art objective quality metrics (28 full-reference (FR) metrics, one reduced-reference (RR) metric, and four no-reference (NR) metrics) for benchmarking.
The tested FR metrics are PSNR, structural similarity index (SSIM) \cite{wang2004image}, multi-scale structural similarity (MS-SSIM) \cite{wang2003multiscale}, visual signal-to-noise ratio (VSNR) \cite{chandler2007vsnr}, VIF \cite{sheikh2006image}, universal image quality index (UQI) \cite{wang2002universal}, information fidelity criterion (IFC) \cite{sheikh2005information}, noise quality measure (NQM) \cite{damera2000image}, weighted signal to noise ratio (WSNR) \cite{damera2000image}, modified versions of PSNR (PSNR-HVS \cite{egiazarian2006new}, PSNR-HVS-M \cite{ponomarenko2007between}, PSNR-HMA, PSNR-HA, PSNR-HMA-C, and PSNR-HA-C \cite{ponomarenko2011modified}), optimal scale selection (OSS)-PSNR and OSS-SSIM \cite{gu2015quality}, information content weighted SSIM (IW-SSIM) \cite{wang2011information}, feature similarity index (FSIM) and chrominance extension of FSIM (FSIM-C) \cite{zhang2011fsim}, gradient magnitude similarity deviation (GMSD) \cite{xue2014gradient}, most apparent distortion (MAD) \cite{larson2010most}, ADM \cite{li2011image}, analysis of distortion distribution-based (ADD)-SSIM \cite{gu2016analysis}, ADD-gradient similarity index (ADD-GSIM) \cite{gu2016analysis}, a visual saliency-induced index (VSI) \cite{zhang2014vsi}, image quality assessment based on gradient similarity (GSM) \cite{liu2012image}, and perceptual similarity (PSIM) \cite{PSIM}.
The RR metric is reduced reference entropic differencing index (RRED) \cite{soundararajan2012rred}, and the NR metrics are spatial-spectral entropy-based quality (SSEQ) \cite{liu2014no}, oriented gradients image quality assessment (OG-IQA) \cite{liu2016blind}, blind image integrity notator using DCT statistics (BLIINDS2) \cite{saad2012blind}, and accelerated screen image quality evaluator (ASIQE) \cite{ASIQE}.

\section{Use case 1 : Benchmarking of objective metrics}
\label{use1}

Objective quality metrics that can automatically predict perceived quality of visual content are a key component of quality-optimized multimedia systems.
For instance, a method enhancing a given degraded image requires an objective metric as a criterion with respect to which the image is enhanced.
Therefore, it is critical to identify a quality metric that mimics the human visual system as closely as possible, so that the results of optimization based on the metric are also optimal for human viewers.
In this context, benchmarking studies of objective quality metrics have been conducted extensively in literature, e.g., \cite{cheon2017subjective, lee2012comparison, hanhart2013benchmarking, tian19benchmark}.
In these studies, as mentioned in the introduction, the prediction accuracy of existing metrics is considered as the most important performance index, which is typically measured in terms of PLCC, SROCC, OR, and RMSE.
However, different metrics have different levels of ambiguity, which can be captured by the proposed approach.
The use case presented in this section demonstrates how such information can be effectively used in the benchmarking.

In this use case, we use the LIVE database.
The accuracy performance of the 33 state-of-the-art objective metrics is measured by PLCC between the ground truth subjective quality scores and the predicted quality scores\footnote{Other measures such as SROCC, OR, and RMSE can be also used, but we use only PLCC for conciseness of presentation.}.
In particular, PLCC is computed after nonlinear regression using the monotonic logistic function:
\begin{equation}\label{eq:oqf2}
{ Q }^{ ' }\quad =\quad { \beta  }_{ 1 }+\frac { { \beta  }_{ 2 }-{ \beta  }_{ 1 } }{ 1+{ e }^{ -\left( \frac { Q-{ \beta  }_{ 3 } }{ { \beta  }_{ 4 } }  \right)  } } 
\end{equation}
to fit the objective scores outputted by a metric to the subjective quality scores, as described in the recommendation \cite{video2000final}.
Here, $Q$ and $Q'$ denote the objective scores before and after regression, respectively.
The initial values of the parameters ($\beta_1$ to $\beta_4$) are set as suggested in \cite{video2000final}.
In addition, the statistical tests are also conducted \cite{itu_t_p1401}, i.e., Z-tests are performed using the Fisher z-transformation for PLCC.
The ambiguity performance of the metrics is evaluated based on the proposed approach.
The mean, maximum, and standard deviation of the widths of the ambiguity intervals are obtained.
In addition, non-parametric Wilcoxon-Mann-Whitney tests are conducted to statistically compare the ambiguity intervals of different metrics.

\begin{figure*} %
\small
\hspace{-2.5cm}
\begin{tabular}{cc}
\includegraphics[trim=0cm 0cm 0cm 0cm, scale = 0.35]{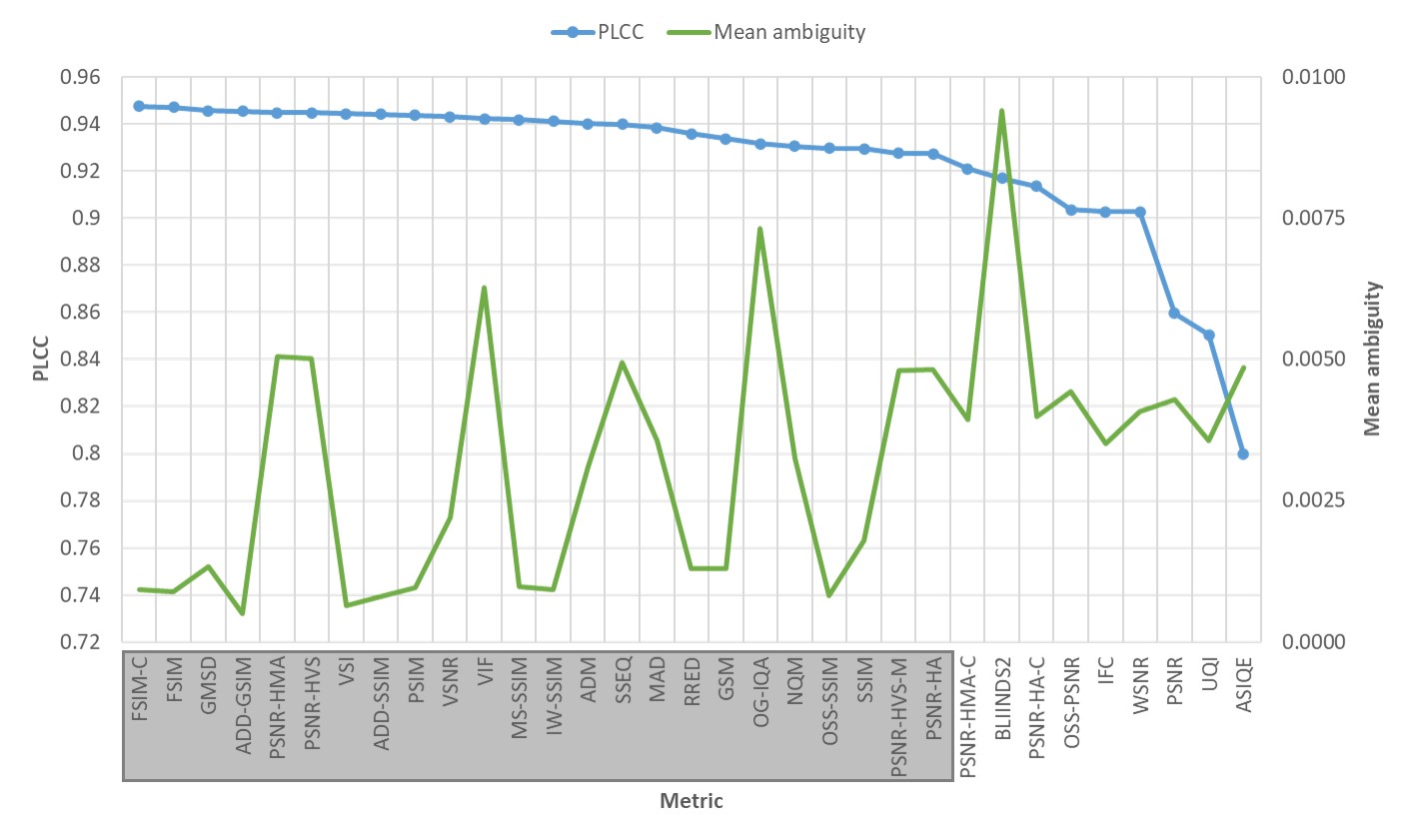}&
\includegraphics[trim=0cm 0cm 0cm 0cm, scale = 0.35]{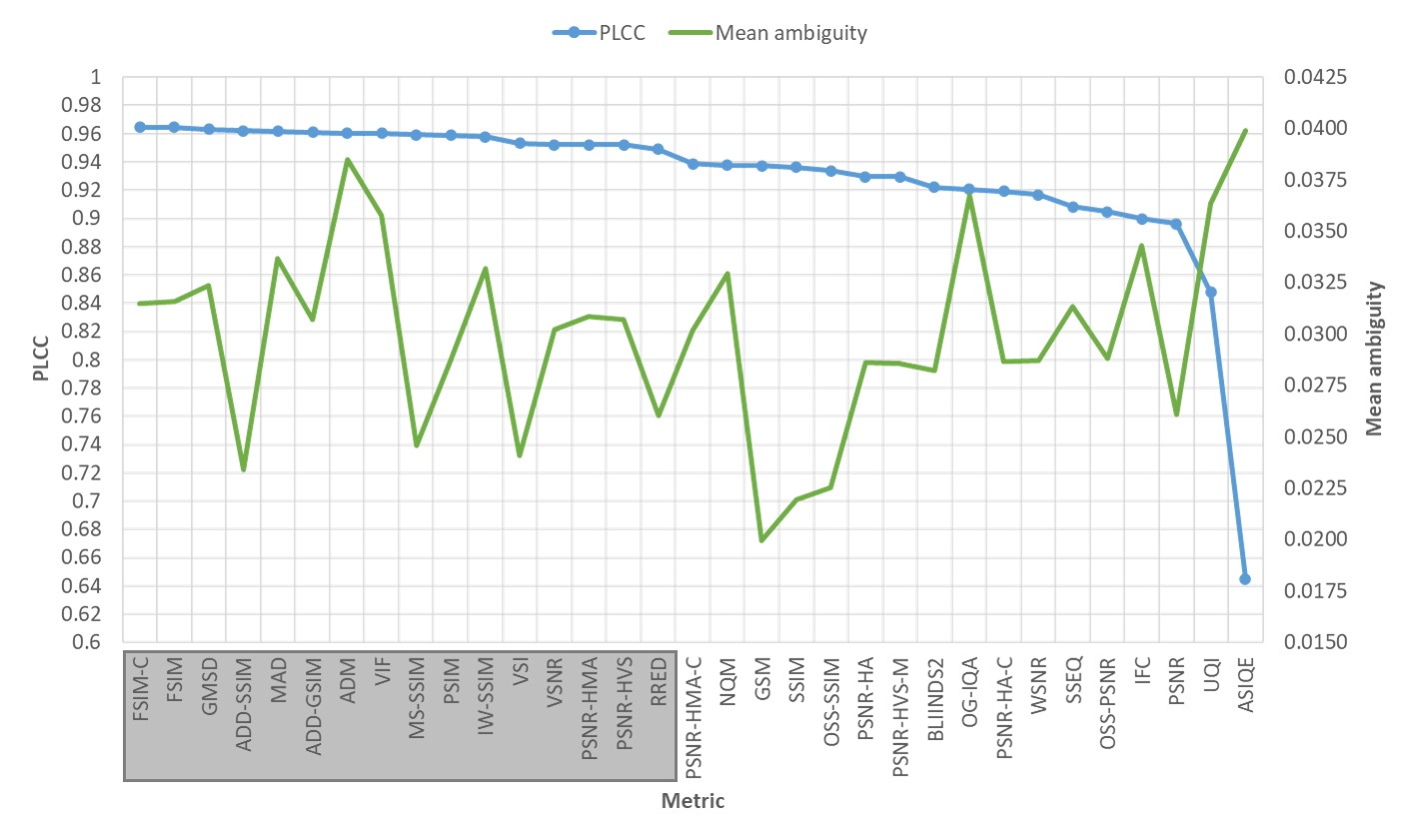}\\
(a) JPEG&(b) JPEG2K\\\\
\includegraphics[trim=0cm 0cm 0cm 0cm, scale = 
0.35]{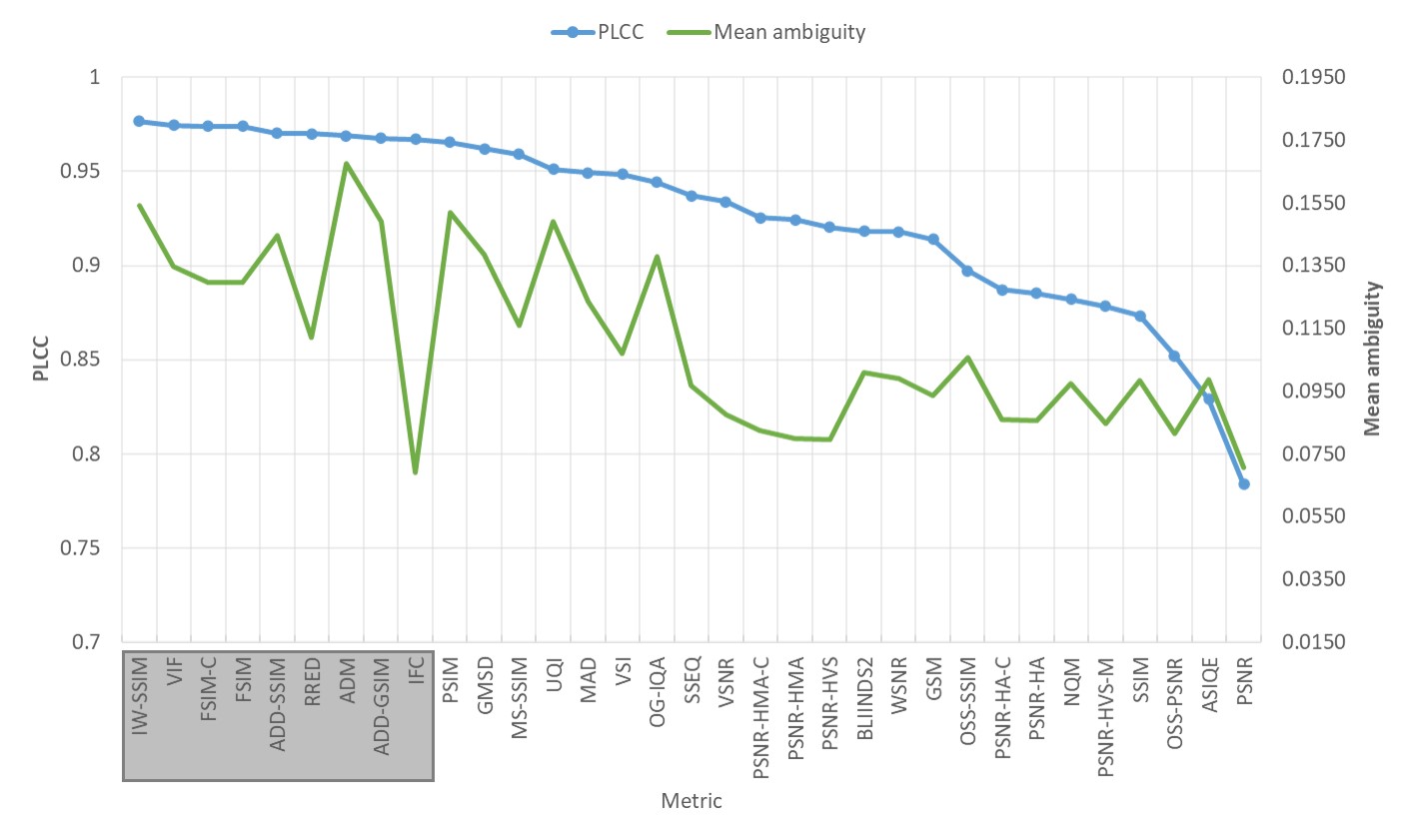}&
\includegraphics[trim=0cm 0cm 0cm 0cm, scale = 
0.35]{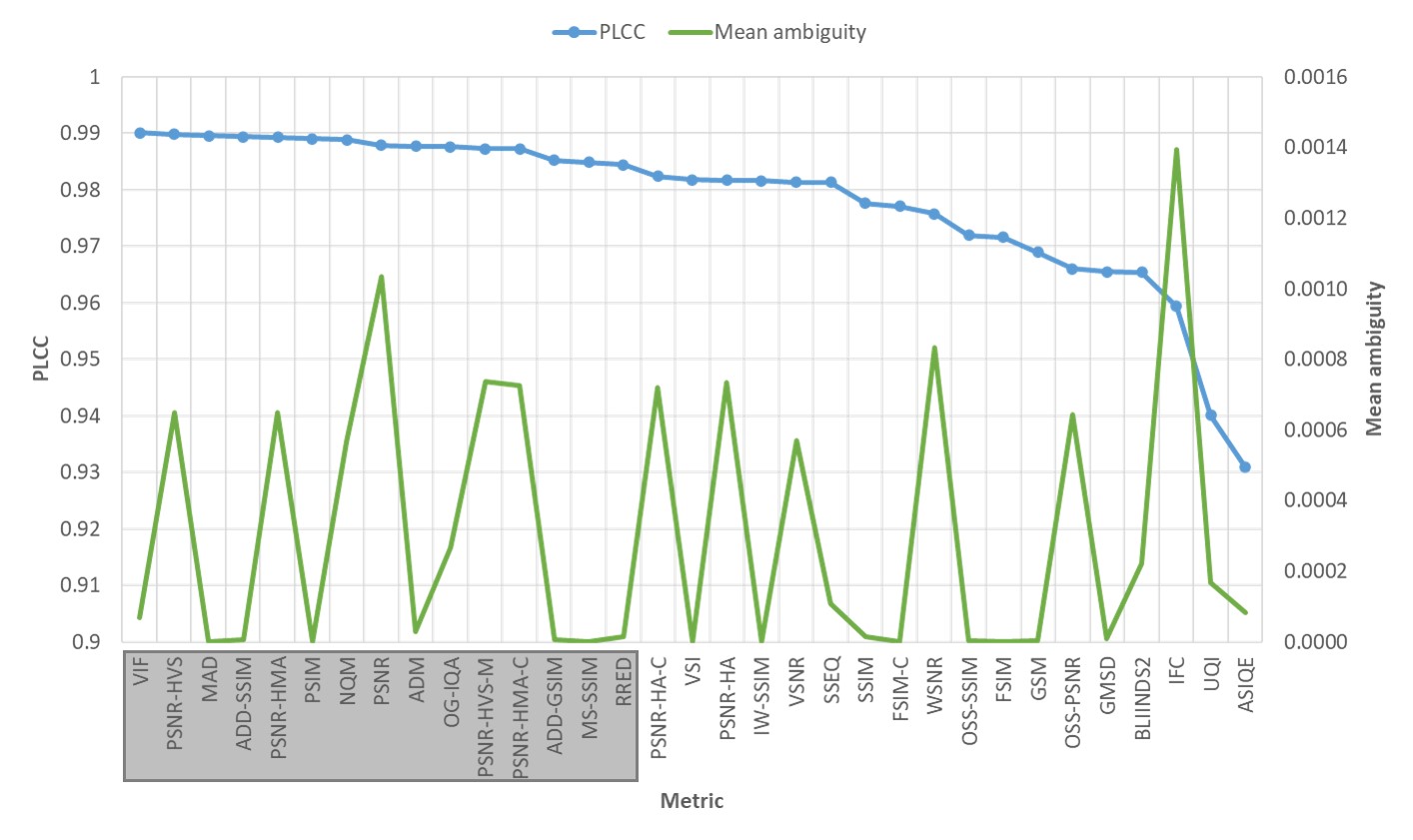}\\
(c) GB&(d) WN\\
\end{tabular}
\caption{\label{fig:accamb-livejpeg}Performance of the objective metrics in terms of Pearson's linear correlation coefficient (PLCC) scores (blue) and mean of ambiguity intervals (green) for the LIVE database. (a) JPEG, (b) JPEG2K, (c) GB, and (d) WN. The metrics are listed in a descending order of the PLCC scores. The statistically equivalent metrics with the best metric for PLCC are marked in a gray box.}
\end{figure*}

Figure \ref{fig:accamb-livejpeg} summarizes the PLCC values and the mean ambiguity intervals of the 33 metrics.
The results for the four distortion types are shown separately, and the metrics are listed in a descending order of the PLCC values.
In the figure, the metrics showing statistically equivalent performance with the best metric in terms of PLCC are marked in the gray box.
We can observe that the superiority of a metric over the others in terms of accuracy may not coincide with its superiority in terms of ambiguity, and vice versa.
For instance, in Figure \ref{fig:accamb-livejpeg}(b), the best metric in terms of accuracy is FSIM-C, but GSM, which is statistically significantly inferior to FSIM-C, is the best in terms of ambiguity.

Many metrics predict perceived image quality with high accuracy.
For instance, the best metric in terms of PLCC for JPEG in Figure \ref{fig:accamb-livejpeg}(a), i.e., FSIM-C, which shows PLCC of about 0.95, is not statistically different with PSNR-HA, which ranks 24th.
Thus, it would be difficult to distinguish the superiority between these metrics.
At this point, we can apply the results of the ambiguity analysis.
Among the top 24 metrics, ADD-GSIM has the smallest mean width of the ambiguity intervals, which is revealed to be significantly smaller than the second smallest one (VSI) by the statistical test ($p<0.01$).
For the other types of distortion, similar trends are also observed, i.e., a number of the metrics show similar performance in terms of PLCC and their performance is not statistically different from that of the best metric, and we can use ambiguity intervals in order to choose the best metric for these cases.
From this approach, ADD-SSIM, IFC, and MAD are selected as the best metrics for JPEG2K, GB, and WN, respectively.

The mean ambiguity interval widths are different depending on the distortion type.
The average values for all metrics are 0.0032, 0.0300, 0.1104, and 0.0003 for JPEG, JPEG2K, GB, and WN, respectively.
The smallest ambiguity intervals are produced for WN, because changes of the amount of white noise can be easily detected compared to the other types of distortion.
The GB distortion yields the largest ambiguity intervals because the change of the strength of GB is relatively hard to distinguish.
JPEG2K also has relatively large ambiguity intervals because the introduced artifacts in the images are quite similar to those by GB.

\begin{figure}
\small
\centering
\begin{tabular}{c}
\includegraphics[trim=2cm 2cm 2cm 0cm, scale = 0.25]{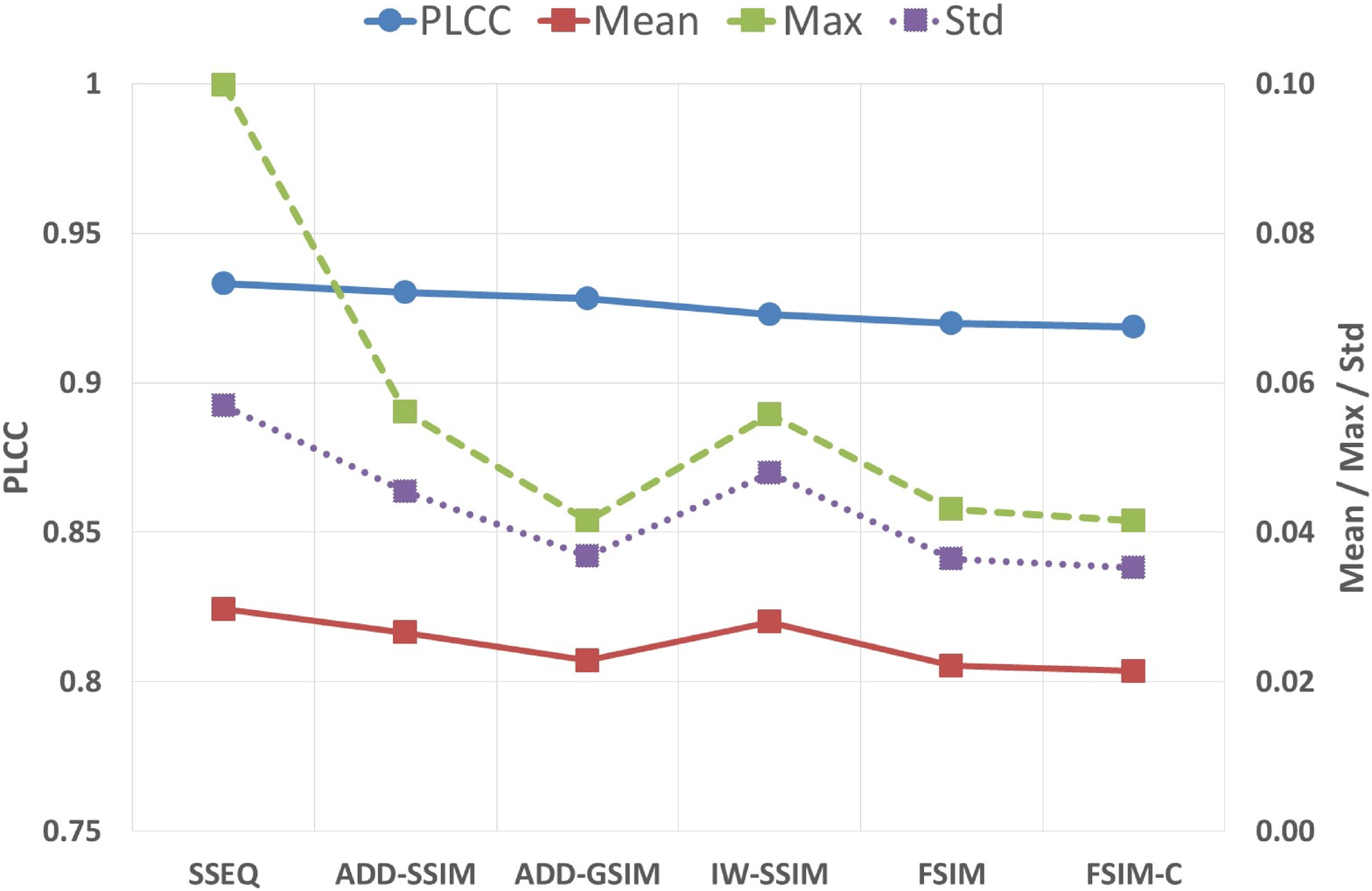}
\end{tabular}
\caption{\label{fig:accamb-live-all-ex} Performance of the top-performing objective metrics showing statistically equivalent PLCC values for all data of the LIVE database. PLCC scores and the mean, maximum, and standard deviation values of ambiguity intervals are shown.}
\end{figure}

In addition to the mean width of the ambiguity intervals, the maximum and standard deviation of the intervals can be also considered in order to analyze the performance of metrics and find the superiority between them.
Figure \ref{fig:accamb-live-all-ex} shows the performance of the objective metrics for all distortion types, which have statistically equivalent performance in terms of PLCC.
When we compare the metrics, e.g., SSEQ and IW-SSIM, the two metrics show similar performance based on PLCC and the mean ambiguity interval.
The maximum and standard deviation of the ambiguity intervals are smaller for IW-SSIM, which can be regarded as a better metric; a low standard deviation of the ambiguity intervals means that it has a uniform quality resolution (or ambiguity) for all quality ranges, which is useful in applications where the metric needs to operate in a wide range of quality.
As another example, ADD-SSIM, ADD-GSIM, FSIM, and FSIM-C show statistically equivalent performance in terms of the mean ambiguity intervals, showing mean ambiguity intervals of only about 2.0-2.5\% of the whole quality range.
However, the maximum and standard deviation of the ambiguity intervals of ADD-SSIM are larger than those of the other three metrics, and thus it may be less preferable.
Therefore, considering all the ambiguity measures, ADD-GSIM, FSIM, and FSIM-C can be regarded as the best metrics.

\section{Use case 2 : Viewing distance vs. ambiguity}
\label{use2}

The viewing distance is one of the most important factors that influence visual quality perception of human viewers. 
As the distance from a viewer to an image gets large, less and less details in the image are distinguished, changes or artifacts in the image become less noticeable, and the viewer's quality perception becomes less reliable.
The proposed approach incorporates this tendency by employing the VDP method that considers the viewing environment including the viewing distance.

\begin{figure} %
\small
\centering
\begin{tabular}{c}
\includegraphics[trim=0cm 0cm 0cm 0cm, scale = 0.6]{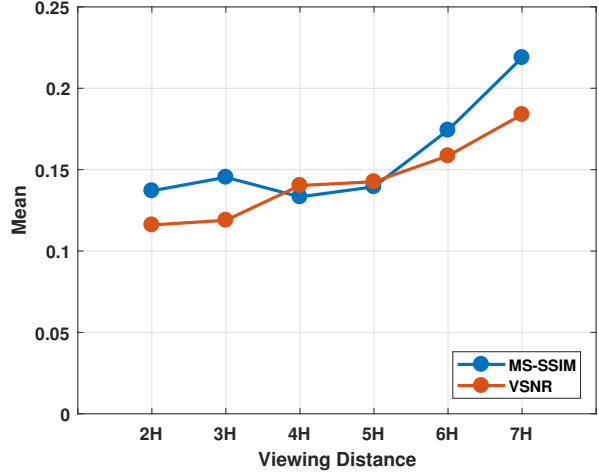}\\
\end{tabular}
\caption{\label{fig:vdid-metric-ex-gm12}Examples of ambiguity intervals of two objective metrics, MS-SSIM (blue) and VSNR (orange), with respect to the viewing distance (in multiples of the display height) for GB of the VDID database. The superiority of a metric against the other varies depending on the viewing distance.}
\end{figure}

However, due to the difference in underlying mechanisms of different objective metrics, they may show different ambiguity patterns with respect to the viewing distance.
For instance, the superiority of the metrics in terms of the ambiguity interval may change depending on the viewing distance.
Figure \ref{fig:vdid-metric-ex-gm12} shows the mean ambiguity intervals of two metrics, MS-SSIM and VSNR, for GB of the VDID database with respect to the viewing distance.
When the viewing distance is 4 or 5 times the display height (i.e., 4H or 5H), MS-SSIM shows slightly smaller ambiguity intervals than VSNR, whereas VSNR shows smaller intervals than MS-SSIM for the other viewing distances.
Thus, the viewing distance should be considered carefully when the ambiguity of a metric is evaluated.
In general, it is preferable for a metric not only to have high accuracy and low ambiguity for a particular viewing distance, but also to show consistent performance over various viewing distances in terms of both accuracy and ambiguity.

In this section, we demonstrate that the ambiguity behavior of metrics with respect to the viewing distance can be used to compare the reliability performance of the metrics, which can be seen as an extension of the benchmarking in the previous section, and to identify proper viewing distances for which a metric can be used reliably.
The VDID and CIDIQ databases are used.

\begin{figure}
\small
\centering
\begin{tabular}{c}
\includegraphics[trim=0cm 0cm 0cm 0cm, scale = 0.5]{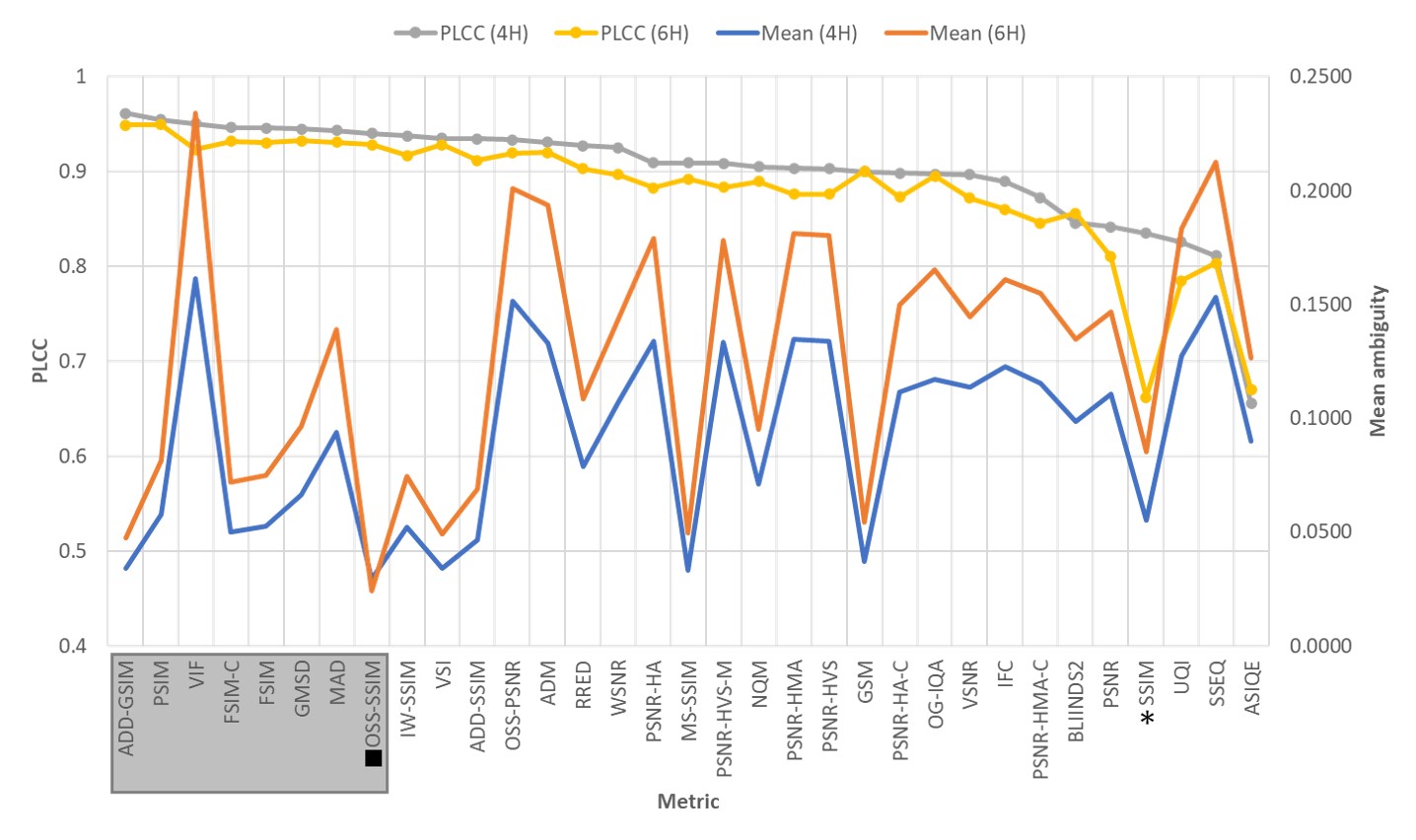}\\
(a) VDID \\\\
\includegraphics[trim=0cm 0cm 0cm 0cm, scale = 0.5]{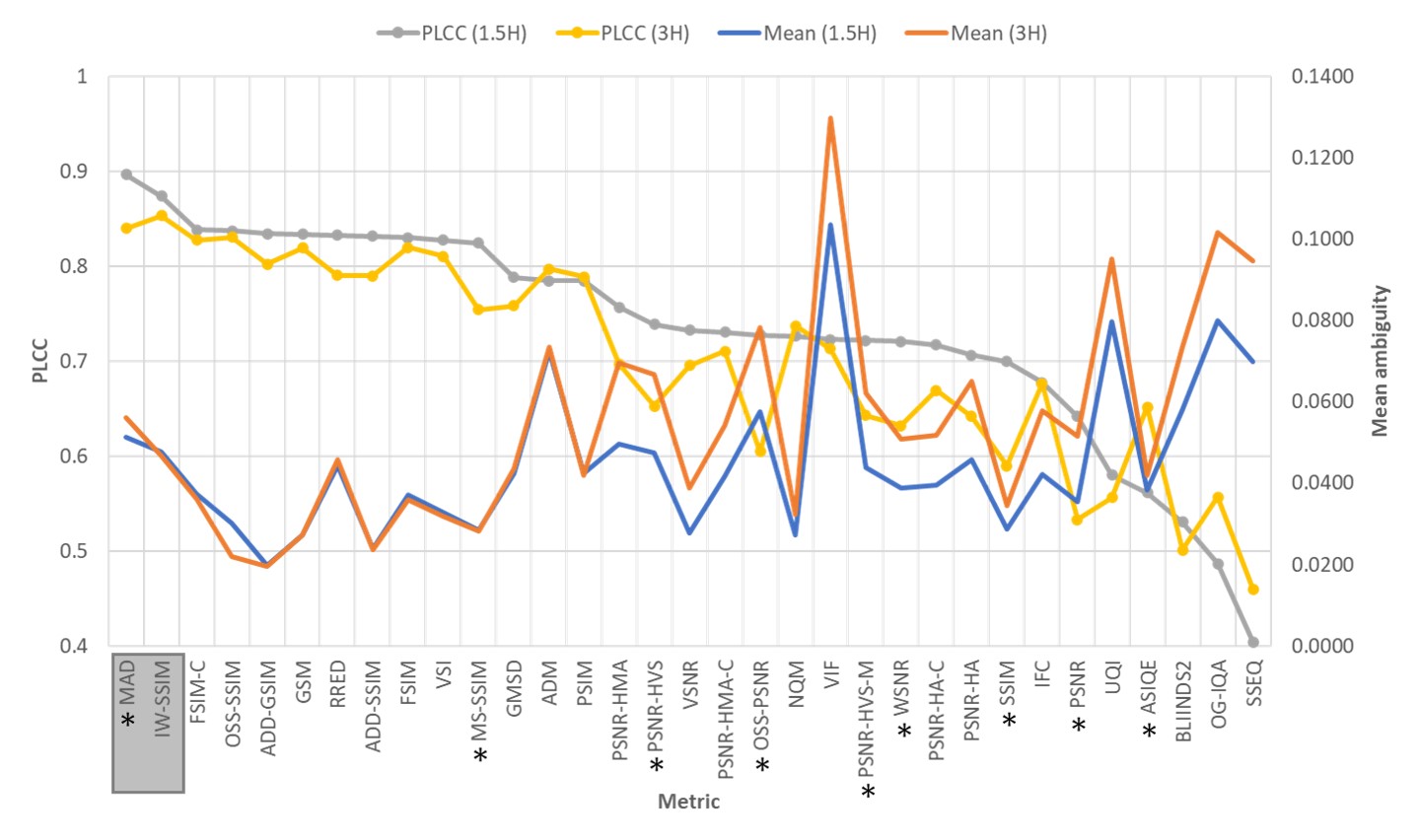}\\
(b) CIDIQ\\
\end{tabular}
\caption{\label{fig:accamb-vdidcidiq-all} Performance of the metrics for two viewing distances in terms of PLCC scores and mean of ambiguity intervals for the (a) VDID and (b) CIDIQ databases. The metrics are listed in a descending order of the PLCC for the short viewing distances (4H for VDID and 1.5H for CIDIQ). The metrics having statistically different accuracy between the two viewing distances are marked with asterisks. The statistically equivalent metrics with the best metric in terms of PLCC for the short viewing distance are marked with a gray box. The metric having statistically equivalent ambiguity interval widths for the two viewing distances is marked with a square.}
\end{figure}

Performance of the metrics for two viewing distances in terms of PLCC and mean of the ambiguity intervals is shown in Figure \ref{fig:accamb-vdidcidiq-all}. %
Most of the metrics show statistically equivalent PLCC scores for the two viewing distances; only one and nine metrics show significantly different accuracy scores for VDID and CIDIQ, respectively (which are marked with asterisks in Figure \ref{fig:accamb-vdidcidiq-all}).
However, for VDID, all metrics except for OSS-SSIM (marked with a square in Figure \ref{fig:accamb-vdidcidiq-all}(a)) show significantly different ambiguity interval widths for the two viewing distances.
Furthermore, OSS-SSIM shows high accuracy, i.e., it is included in the group of top-performing metrics (showing statistically equivalent PLCC scores with the best one for the short distance), and shows the smallest mean ambiguity intervals for both viewing distances (which are statistically equivalent).
Thus, we can choose OSS-SSIM as the best metric considering both the accuracy and the ambiguity for different viewing distances.
OSS-SSIM explicitly considers the effect of the viewing distance, which seems to be the reason for the consistency of its ambiguity performance.
In the case of CIDIQ, all metrics have significantly different results of ambiguity intervals.
MAD and IW-SSIM are two top-performing metrics in terms of accuracy for the short distance.
However, these metrics have relatively lower performance in terms of ambiguity (i.e., larger mean interval widths) than the following ones (in the ranking of accuracy), e.g.,  OSS-SSIM and ADD-GSIM.
If we accept a slight loss in terms of accuracy, it would be a better choice to select ADD-GSIM or OSS-SSIM as the best metric with consideration of both the accuracy and ambiguity for the two viewing distances.

\begin{figure}
\small
\centering
\begin{tabular}{cc}
\includegraphics[trim=0cm 0cm 0cm 0cm, scale = 0.45]{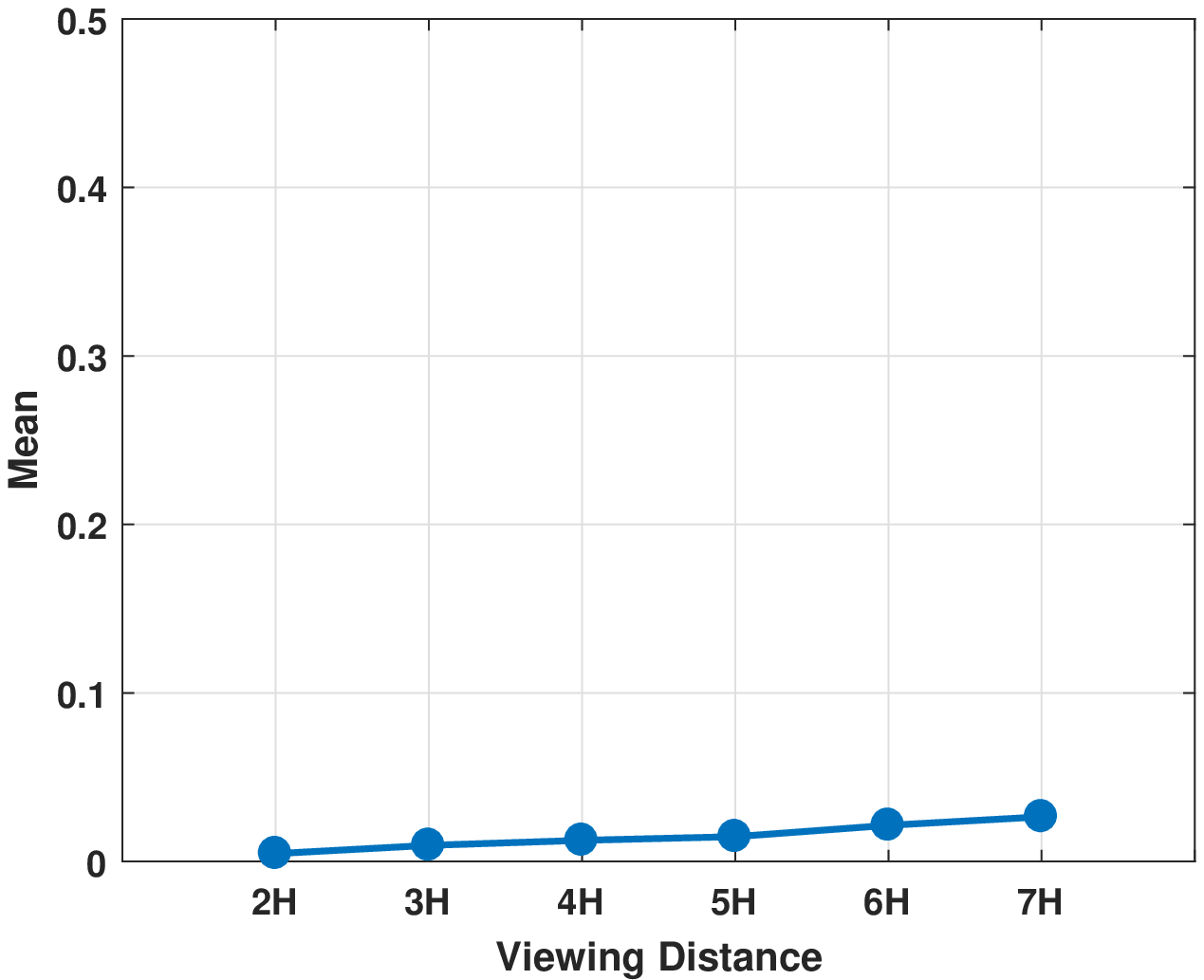}&
\includegraphics[trim=0cm 0cm 0cm 0cm, scale = 0.45]{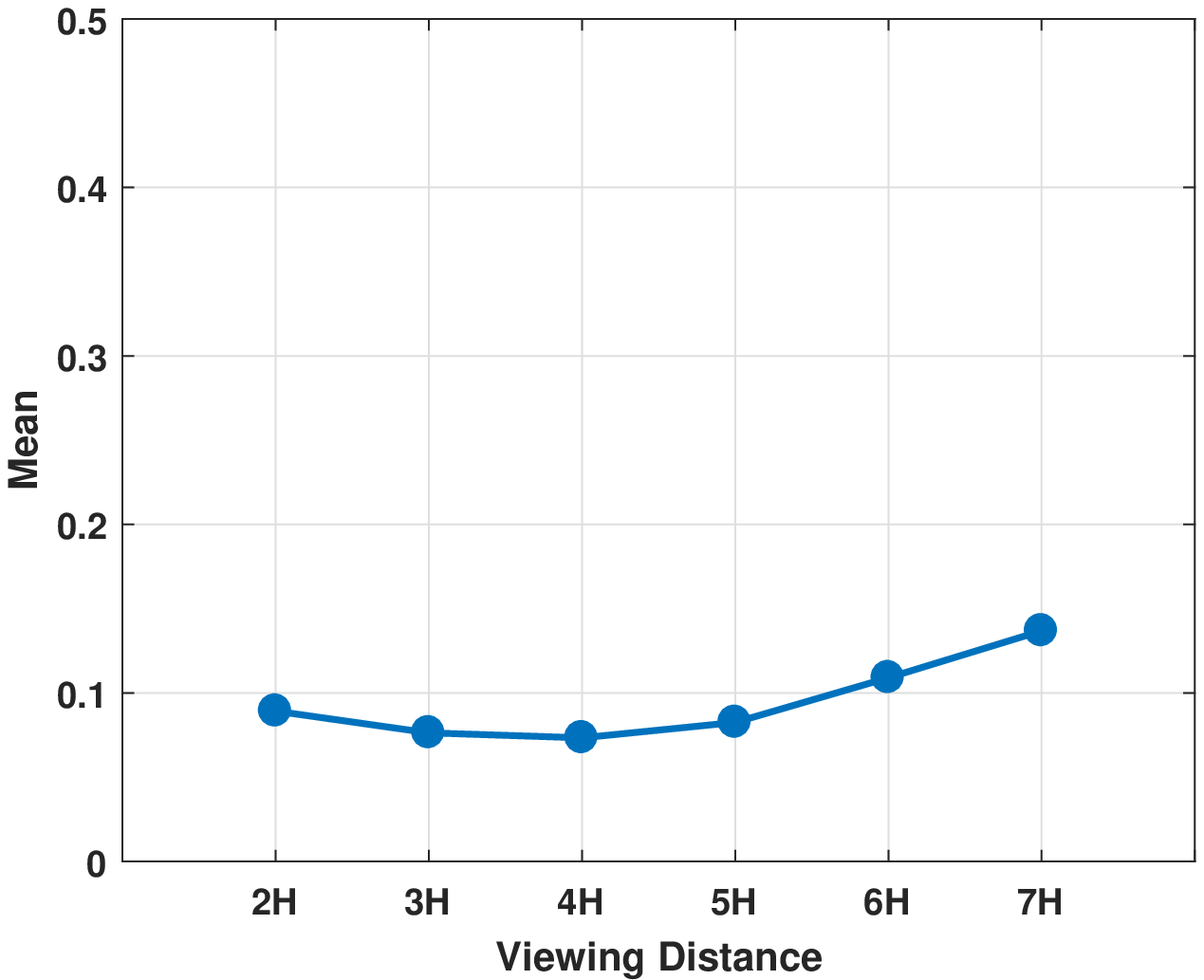}\\
(a) JPEG&(b) JPEG2K\\
\includegraphics[trim=0cm 0cm 0cm 0cm, scale = 0.45]{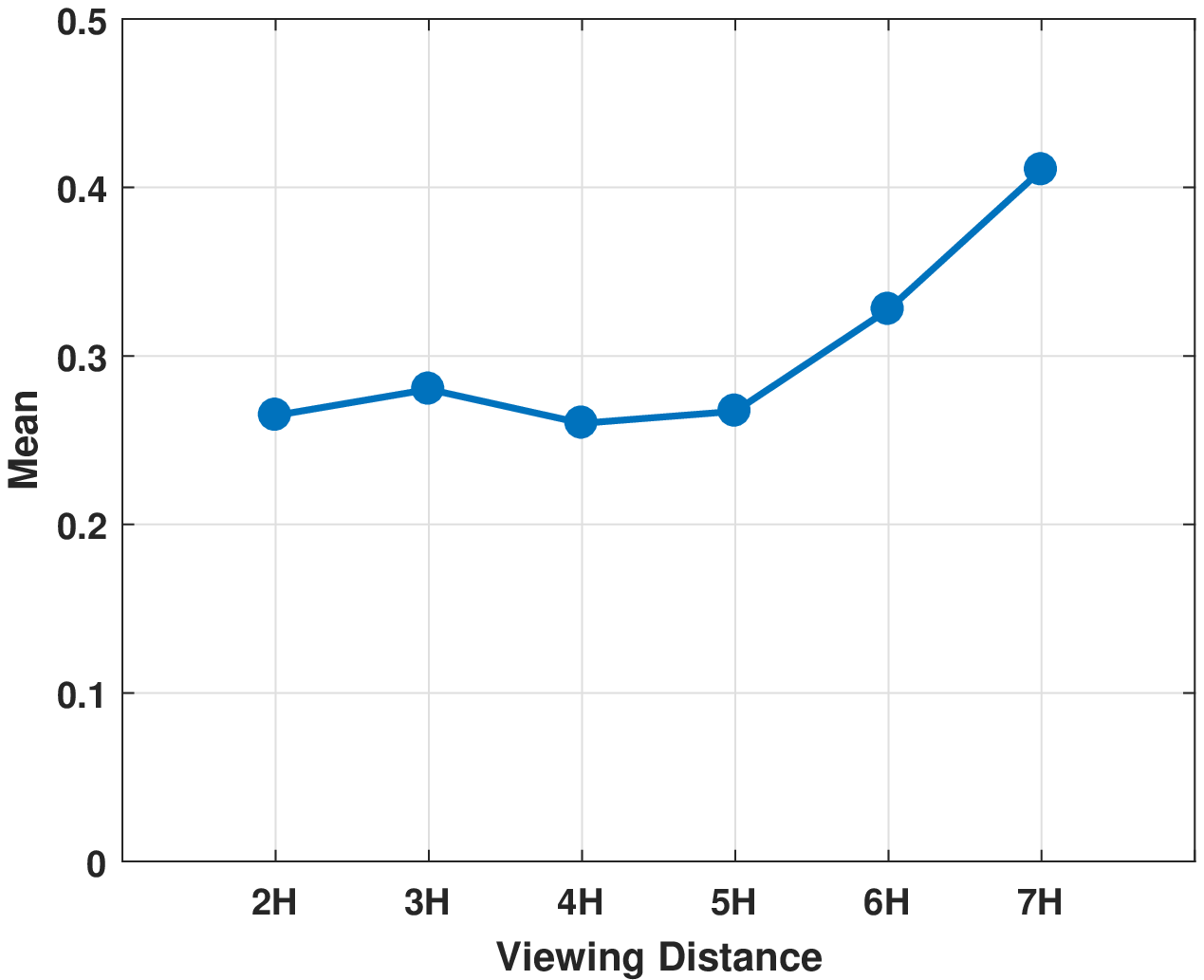}&
\includegraphics[trim=0cm 0cm 0cm 0cm, scale = 0.45]{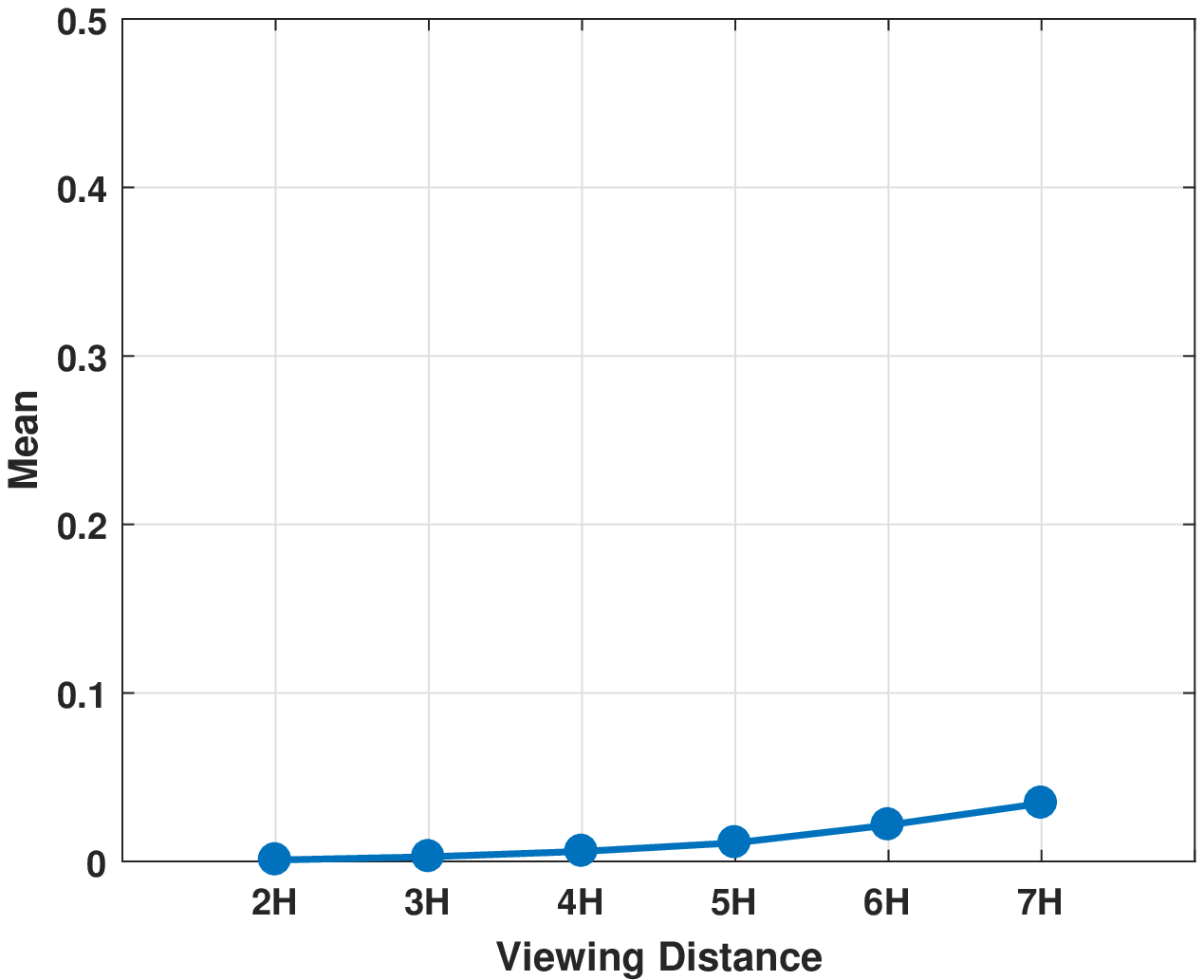}\\
(c) GB&(d) WN\\
\end{tabular}
\caption{\label{fig:vdid-metric-ex-addgsim}Mean widths of the ambiguity intervals of ADD-GSIM for the VDID database (a) JPEG, (b) JPEG2K, (c) GB, and (d) WN. The distortion type influences the slopes of the curves.}
\end{figure}

Next, we analyze patterns of the ambiguity intervals over various viewing distances.
As an example, Figure \ref{fig:vdid-metric-ex-addgsim}  shows the mean widths of the ambiguity intervals of ADD-GSIM for each of the four distortion types of VDID.
As aforementioned, as the viewing distance increases, the ability of human viewers to distinguish the details in images decreases.
The ambiguity intervals obtained by our approach also tend to increase with the increasing viewing distance.
A gradual increase of the ambiguity intervals due to increase of the viewing distance  is acceptable, but a sudden increase of the slope would not be desirable.
For instance, in Figure \ref{fig:vdid-metric-ex-addgsim}(c), the slope for GB increases suddenly after 5H, thus, care must be taken when the metric is used for viewing distances larger than 5H.

\begin{figure*}
\small
\hspace{-1cm}
\begin{tabular}{ccccc}
\includegraphics[trim=0cm 0cm 0cm 0cm, scale = \figscale]{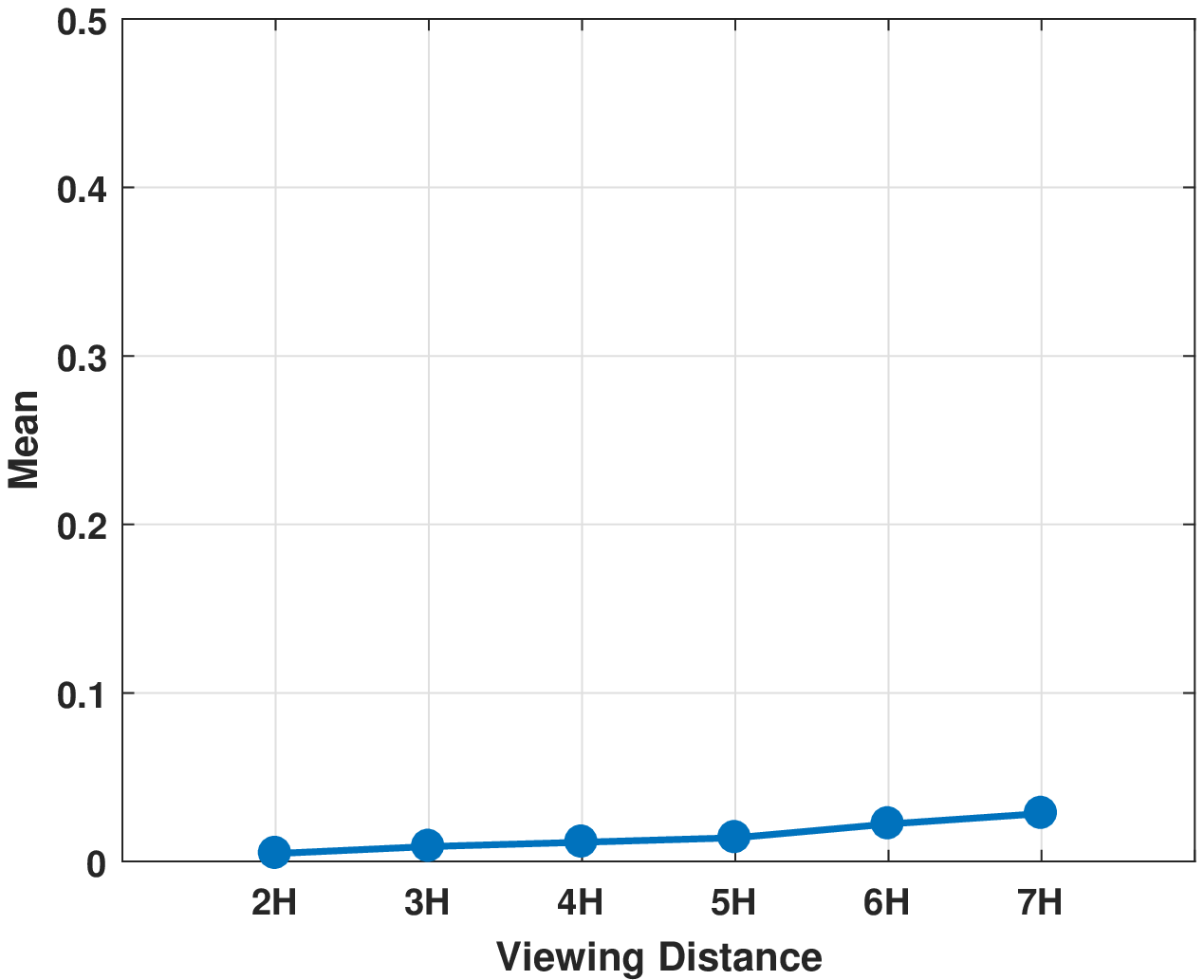}&
\includegraphics[trim=0cm 0cm 0cm 0cm, scale = \figscale]{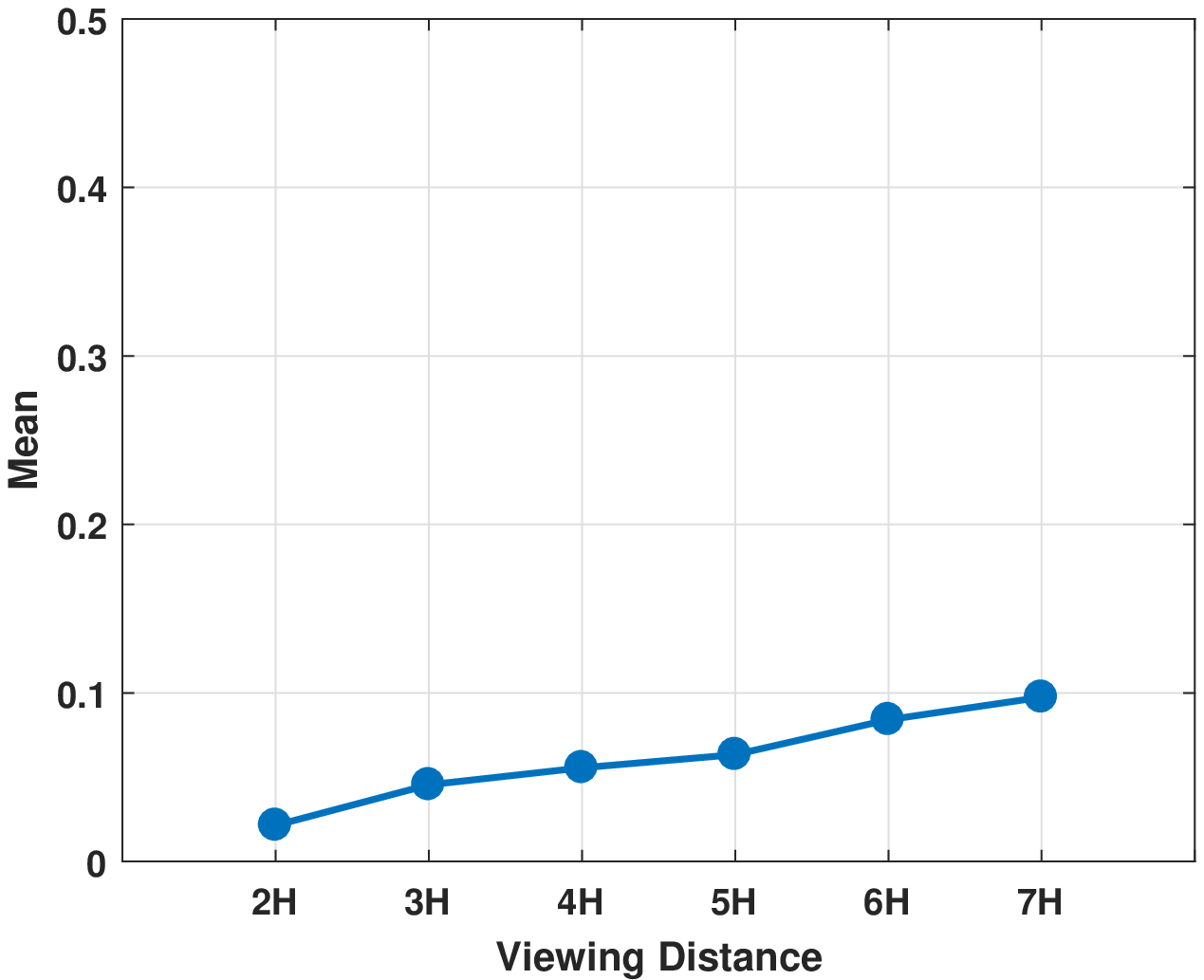}&
\includegraphics[trim=0cm 0cm 0cm 0cm, scale = \figscale]{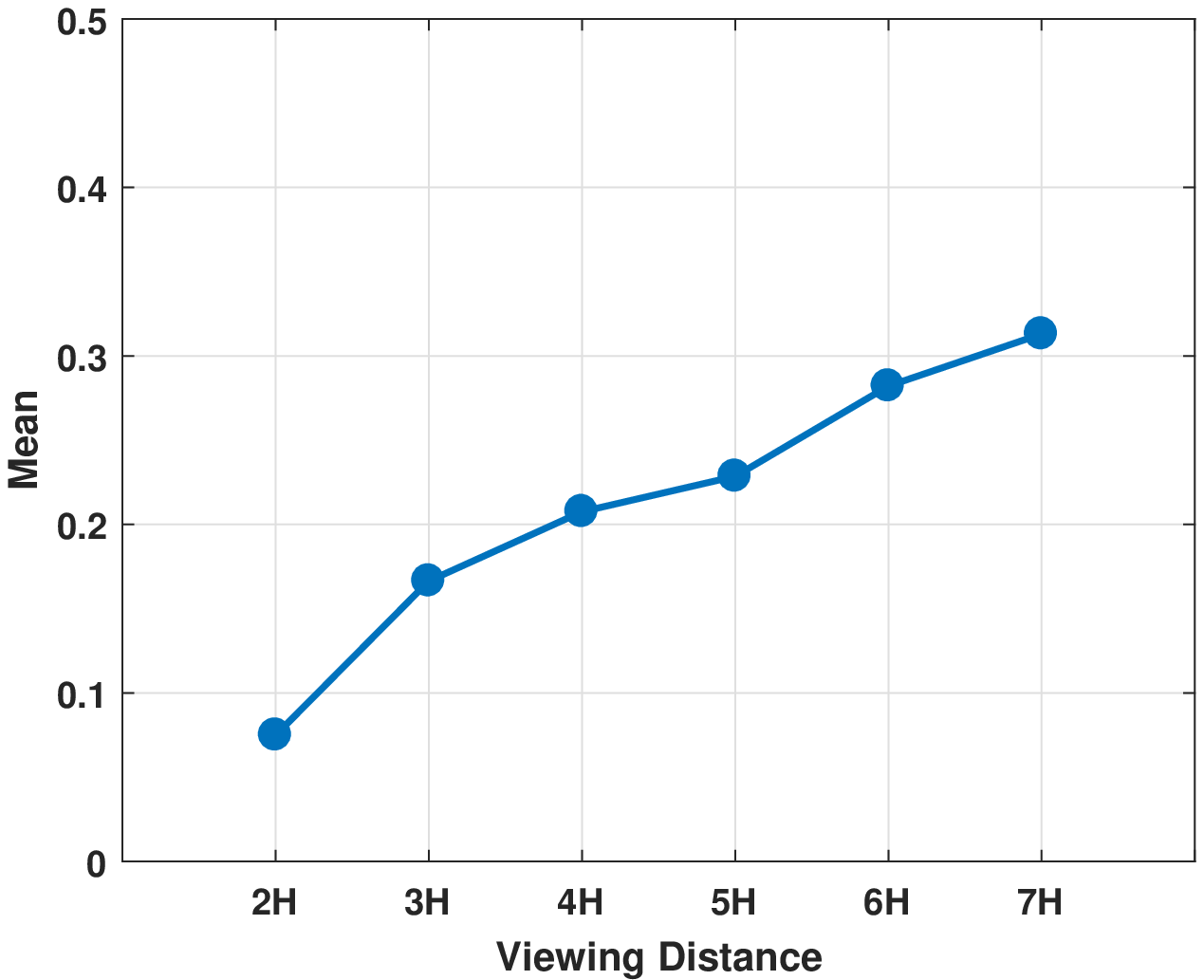}&
\includegraphics[trim=0cm 0cm 0cm 0cm, scale = \figscale]{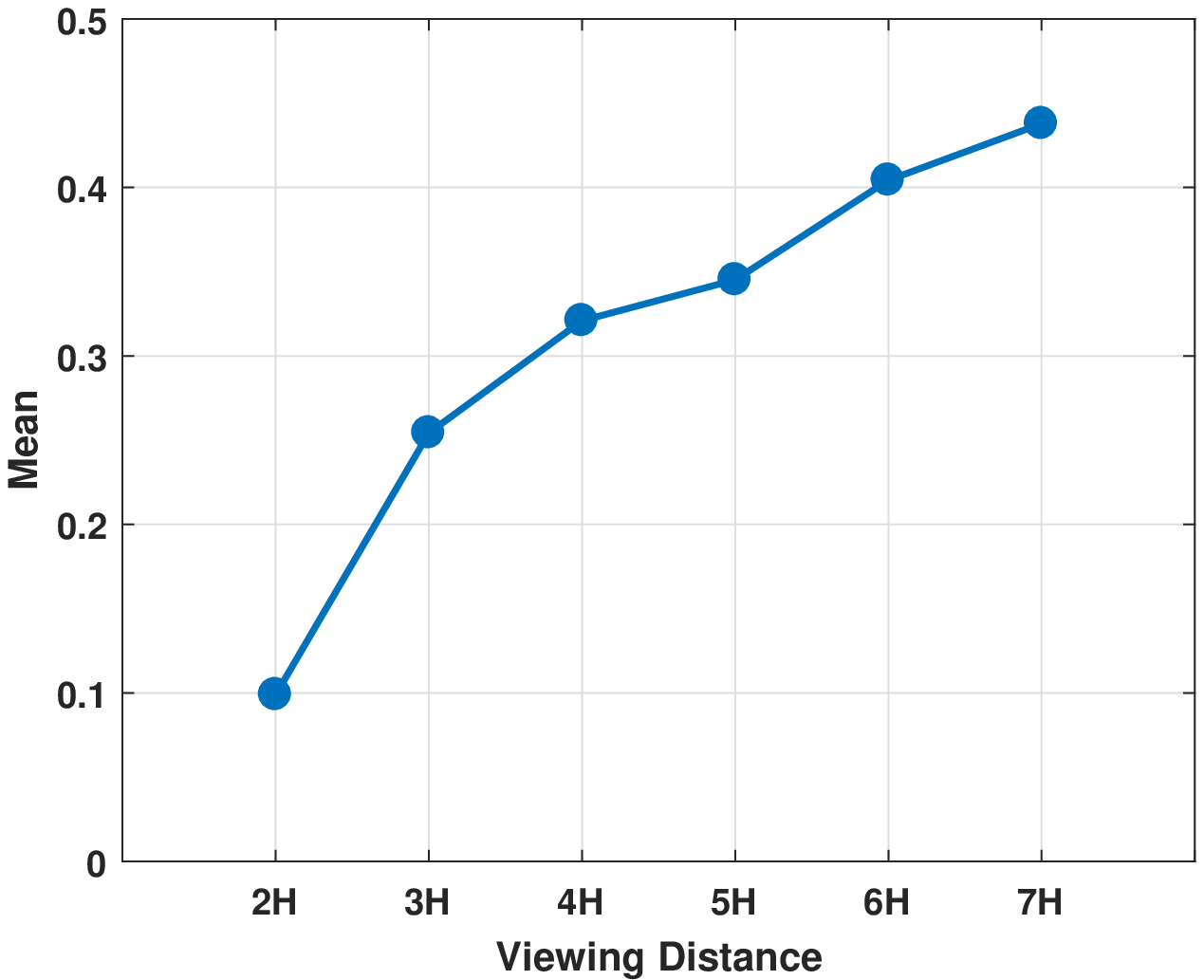}\\
(a1) OSS-SSIM&(a2) GSM&(a3) WSNR&(a4) PSNR\\\\
\includegraphics[trim=0cm 0cm 0cm 0cm, scale = \figscale]{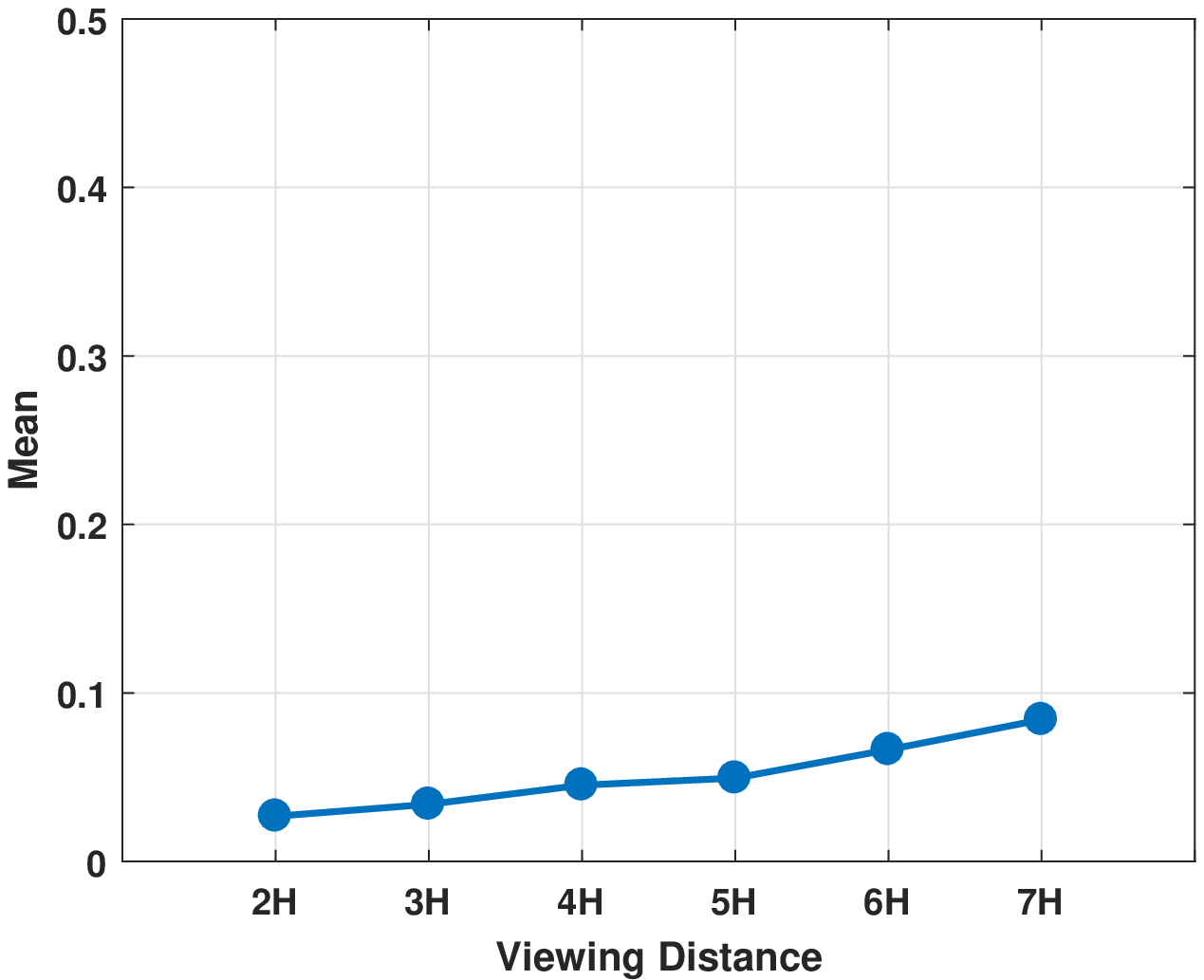}&
\includegraphics[trim=0cm 0cm 0cm 0cm, scale = \figscale]{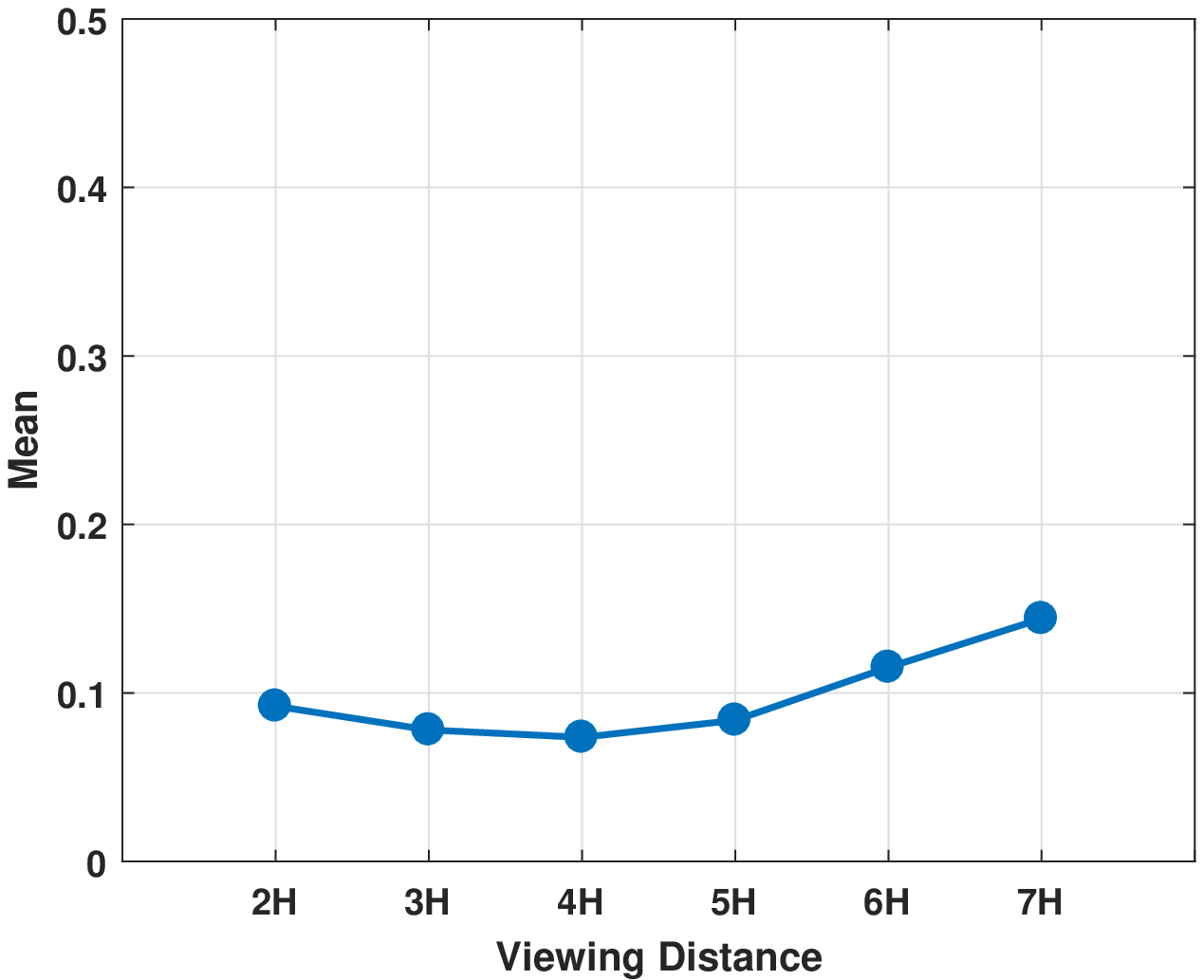}&
\includegraphics[trim=0cm 0cm 0cm 0cm, scale = \figscale]{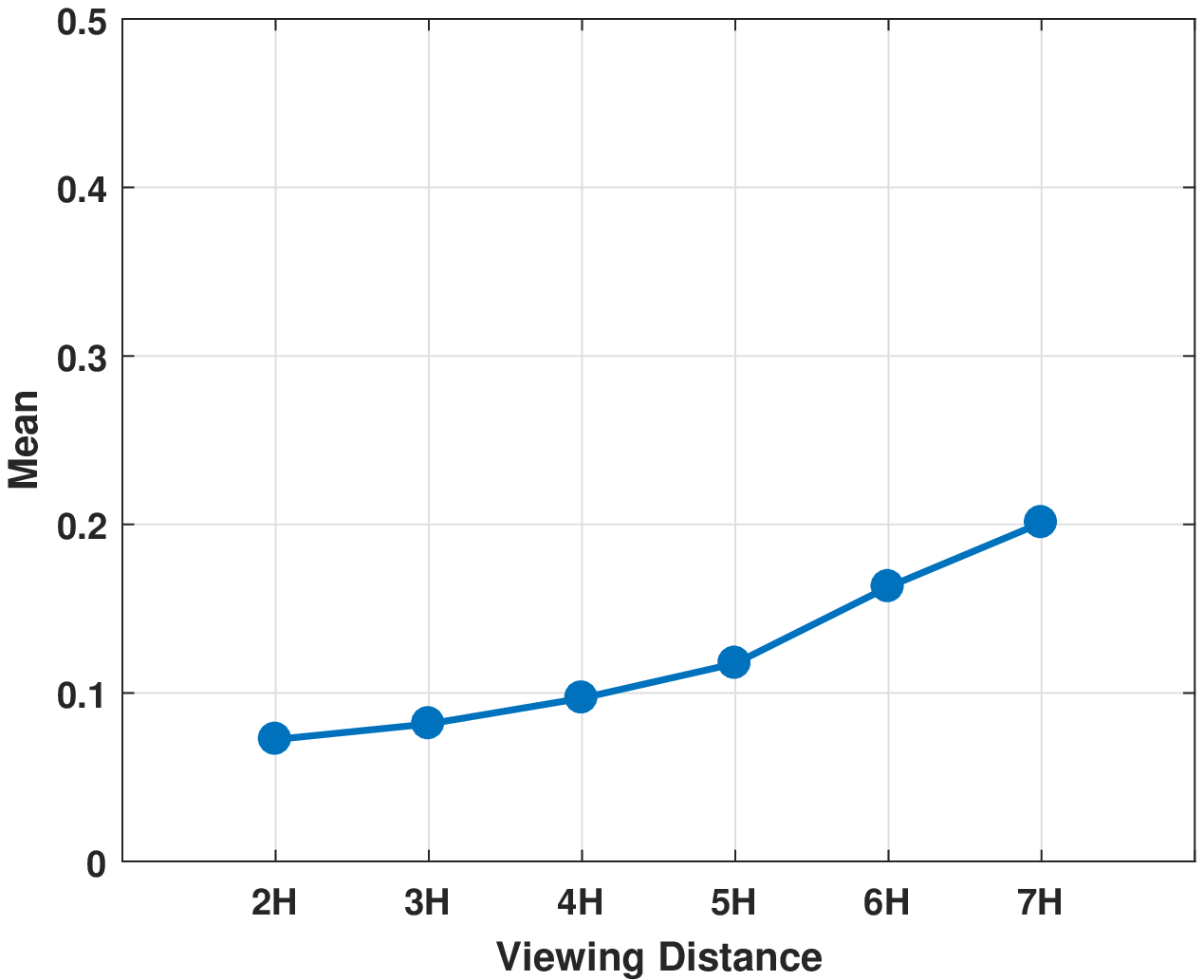}&
\includegraphics[trim=0cm 0cm 0cm 0cm, scale = \figscale]{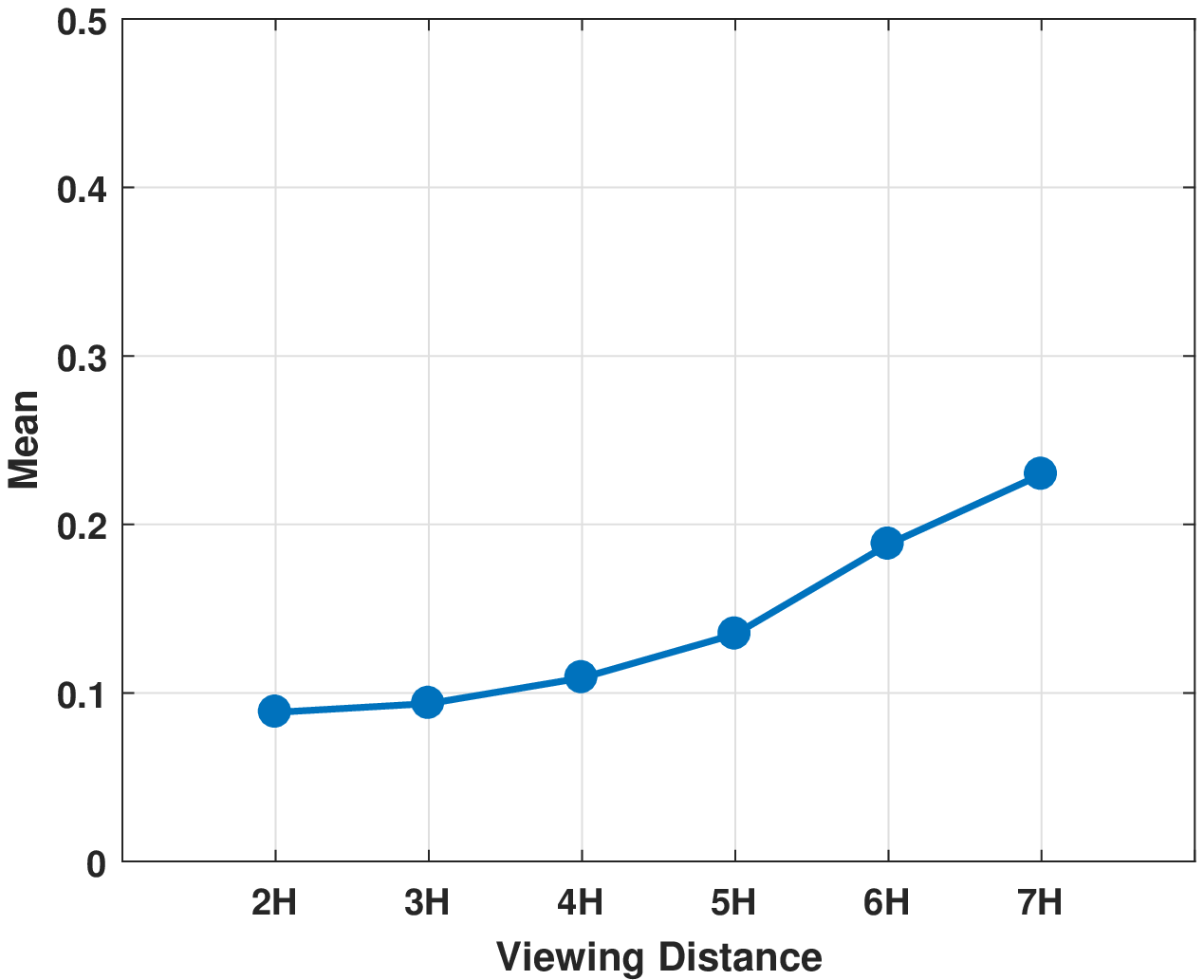}\\
(b1) VSNR&(b2) FSIM&(b3) PSNR-HA&(b4) VIF\\\\
\includegraphics[trim=0cm 0cm 0cm 0cm, scale = \figscale]{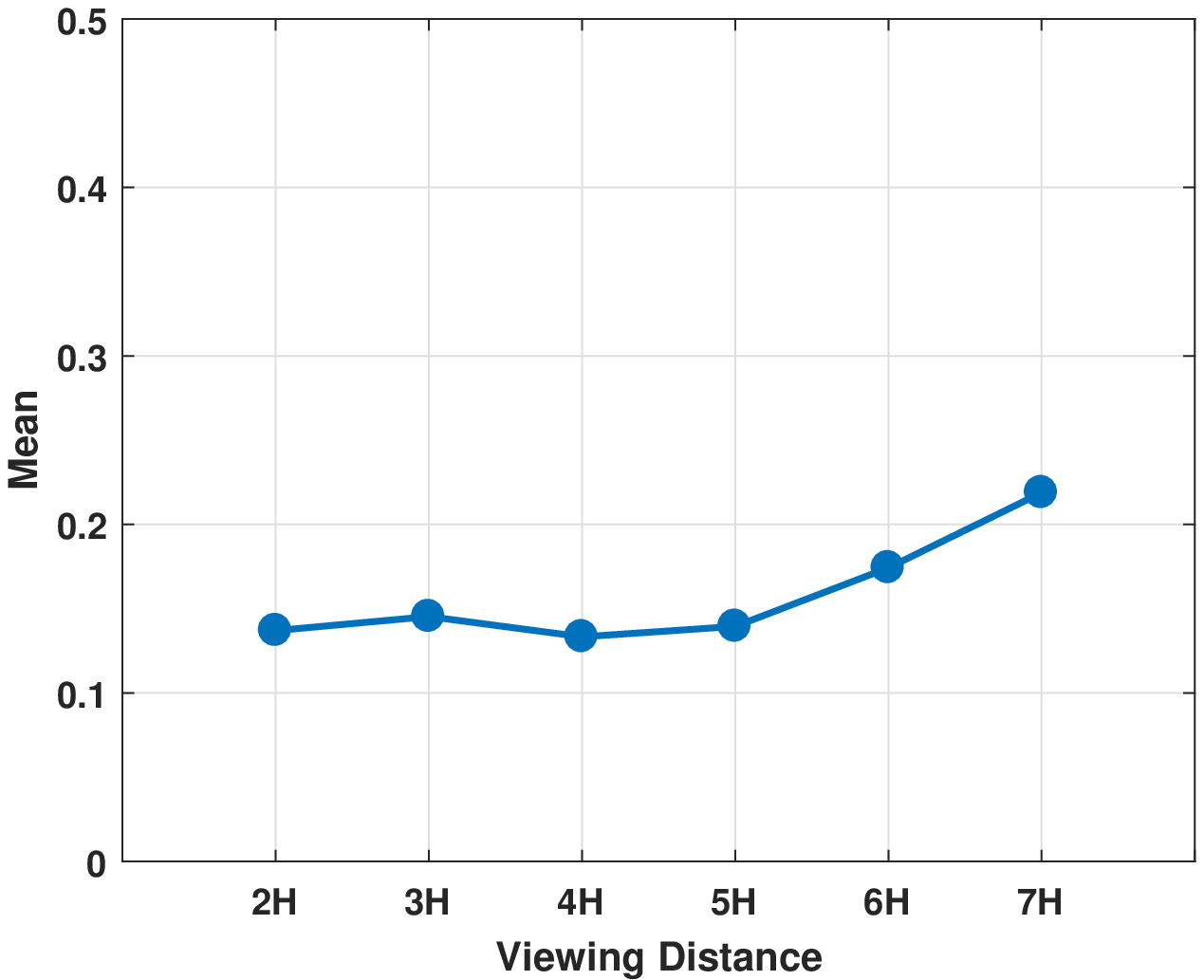}&
\includegraphics[trim=0cm 0cm 0cm 0cm, scale = \figscale]{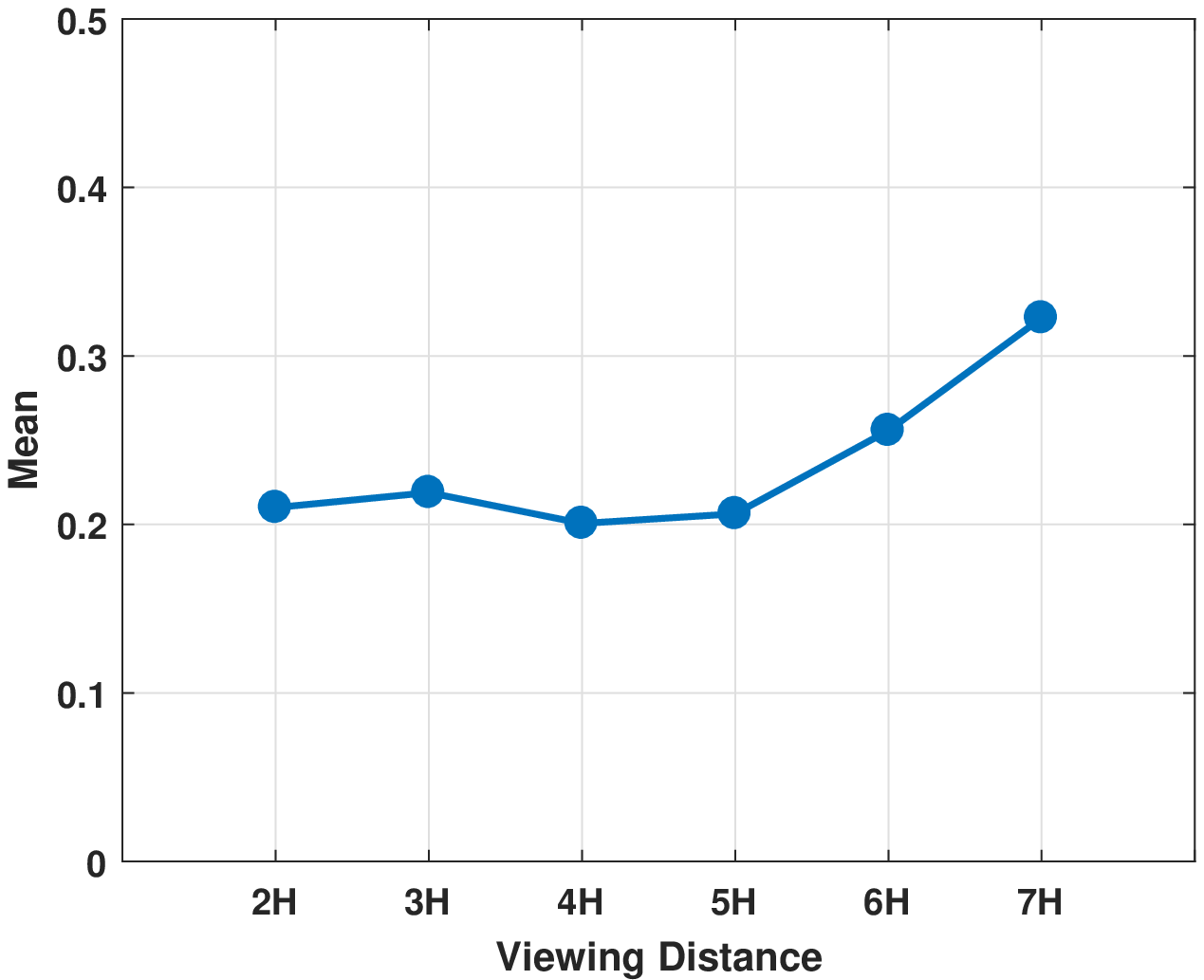}&
\includegraphics[trim=0cm 0cm 0cm 0cm, scale = \figscale]{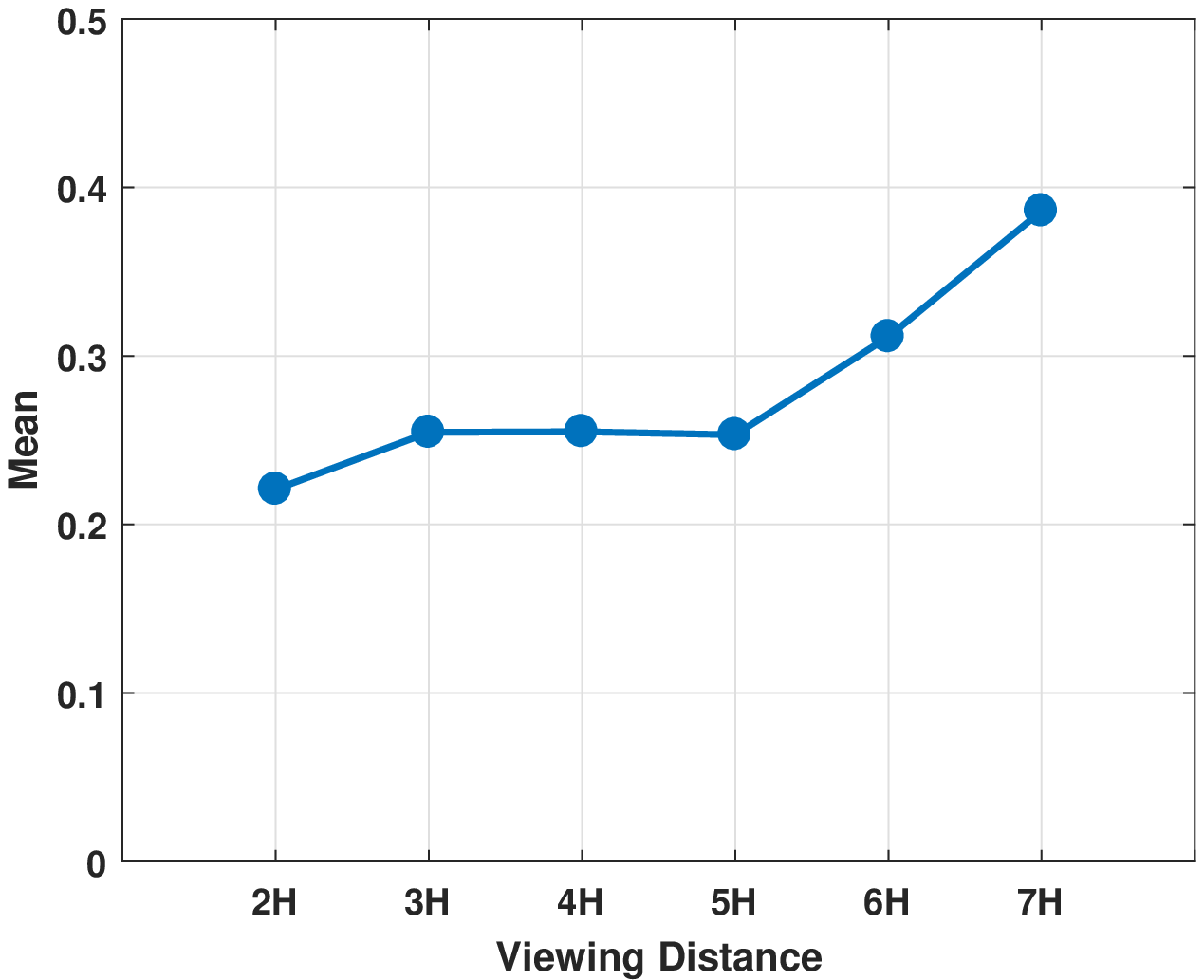}&
\includegraphics[trim=0cm 0cm 0cm 0cm, scale = \figscale]{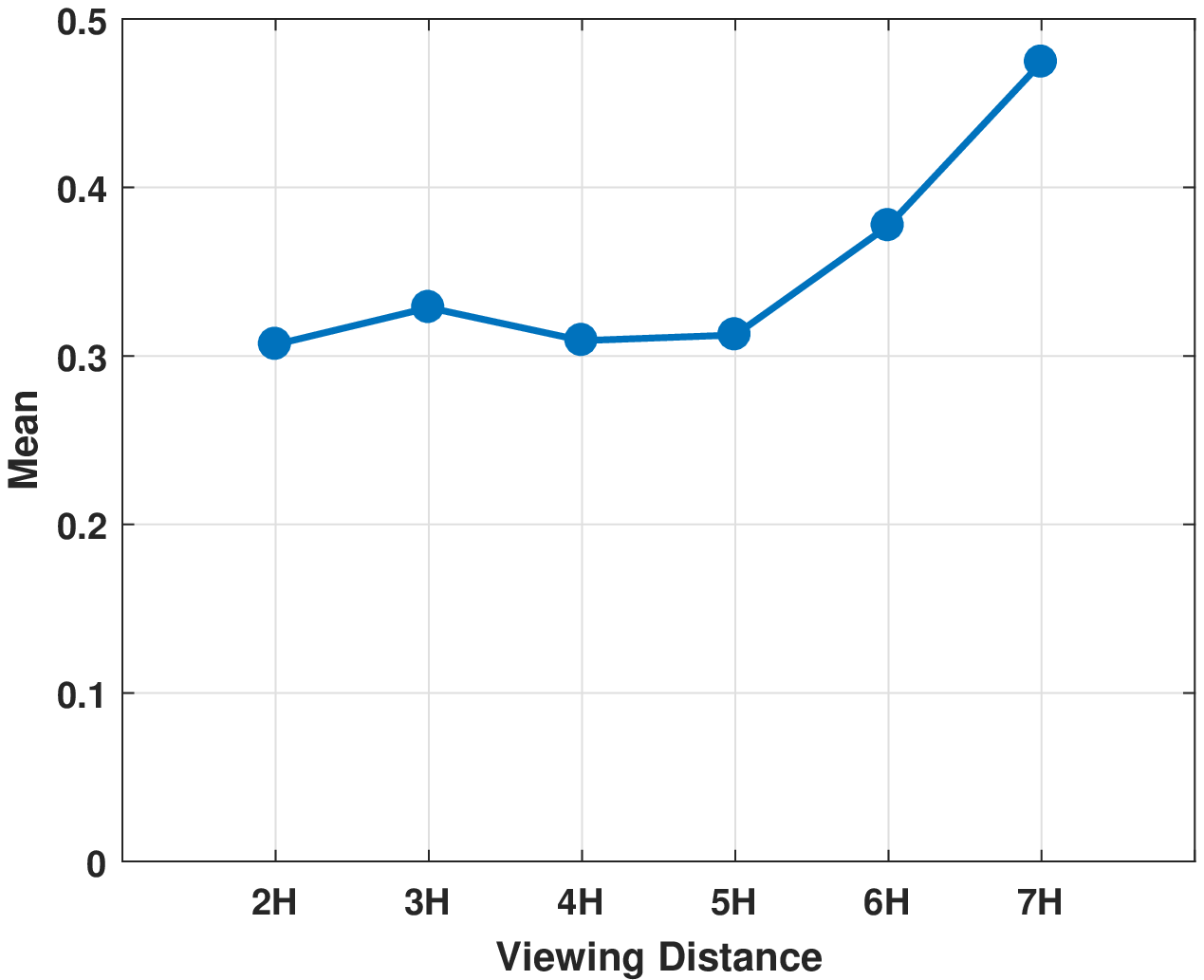}\\
(c1) MS-SSIM&(c2) FSIM&(c3) BLIINDS2&(c4) ADM\\\\
\includegraphics[trim=0cm 0cm 0cm 0cm, scale = \figscale]{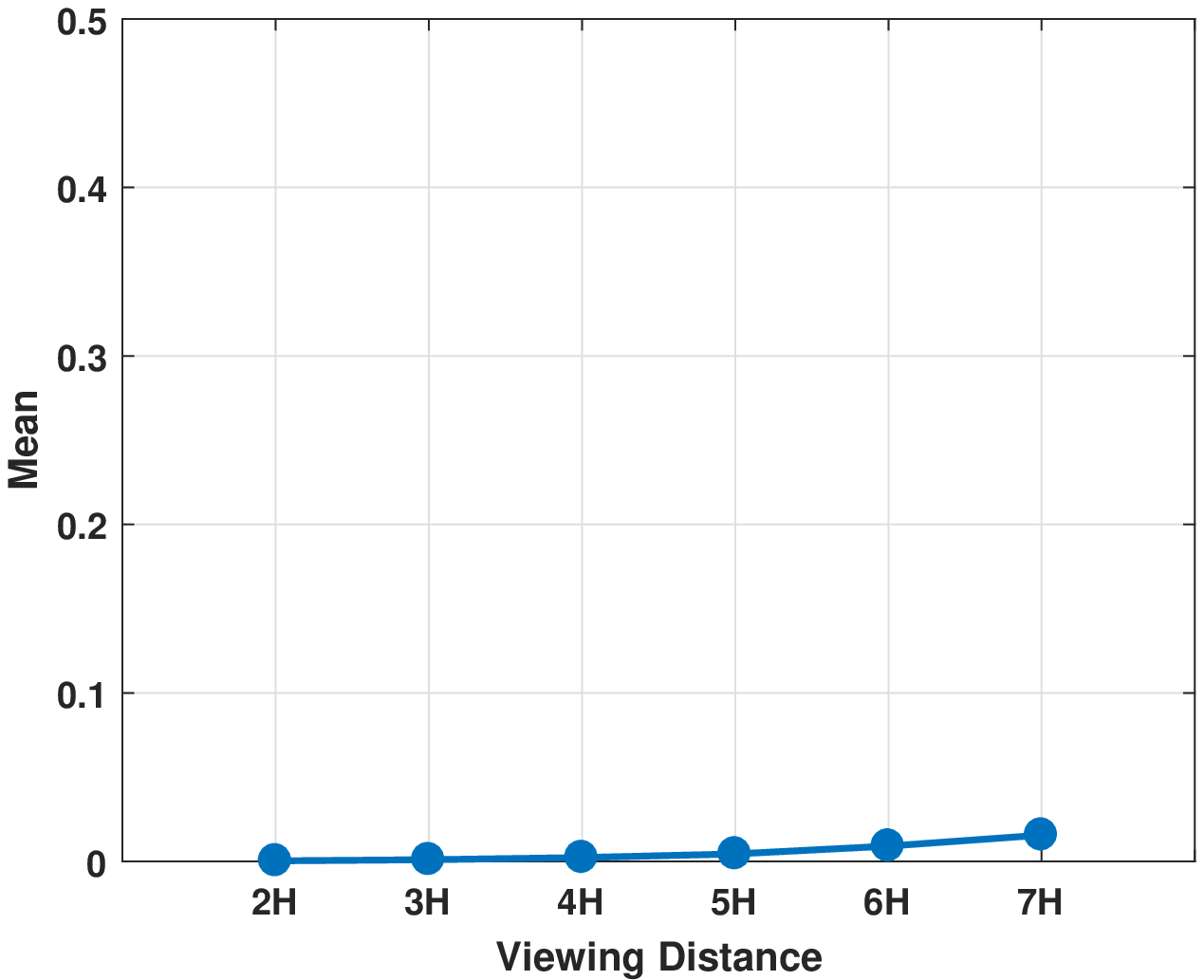}&
\includegraphics[trim=0cm 0cm 0cm 0cm, scale = \figscale]{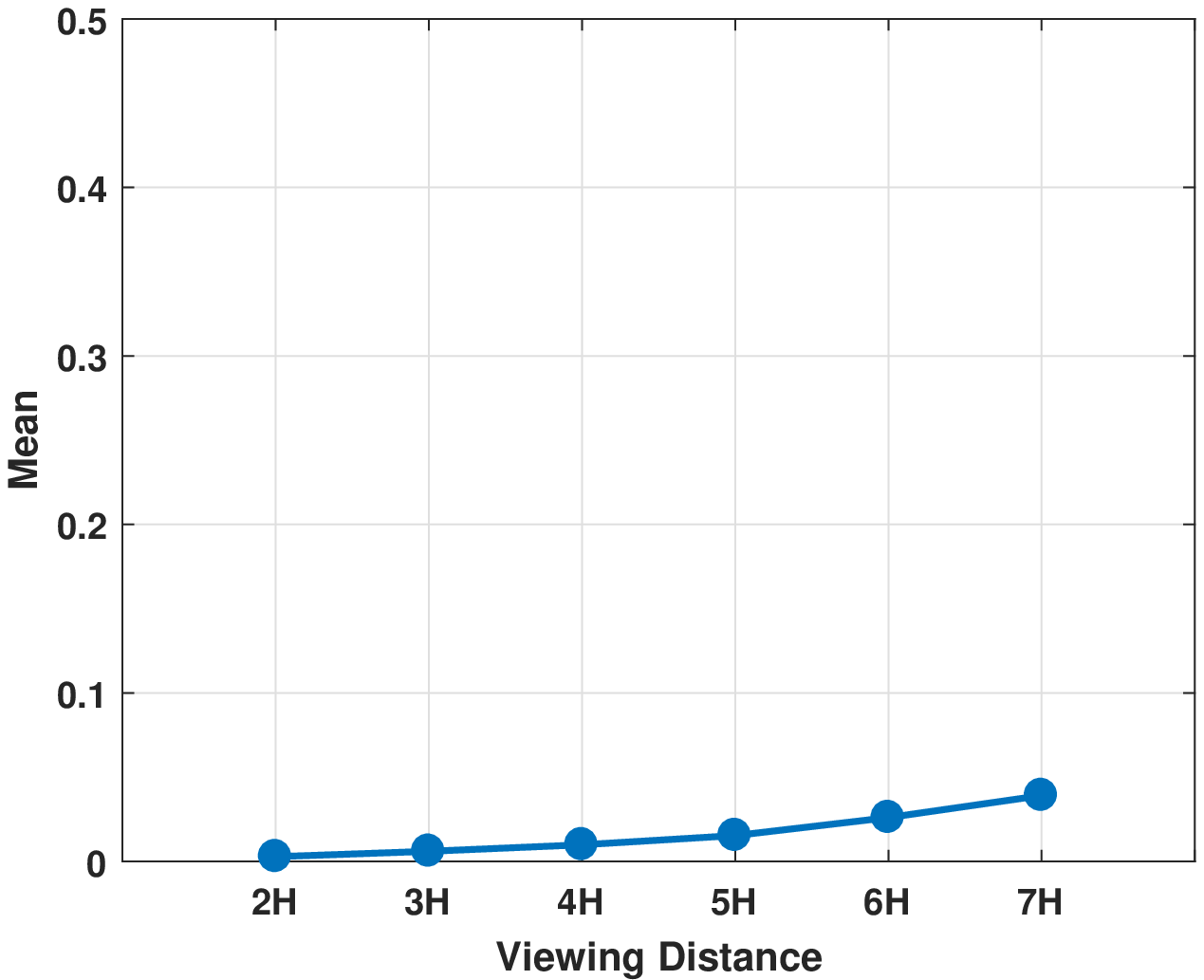}&
\includegraphics[trim=0cm 0cm 0cm 0cm, scale = \figscale]{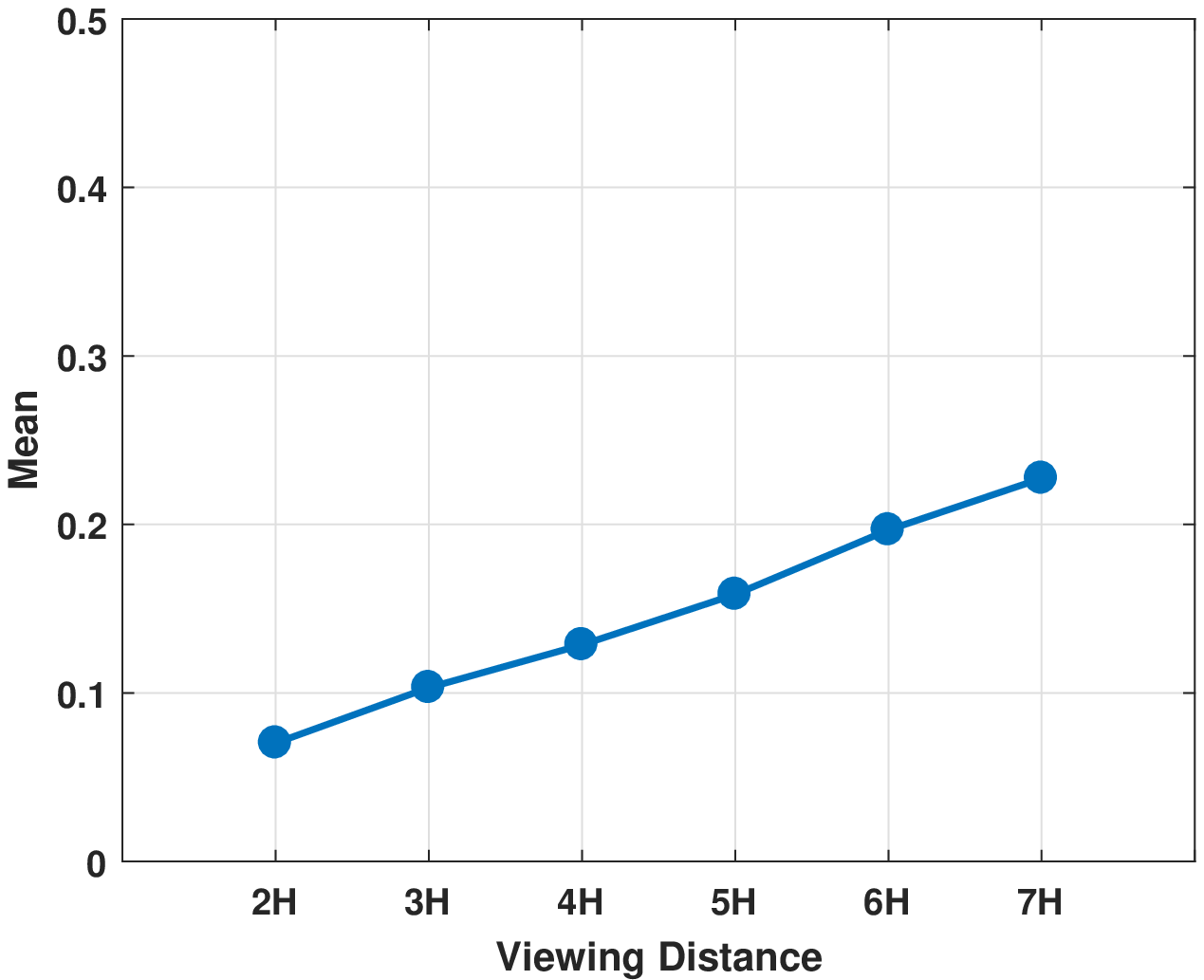}&
\includegraphics[trim=0cm 0cm 0cm 0cm, scale = \figscale]{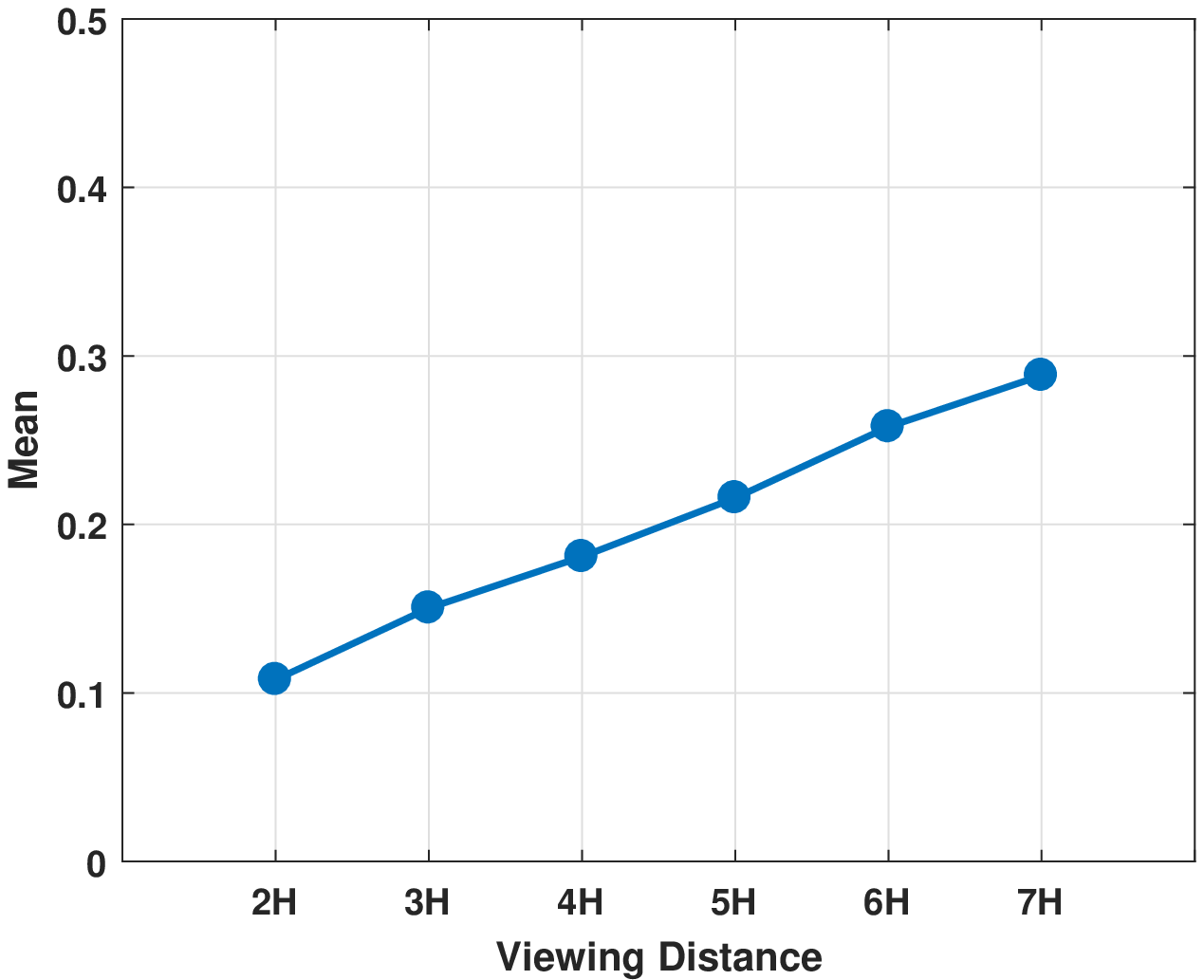}\\
(d1) OSS-SSIM&(d2) RRED&(d3) NQM&(d4) IFC\\\\
\end{tabular}
\caption{\label{fig:vdid-metric-ex} Mean widths of the ambiguity intervals of the objective metrics for the VDID database. For each distortion type, metrics in the first, second, third, and last quarters in the ascending order of the mean ambiguity interval width for 4H are shown from left to right. (a1)-(a4) JPEG, (b1)-(b4) JPEG2K, (c1)-(c4) GB, and (d1)-(d4) WN}
\end{figure*}

\begin{figure*}
\small
\hspace{-1cm}
\begin{tabular}{ccccc}
\includegraphics[trim=0cm 0cm 0cm 0cm, scale = \figscale]{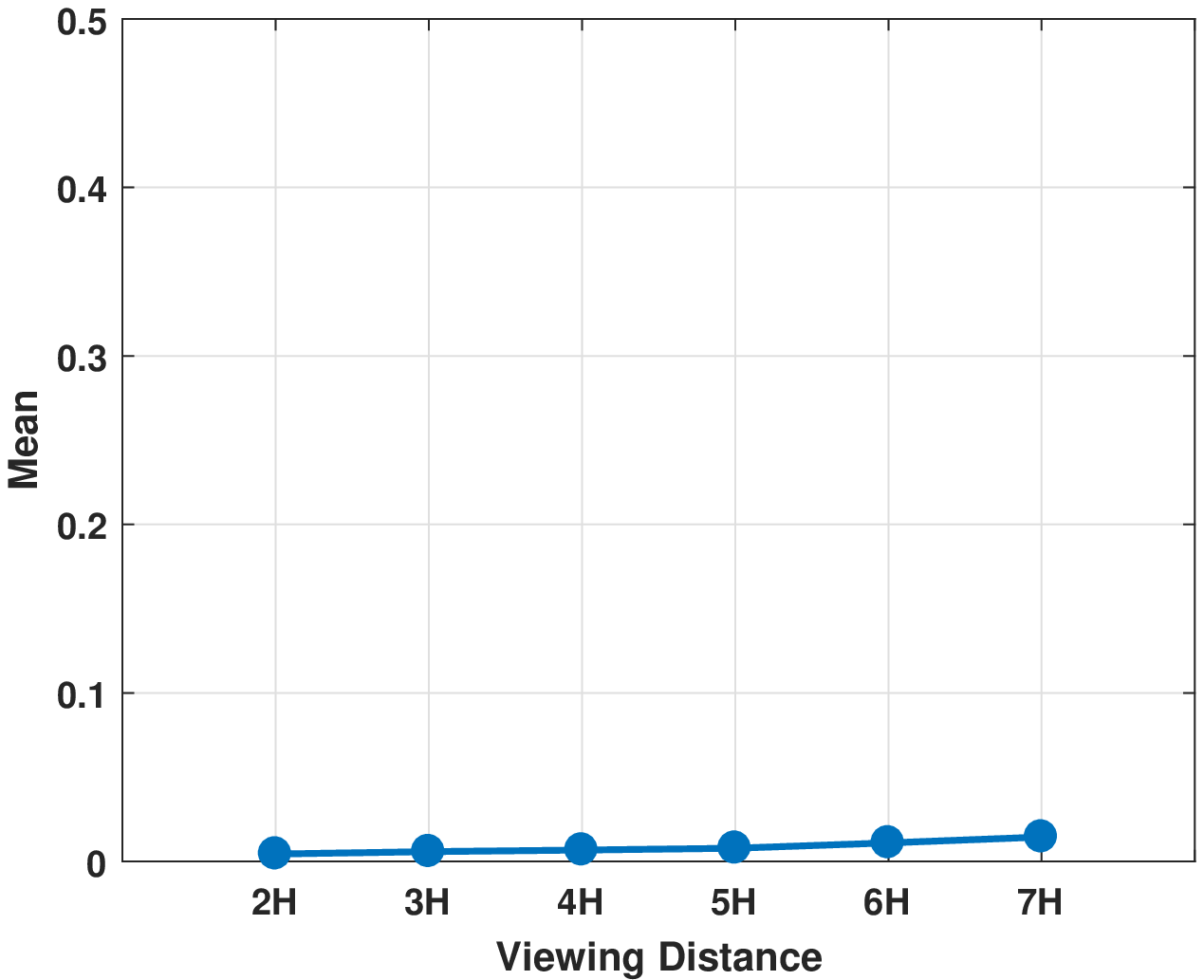}&
\includegraphics[trim=0cm 0cm 0cm 0cm, scale = \figscale]{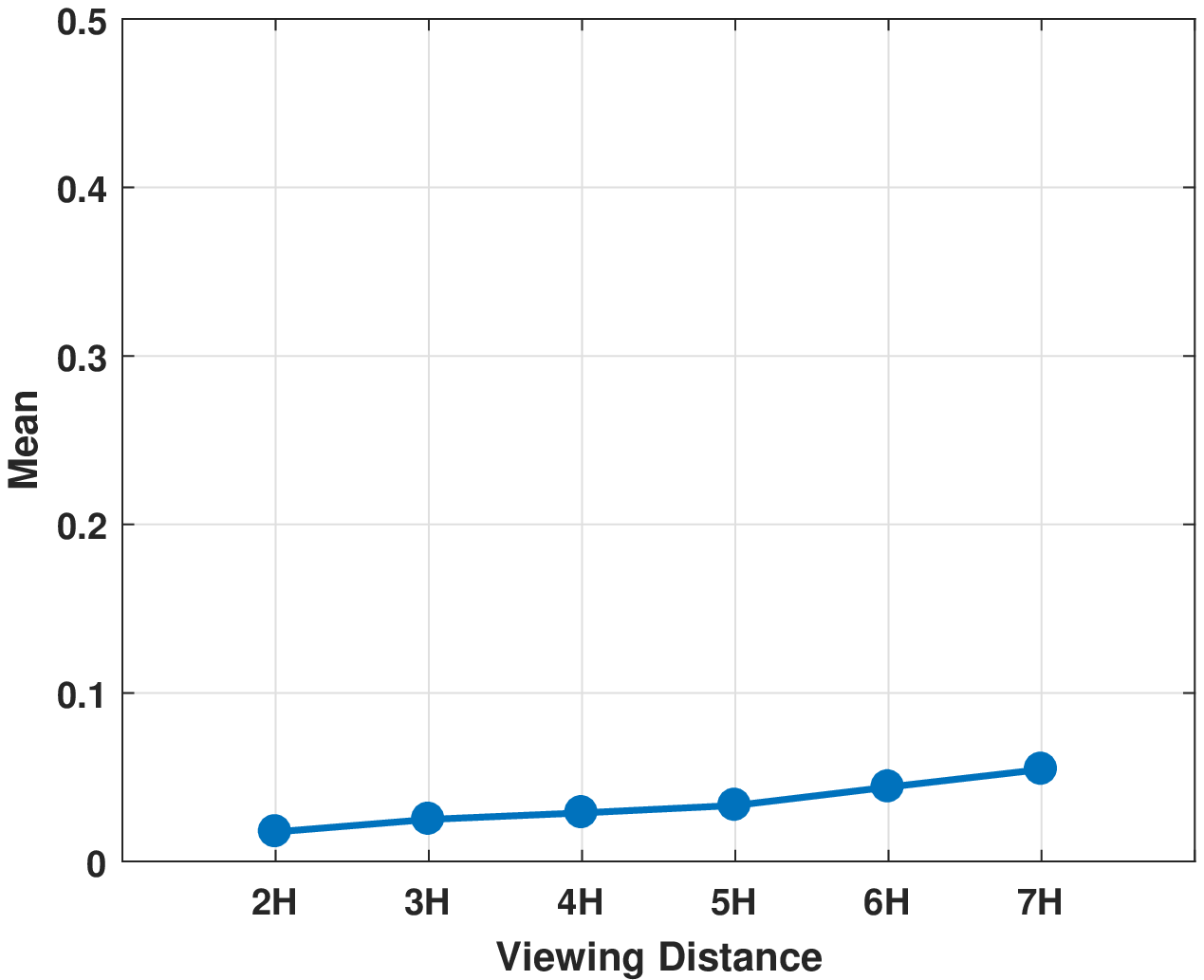}&
\includegraphics[trim=0cm 0cm 0cm 0cm, scale = \figscale]{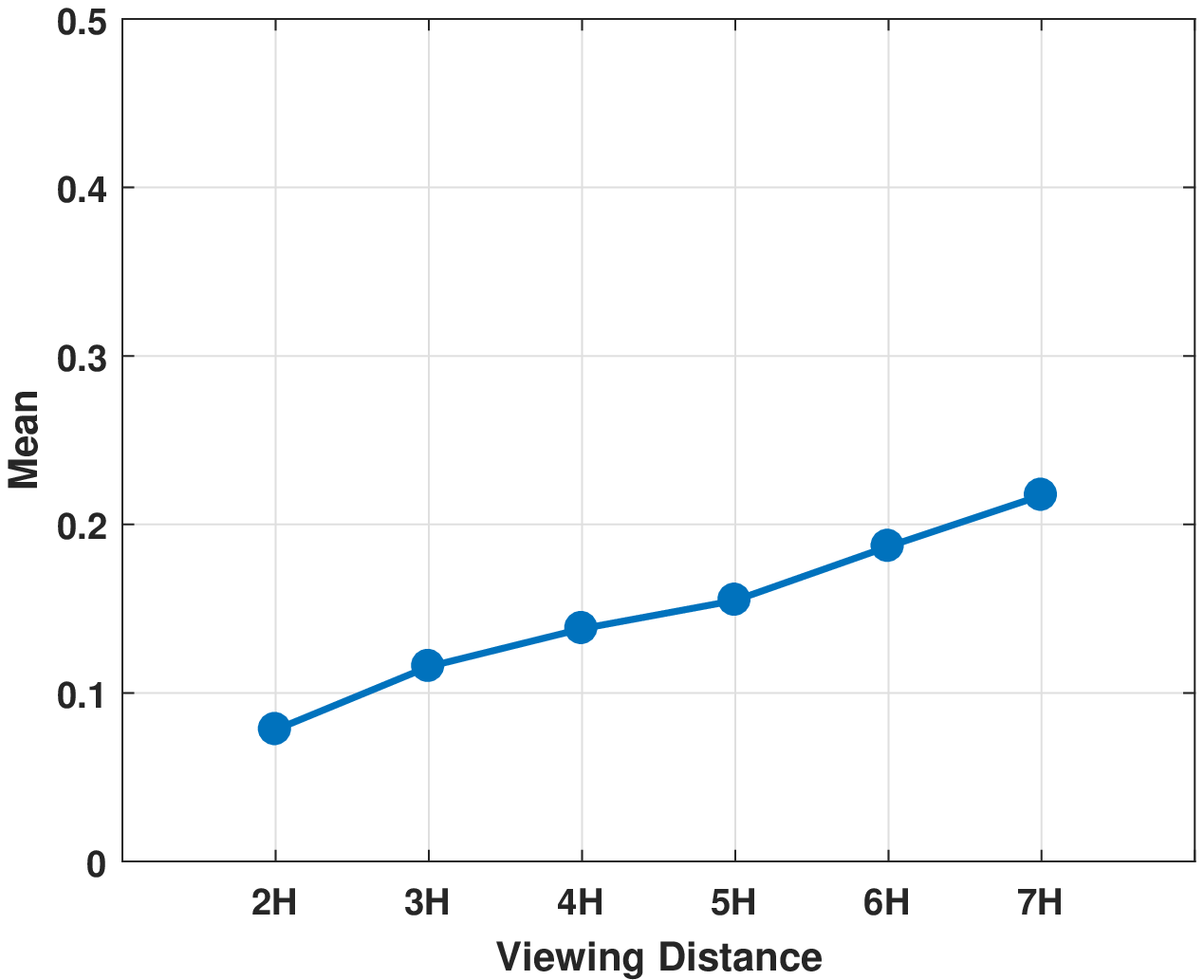}&
\includegraphics[trim=0cm 0cm 0cm 0cm, scale = \figscale]{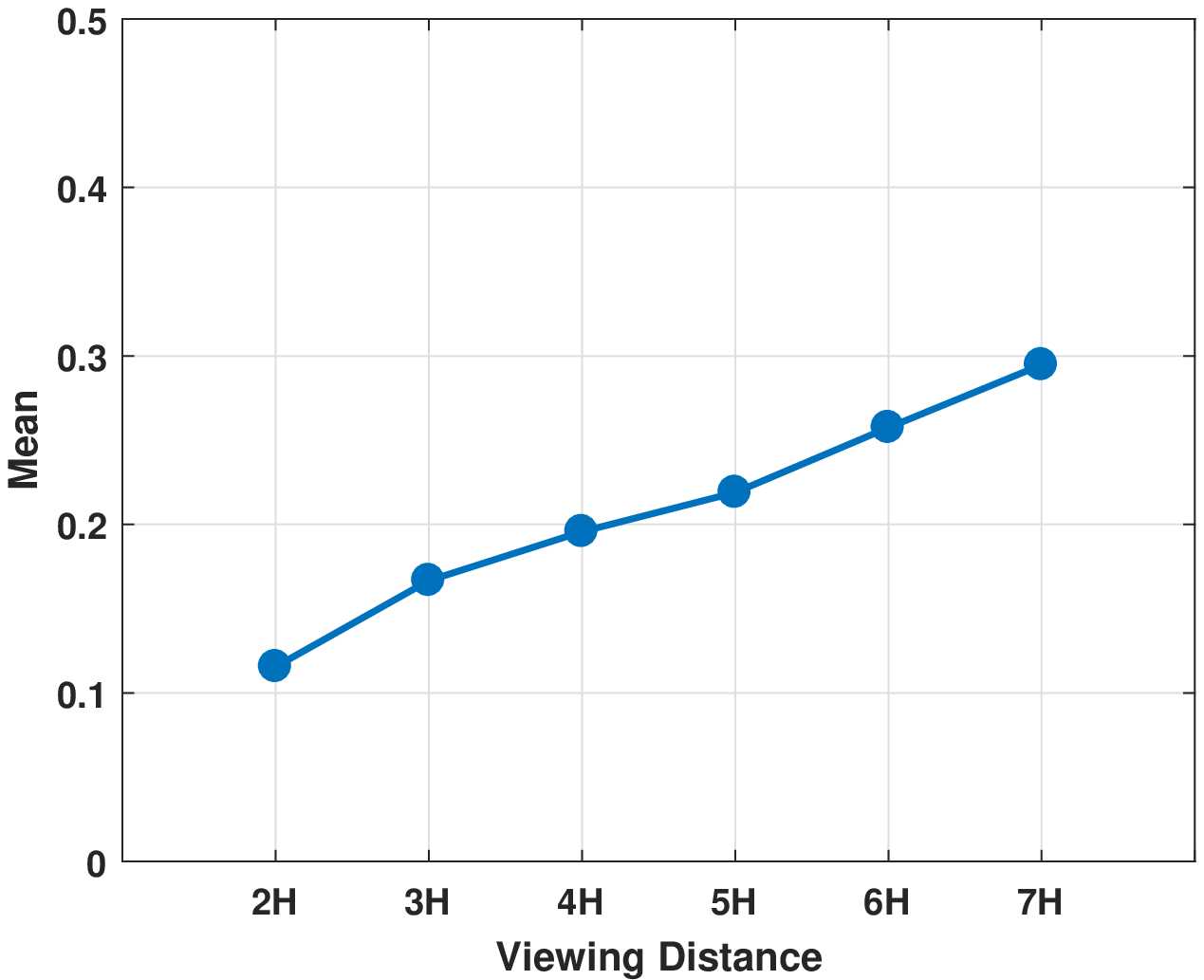}\\
(a1) ADD-GSIM&(a2) RRED&(a3) WSNR&(a4) OSS-PSNR\\\\
\includegraphics[trim=0cm 0cm 0cm 0cm, scale = \figscale]{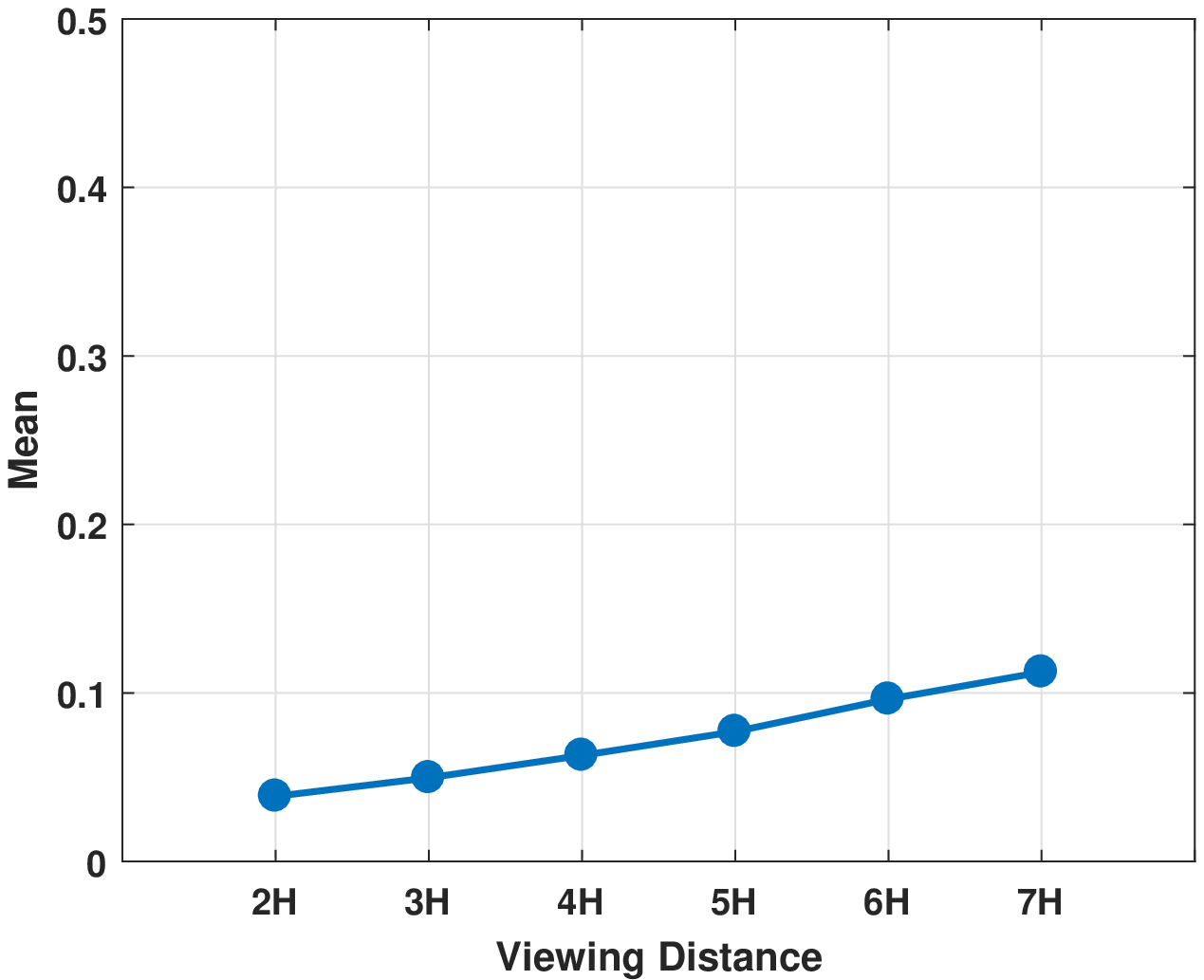}&
\includegraphics[trim=0cm 0cm 0cm 0cm, scale = \figscale]{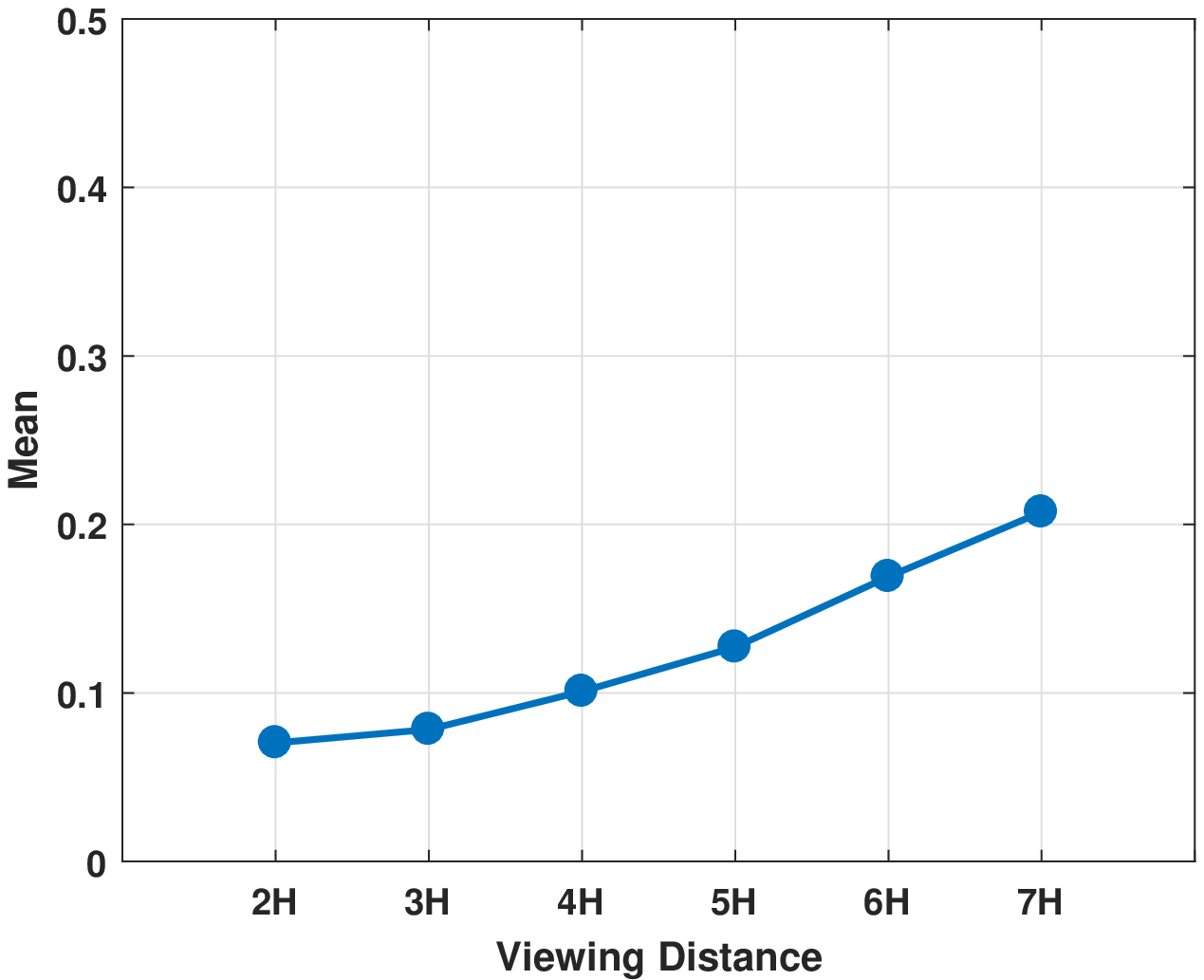}&
\includegraphics[trim=0cm 0cm 0cm 0cm, scale = \figscale]{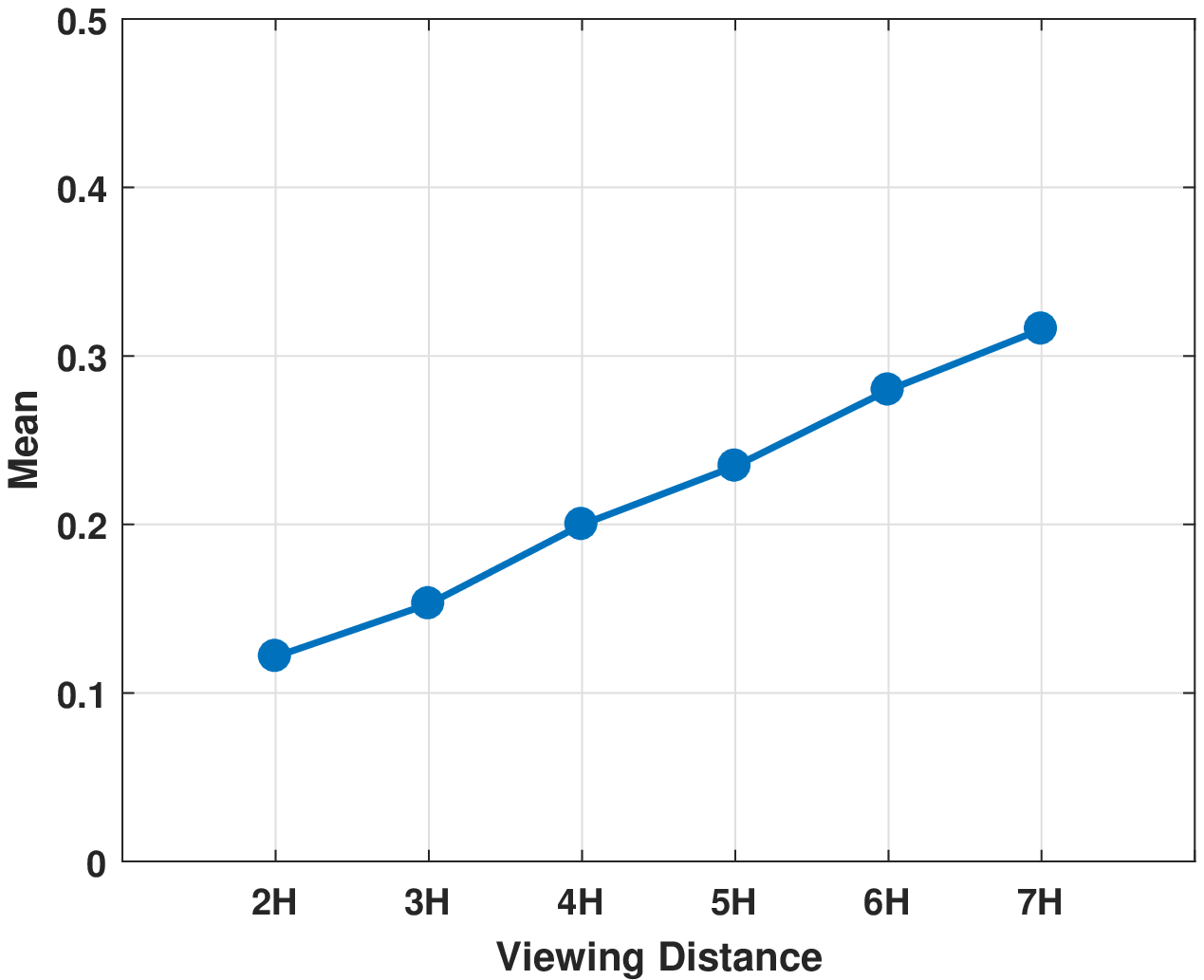}&
\includegraphics[trim=0cm 0cm 0cm 0cm, scale = \figscale]{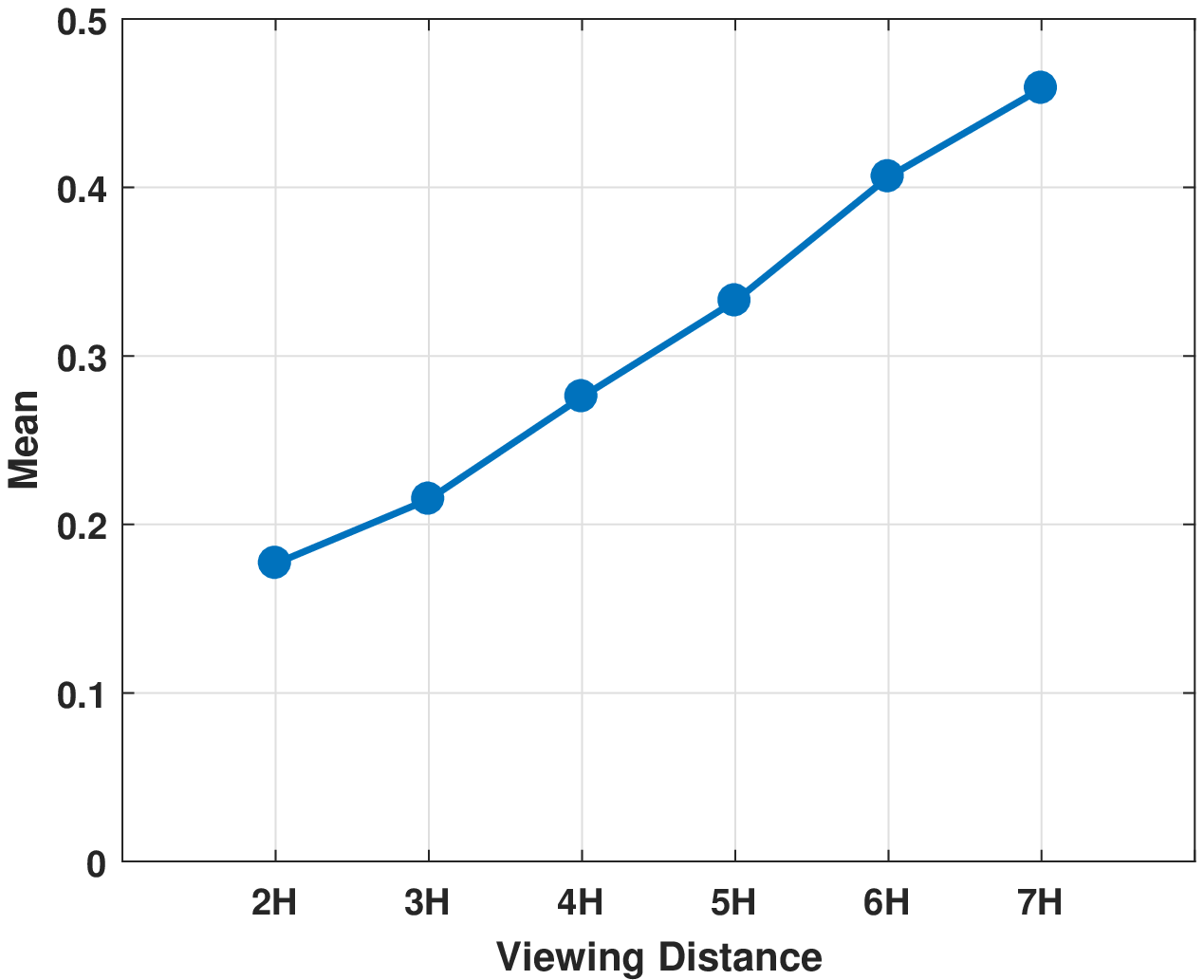}\\
(b1) VSNR&(b2) GMSD&(b3) UQI&(b4) VIF\\\\
\includegraphics[trim=0cm 0cm 0cm 0cm, scale = \figscale]{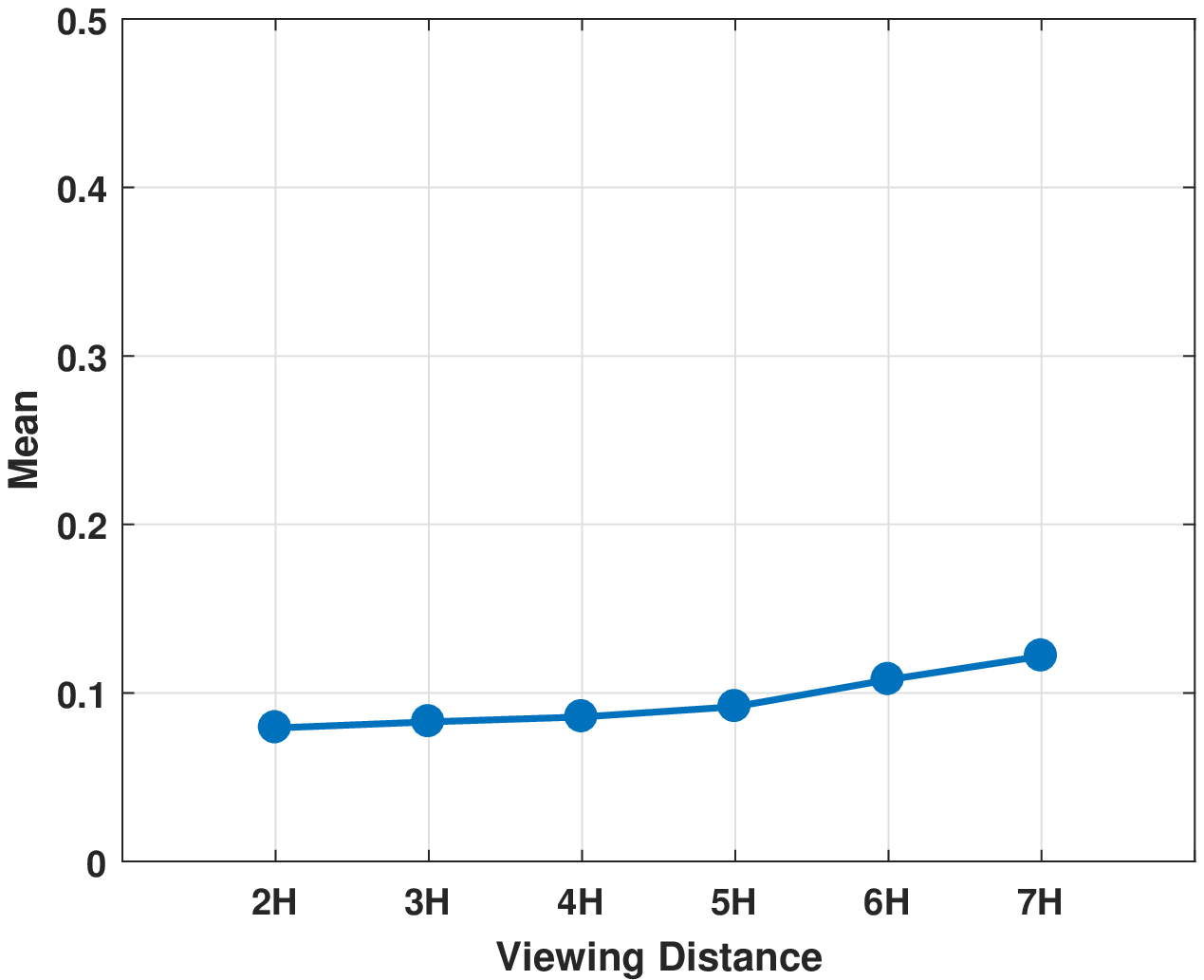}&
\includegraphics[trim=0cm 0cm 0cm 0cm, scale = \figscale]{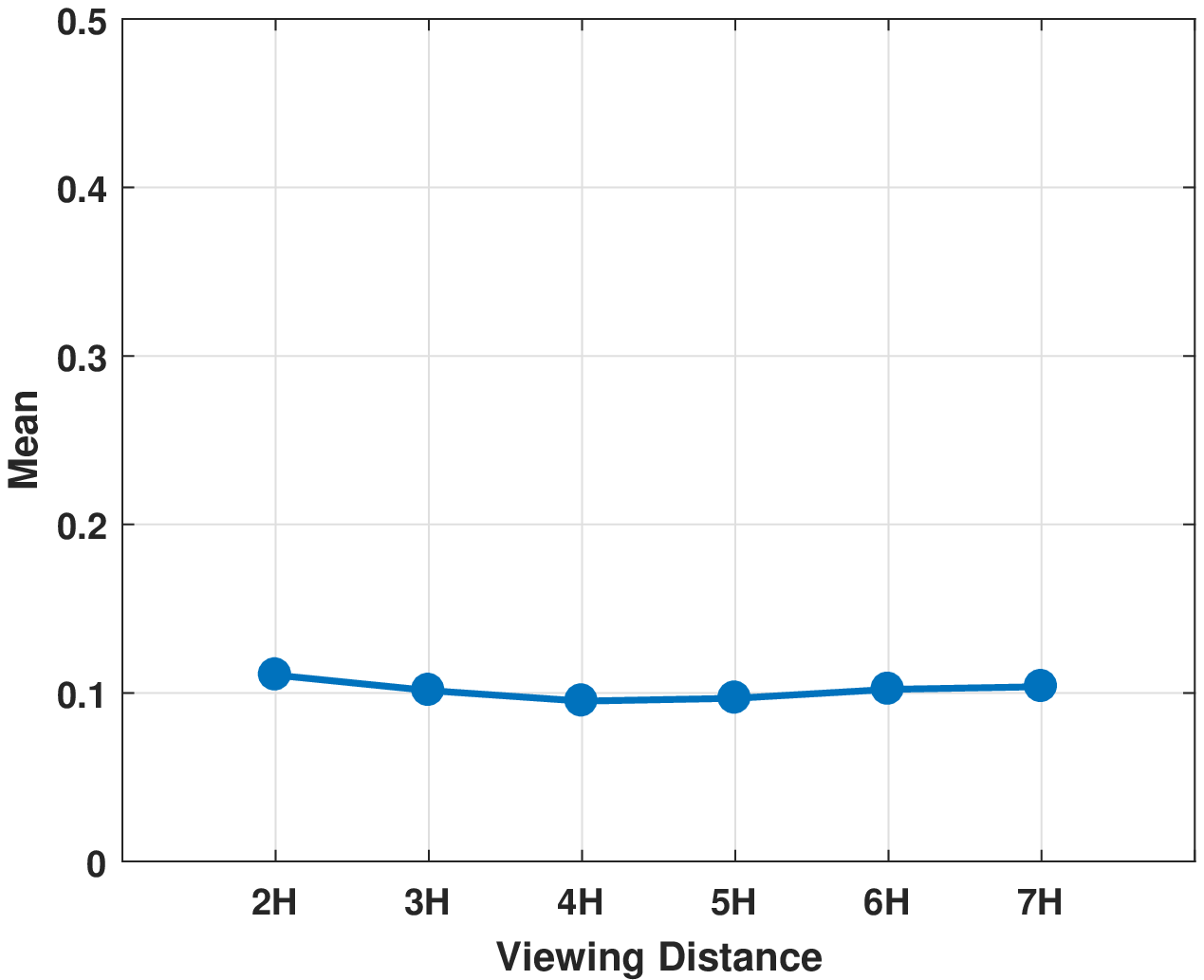}&
\includegraphics[trim=0cm 0cm 0cm 0cm, scale = \figscale]{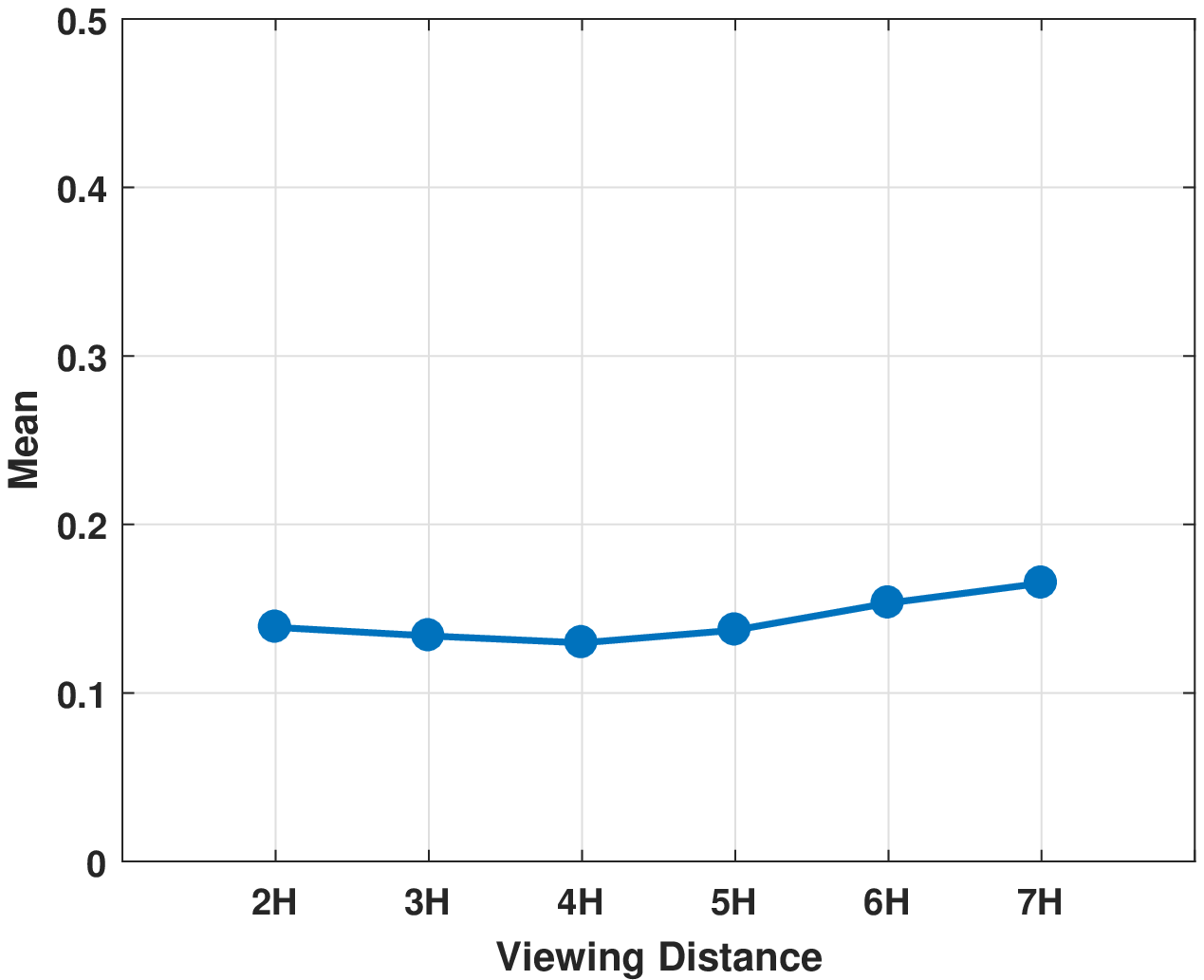}&
\includegraphics[trim=0cm 0cm 0cm 0cm, scale = \figscale]{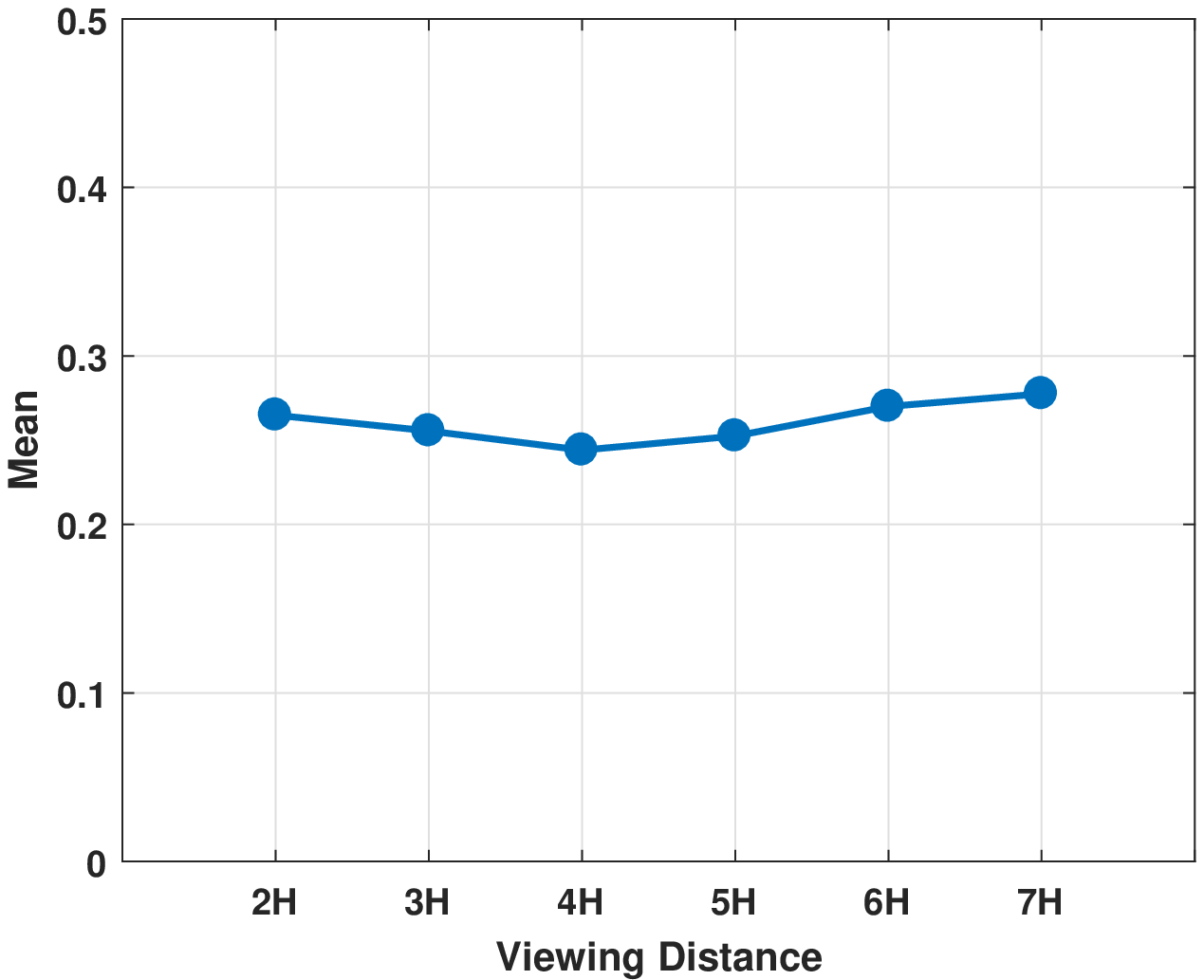}\\
(c1) VSNR&(c2) GSM&(c3) GMSD&(c4) ADM\\\\
\includegraphics[trim=0cm 0cm 0cm 0cm, scale = \figscale]{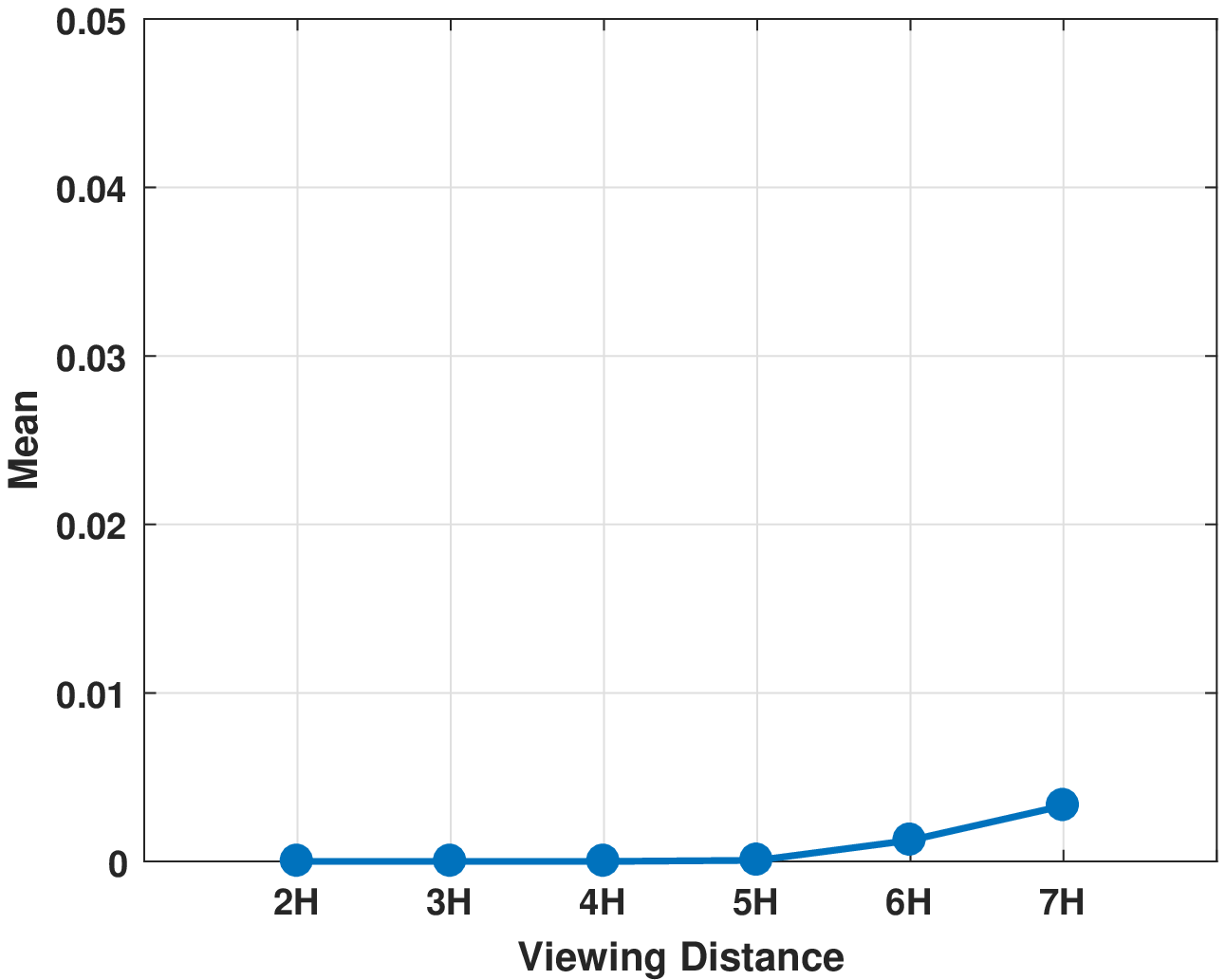}&
\includegraphics[trim=0cm 0cm 0cm 0cm, scale = \figscale]{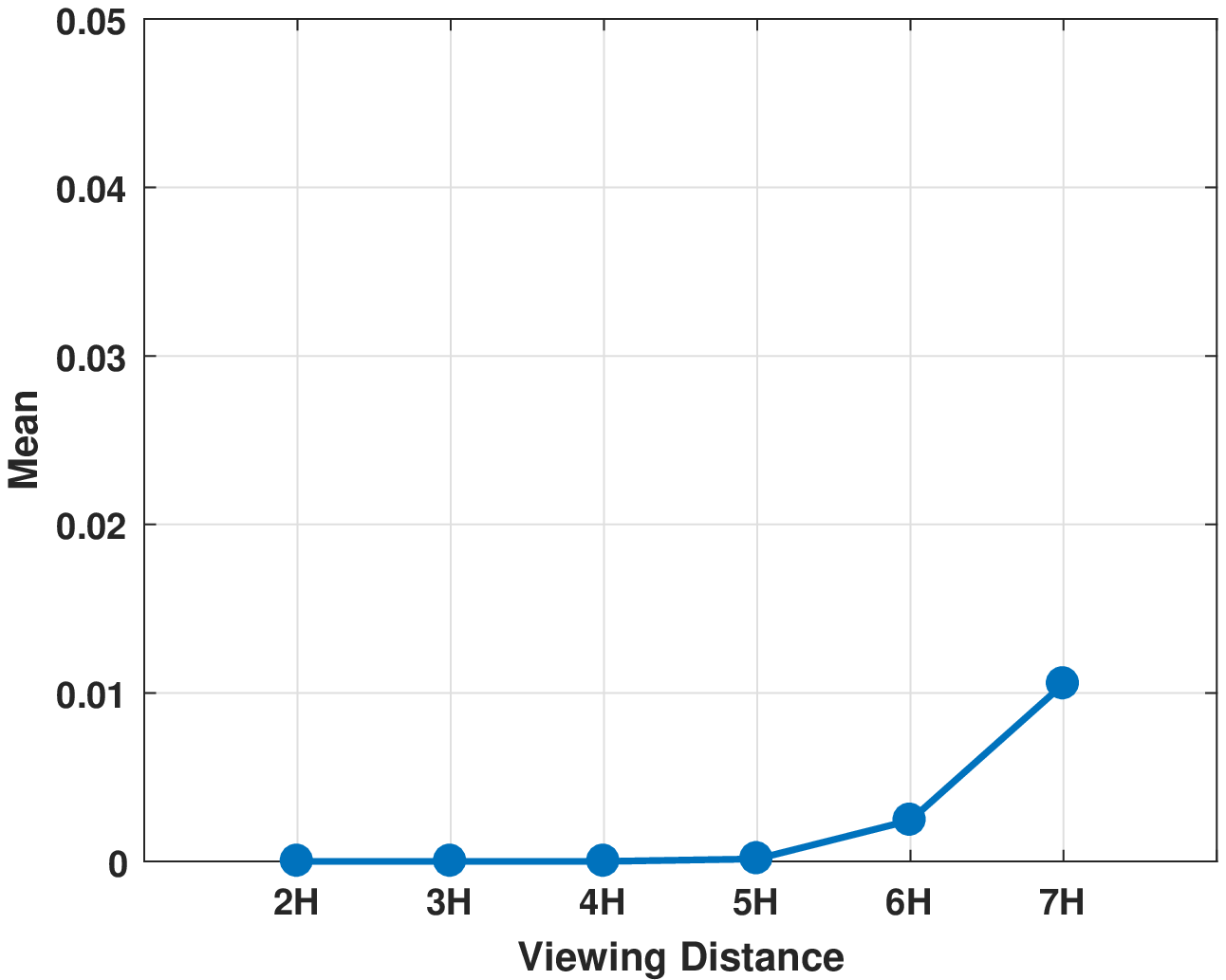}&
\includegraphics[trim=0cm 0cm 0cm 0cm, scale = \figscale]{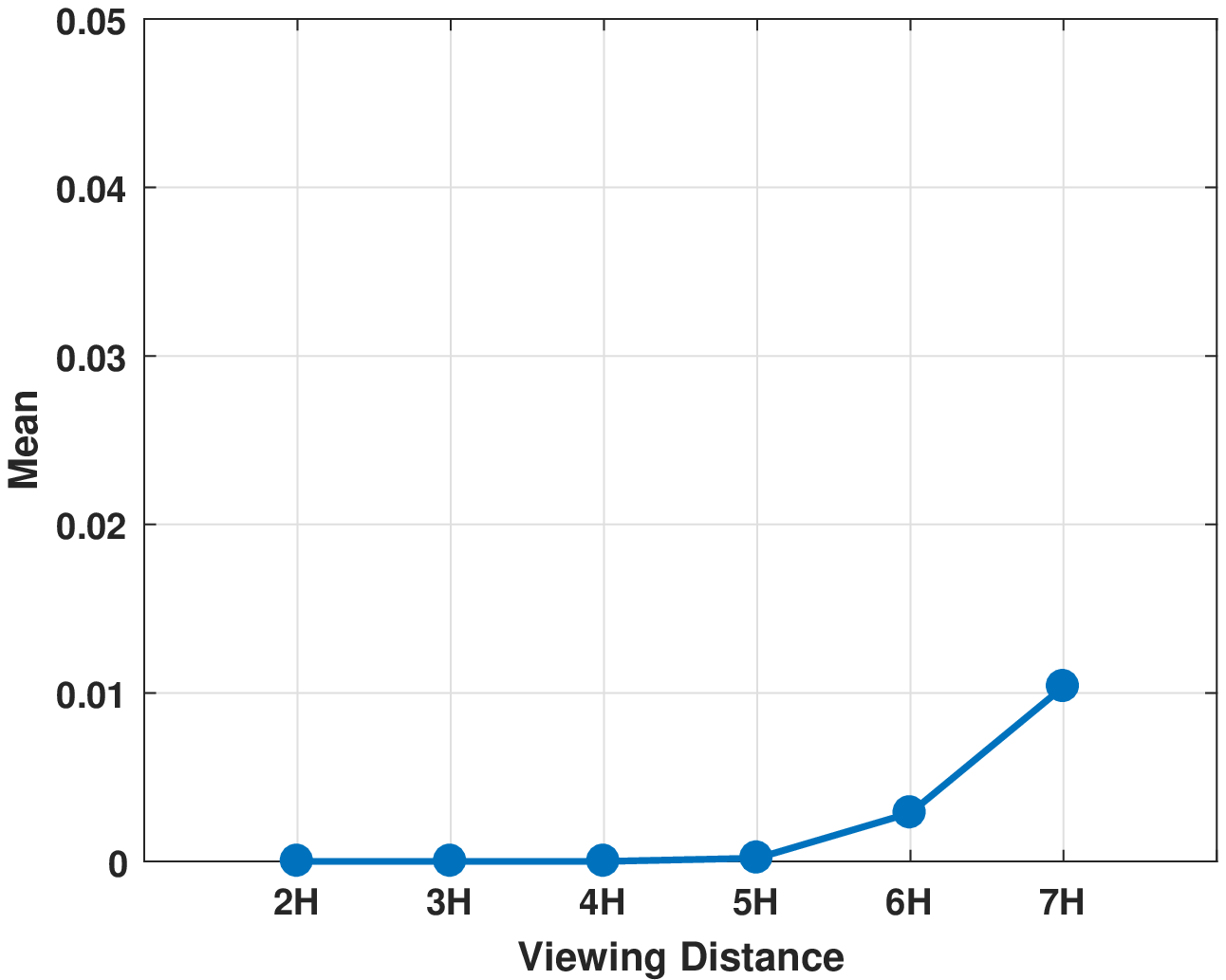}&
\includegraphics[trim=0cm 0cm 0cm 0cm, scale = \figscale]{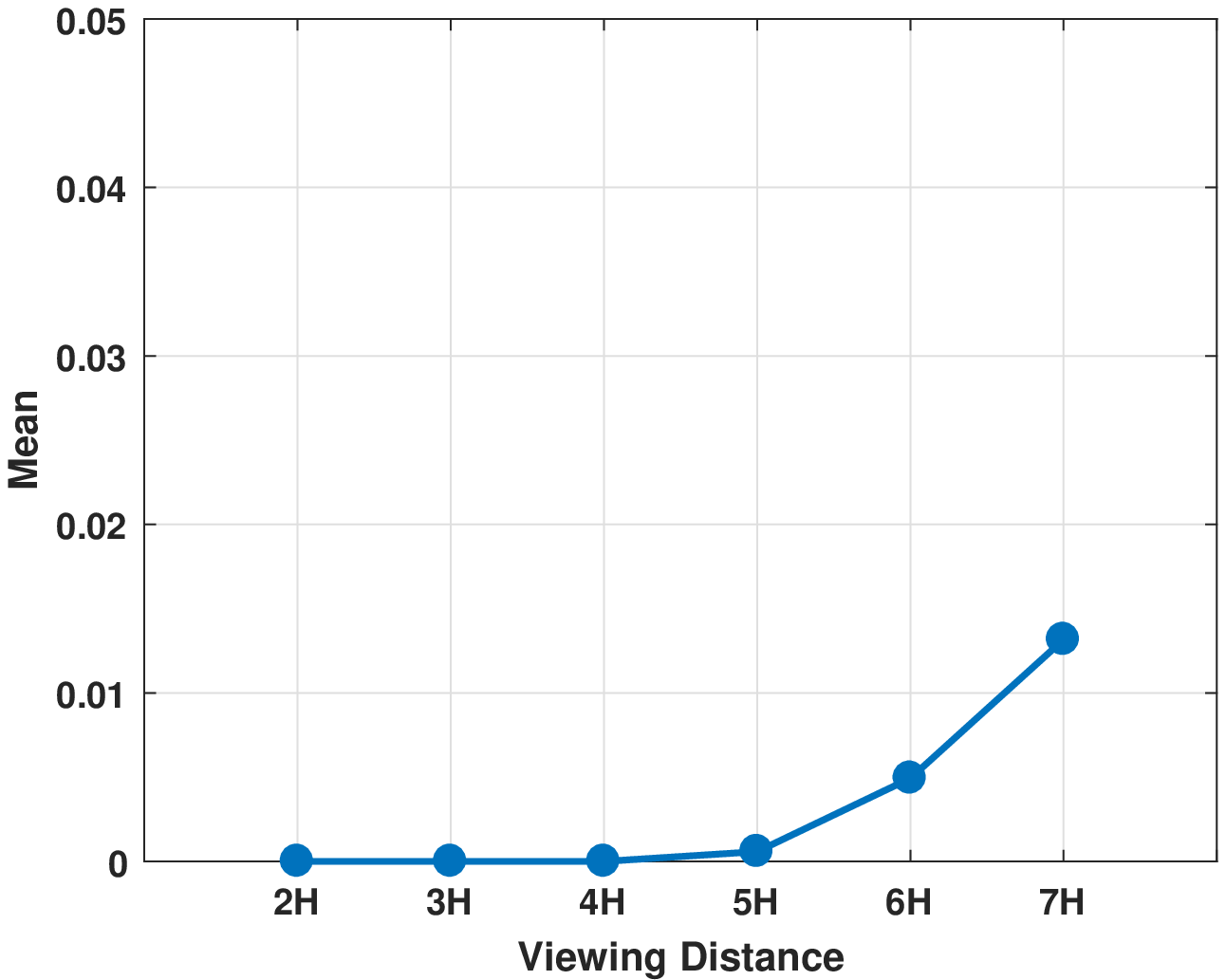}\\
(d1) BLIINDS2&(d2) PSNR-HMA&(d3) PSNR&(d4) OG-IQA\\\\
\end{tabular}
\caption{\label{fig:cidiq-metric-ex} Mean widths of the ambiguity intervals of the objective metrics for the CIDIQ database. For each distortion type, metrics in the first, second, third, and last quarters in the ascending order of the mean ambiguity interval width for 1.5H are shown from left to right. (a1)-(a4) JPEG, (b1)-(b4) JPEG2K, (c1)-(c4) GB, and (d1)-(d4) PN}
\end{figure*}

Figures \ref{fig:vdid-metric-ex} and \ref{fig:cidiq-metric-ex} show the mean ambiguity interval widths of the metrics with respect to the viewing distance for VDID and CIDIQ, respectively.
For each distortion type, the metrics are sorted with respect to the mean ambiguity interval width for the short viewing distances, and the result of a metric in each quarter is presented.
We can observe that the overall slopes of the mean ambiguity interval due to the viewing distance change are different depending on the objective metrics, distortion types, and databases.

When we compare the metrics for the same distortion types (i.e., the four panels in each row), the worse the performance of the metric in terms of ambiguity is, the larger the slope of the graph is.
This tendency is observed clearly except for GB of CIDIQ (Figure \ref{fig:cidiq-metric-ex}(c)).
Therefore, choosing a metric showing good performance in terms of ambiguity for a particular viewing distance is useful also for its reliable usage over different viewing distances.

In Figures \ref{fig:vdid-metric-ex} and \ref{fig:cidiq-metric-ex}, it is also observed that the shapes of the graphs are different depending on the distortion type.
The cases of JPEG and JPEG2K for both databases show mostly monotonically increasing patterns.
Monotonic increases are also observed for the two types of noise (WN of VDID and PN of CIDIQ), except for the clipping at zero for PN of CIDIQ.
The graphs of GB show different tendencies; the mean widths of the intervals remain almost the same for some ranges of viewing distance (for small distances up to 5H for VDID and all distances considered for CIDIQ).
As mentioned earlier, when the viewing distance increases, an ability to distinguish the details in the image also decreases.
Since GB has already reduced the details in the image, the ambiguity intervals are less affected by the viewing distance change.
In some cases, there exist sudden increases of the ambiguity interval widths (GB of VDID and PN of CIDIQ, both after 5H), which needs to be carefully considered when a system using a quality metric operates for a wide range of viewing distances.

\section{Conclusion}
\label{conclusion}

In this paper, we have proposed a new way to measure performance of objective image quality metrics in the viewpoint of the quality resolution.
The procedure to obtain the ambiguity interval for an objective quality score has been developed.
We have demonstrated that the width and the uniformity of the interval over the quality range are useful as performance measures in addition to the accuracy of quality estimation for comparison of different metrics.
In addition, the need for consideration of the viewing distance has been emphasized when the ambiguity intervals are used.

In addition to the use cases shown in this paper, there are several other potential applications of the proposed method. An example is to construct rate-distortion (R-D) curves having ambiguity intervals for evaluating image compression methods, where the distortion is measured with an objective metric and the ambiguity interval at each rate value is obtained using the proposed method, which is an objective counterpart of the method obtaining R-D curves having intervals based on subjective quality scores \cite{hanhart2014calculation}.

One possible follow-up research question is: Is there a way to combine the two (possibly conflicting) performance dimensions of objective quality metrics (i.e., accuracy and ambiguity) to have a single performance measure?
A general solution to this issue may be very challenging to develop.
However, some guidelines to define the superiority and inferiority among different metrics could be identified depending on the target application, which would be desirable to explore as future work.

\section*{Acknowledgment}
This research was supported by the Ministry of Science and ICT (MSIT), Korea, under the ``ICT Consilience Creative Program'' (IITP-2018-2017-0-01015) supervised by the Institute for Information \& communications Technology Promotion (IITP) and also by the IITP grant funded by the Korea government (MSIT) (R7124-16-0004, Development of Intelligent Interaction Technology Based on Context Awareness and Human Intention Understanding).

\bibliography{draft_aioq}

\begin{thebibliography}{10}
\expandafter\ifx\csname url\endcsname\relax
  \def\url#1{\texttt{#1}}\fi
\expandafter\ifx\csname urlprefix\endcsname\relax\def\urlprefix{URL }\fi
\expandafter\ifx\csname href\endcsname\relax
  \def\href#1#2{#2} \def\path#1{#1}\fi

\bibitem{cheon2016ambiguity}
M.~Cheon, J.-S. Lee, Ambiguity-based evaluation of objective quality metrics
  for image compression, in: Proc. Int. Conf. Quality of Multimedia Experience
  (QoMEX), 2016, pp. 1--6.

\bibitem{wallace1991jpeg}
G.~K. Wallace, The {JPEG} still picture compression standard, Communications of
  the ACM 34~(4) (1991) 30--44.

\bibitem{skodras2001jpeg}
A.~Skodras, C.~Christopoulos, T.~Ebrahimi, The {JPEG 2000} still image
  compression standard, IEEE Signal Processing Magazine 18~(5) (2001) 36--58.

\bibitem{wiegand2003overview}
T.~Wiegand, G.~J. Sullivan, G.~Bjontegaard, A.~Luthra, Overview of the
  {H.264/AVC} video coding standard, IEEE Trans. Circuits and Systems for Video
  Technology 13~(7) (2003) 560--576.

\bibitem{sullivan2012overview}
G.~J. Sullivan, J.~Ohm, W.-J. Han, T.~Wiegand, Overview of the high efficiency
  video coding {(HEVC)} standard, IEEE Trans. Circuits and Systems for Video
  Technology 22~(12) (2012) 1649--1668.

\bibitem{sheikh2006statistical}
H.~R. Sheikh, M.~F. Sabir, A.~C. Bovik, A statistical evaluation of recent full
  reference image quality assessment algorithms, IEEE Trans. Image Processing
  15~(11) (2006) 3440--3451.

\bibitem{chikkerur2011objective}
S.~Chikkerur, V.~Sundaram, M.~Reisslein, L.~J. Karam, Objective video quality
  assessment methods: {A} classification, review, and performance comparison,
  IEEE Trans. Broadcasting 57~(2) (2011) 165--182.

\bibitem{cheon2017subjective}
M.~Cheon, J.-S. Lee, Subjective and objective quality assessment of compressed
  4{K UHD} videos for immersive experience, IEEE Trans. Circuits and Systems
  for Video Technology 28~(7) (2018) 1467--1480.

\bibitem{wang2011applications}
Z.~Wang, Applications of objective image quality assessment methods
  [applications corner], IEEE Signal Processing Magazine 28~(6) (2011)
  137--142.

\bibitem{itu_t_p1401}
ITU-T, {Recommendation} {P}.1401: Methods, metrics and procedures for
  statistical evaluation, qualification and comparison of objective quality
  prediction models, Tech. rep., {ITU-T} (2012).

\bibitem{krasula2016accuracy}
L.~Krasula, K.~Fliegel, P.~Le~Callet, M.~Kl{\'\i}ma, On the accuracy of
  objective image and video quality models: {N}ew methodology for performance
  evaluation, in: Proc. Int. Workshop on Quality of Multimedia Experience
  (QoMEX), 2016, pp. 1--6.

\bibitem{jayant1993signal}
N.~Jayant, J.~Johnston, R.~Safranek, Signal compression based on models of
  human perception, Proc. IEEE 81~(10) (1993) 1385--1422.

\bibitem{cheon2015ambiguity}
M.~Cheon, J.-S. Lee, On ambiguity of objective image quality assessment,
  Electronics Letters 52~(1) (2016) 34--35.

\bibitem{itu_r_bt500}
ITU-R, {Recommendation} {BT}.500-13: {M}ethodology for the subjective
  assessment of the quality of television, Tech. rep., {ITU-R} (2012).

\bibitem{lee2014on}
J.-S. Lee, On designing paired comparison experiments for subjective multimedia
  quality assessment, IEEE Trans. Multimedia 16~(2) (2014) 564--571.

\bibitem{narwaria2015hdr}
M.~Narwaria, R.~K. Mantiuk, M.~P. Da~Silva, P.~Le~Callet, {HDR-VDP-2.2}: {A}
  calibrated method for objective quality prediction of high-dynamic range and
  standard images, Journal of Electronic Imaging 24~(1) (2015) 1--3.

\bibitem{sheikh2006image}
H.~R. Sheikh, A.~C. Bovik, Image information and visual quality, IEEE Trans.
  Image Processing 15~(2) (2006) 430--444.

\bibitem{wang2004image}
Z.~Wang, A.~C. Bovik, H.~R. Sheikh, E.~P. Simoncelli, Image quality assessment:
  from error visibility to structural similarity, IEEE Trans. Image Processing
  13~(4) (2004) 600--612.

\bibitem{daly1992visible}
S.~J. Daly, Visible differences predictor: an algorithm for the assessment of
  image fidelity, in: Proc. SPIE/IS\&T Symposium on Electronic Imaging: Science
  and Technology, 1992, pp. 2--15.

\bibitem{mohammadi2015subjective}
P.~Mohammadi, A.~Ebrahimi-Moghadam, S.~Shirani, Subjective and objective
  quality assessment of image: {A} survey, Majlesi Journal of Electrical
  Engineering 9~(1) (2015) 55--83.

\bibitem{lin2011perceptual}
W.~Lin, C.-C.~J. Kuo, Perceptual visual quality metrics: {A} survey, Journal of
  Visual Communication and Image Representation 22~(4) (2011) 297--312.

\bibitem{lee2012comparison}
J.-S. Lee, Comparison of objective quality metrics on the scalable extension of
  {H.264/AVC}, in: Proc. IEEE Int. Conf. Image Processing (ICIP), 2012, pp.
  693--696.

\bibitem{cheon2016evaluation}
M.~Cheon, J.-S. Lee, Evaluation of objective quality metrics for
  multidimensional video scalability, Journal of Visual Communication and Image
  Representation 35 (2016) 132--145.

\bibitem{mantiuk2011hdr}
R.~Mantiuk, K.~J. Kim, A.~G. Rempel, W.~Heidrich, {HDR-VDP-2}: a calibrated
  visual metric for visibility and quality predictions in all luminance
  conditions, ACM Trans. Graphics 30~(4) (2011) 40:1--13.

\bibitem{li2011image}
S.~Li, F.~Zhang, L.~Ma, K.~N. Ngan, Image quality assessment by separately
  evaluating detail losses and additive impairments, IEEE Trans. Multimedia
  13~(5) (2011) 935--949.

\bibitem{gu2015quality}
K.~Gu, M.~Liu, G.~Zhai, X.~Yang, W.~Zhang, Quality assessment considering
  viewing distance and image resolution, IEEE Trans. Broadcasting 61~(3) (2015)
  520--531.

\bibitem{liu2014cid}
X.~Liu, M.~Pedersen, J.~Y. Hardeberg, {CID:IQ}--a new image quality database,
  in: Proc. Int. Conf. Image and Signal Processing, 2014, pp. 193--202.

\bibitem{wang2003multiscale}
Z.~Wang, E.~P. Simoncelli, A.~C. Bovik, Multiscale structural similarity for
  image quality assessment, in: Proc. IEEE Asilomar Conf. Signals, Systems and
  Computers, Vol.~2, 2003, pp. 1398--1402.

\bibitem{chandler2007vsnr}
D.~M. Chandler, S.~S. Hemami, {VSNR: A} wavelet-based visual signal-to-noise
  ratio for natural images, IEEE Trans. Image Processing 16~(9) (2007)
  2284--2298.

\bibitem{wang2002universal}
Z.~Wang, A.~C. Bovik, A universal image quality index, IEEE Signal Processing
  Letters 9~(3) (2002) 81--84.

\bibitem{sheikh2005information}
H.~R. Sheikh, A.~C. Bovik, G.~De~Veciana, An information fidelity criterion for
  image quality assessment using natural scene statistics, IEEE Trans. Image
  Processing 14~(12) (2005) 2117--2128.

\bibitem{damera2000image}
N.~Damera-Venkata, T.~D. Kite, W.~S. Geisler, B.~L. Evans, A.~C. Bovik, Image
  quality assessment based on a degradation model, IEEE Trans. Image Processing
  9~(4) (2000) 636--650.

\bibitem{egiazarian2006new}
K.~Egiazarian, J.~Astola, N.~Ponomarenko, V.~Lukin, F.~Battisti, M.~Carli, New
  full-reference quality metrics based on {HVS}, in: Proc. Int. Workshop on
  Video Processing and Quality Metrics, Vol.~4, 2006.

\bibitem{ponomarenko2007between}
N.~Ponomarenko, F.~Silvestri, K.~Egiazarian, M.~Carli, J.~Astola, V.~Lukin, On
  between-coefficient contrast masking of {DCT} basis functions, in: Proc. Int.
  Workshop on Video Processing and Quality Metrics, Vol.~4, 2007.

\bibitem{ponomarenko2011modified}
N.~Ponomarenko, O.~Ieremeiev, V.~Lukin, K.~Egiazarian, M.~Carli, Modified image
  visual quality metrics for contrast change and mean shift accounting, in:
  Proc. Int. Conf. The Experience of Designing and Application of CAD Systems
  in Microelectronics, 2011, pp. 305--311.

\bibitem{wang2011information}
Z.~Wang, Q.~Li, Information content weighting for perceptual image quality
  assessment, IEEE Trans. Image Processing 20~(5) (2011) 1185--1198.

\bibitem{zhang2011fsim}
L.~Zhang, D.~Zhang, X.~Mou, {FSIM}: a feature similarity index for image
  quality assessment, IEEE Trans. Image Processing 20~(8) (2011) 2378--2386.

\bibitem{xue2014gradient}
W.~Xue, L.~Zhang, X.~Mou, A.~C. Bovik, Gradient magnitude similarity deviation:
  A highly efficient perceptual image quality index, IEEE Trans. Image
  Processing 23~(2) (2014) 684--695.

\bibitem{larson2010most}
E.~C. Larson, D.~M. Chandler, Most apparent distortion: full-reference image
  quality assessment and the role of strategy, Journal of Electronic Imaging
  19~(1) (2010) 1--21.

\bibitem{gu2016analysis}
K.~Gu, S.~Wang, G.~Zhai, W.~Lin, X.~Yang, W.~Zhang, Analysis of distortion
  distribution for pooling in image quality prediction, IEEE Trans.
  Broadcasting 62~(2) (2016) 446--456.

\bibitem{zhang2014vsi}
L.~Zhang, Y.~Shen, H.~Li, {VSI}: A visual saliency-induced index for perceptual
  image quality assessment, IEEE Trans. Image Processing 23~(10) (2014)
  4270--4281.

\bibitem{liu2012image}
A.~Liu, W.~Lin, M.~Narwaria, Image quality assessment based on gradient
  similarity, IEEE Trans. Image Processing 21~(4) (2012) 1500--1512.

\bibitem{PSIM}
K.~{Gu}, L.~{Li}, H.~{Lu}, X.~{Min}, W.~{Lin}, A fast reliable image quality
  predictor by fusing micro- and macro-structures, IEEE Trans. on Industrial
  Electronics 64~(5) (2017) 3903--3912.

\bibitem{soundararajan2012rred}
R.~Soundararajan, A.~C. Bovik, {RRED} indices: Reduced reference entropic
  differencing for image quality assessment, IEEE Trans. Image Processing
  21~(2) (2012) 517--526.

\bibitem{liu2014no}
L.~Liu, B.~Liu, H.~Huang, A.~C. Bovik, No-reference image quality assessment
  based on spatial and spectral entropies, Signal Processing: Image
  Communication 29~(8) (2014) 856--863.

\bibitem{liu2016blind}
L.~Liu, Y.~Hua, Q.~Zhao, H.~Huang, A.~C. Bovik, Blind image quality assessment
  by relative gradient statistics and adaboosting neural network, Signal
  Processing: Image Communication 40 (2016) 1--15.

\bibitem{saad2012blind}
M.~A. Saad, A.~C. Bovik, C.~Charrier, Blind image quality assessment: A natural
  scene statistics approach in the {DCT} domain, IEEE Trans. Image Processing
  21~(8) (2012) 3339--3352.

\bibitem{ASIQE}
K.~{Gu}, J.~{Zhou}, J.~{Qiao}, G.~{Zhai}, W.~{Lin}, A.~C. {Bovik}, No-reference
  quality assessment of screen content pictures, IEEE Trans. on Image
  Processing 26~(8) (2017) 4005--4018.

\bibitem{hanhart2013benchmarking}
P.~Hanhart, P.~Korshunov, T.~Ebrahimi, Benchmarking of quality metrics on
  ultra-high definition video sequences, in: Proc. Int. Conf. Digital Signal
  Processing (DSP), 2013, pp. 1--8.

\bibitem{tian19benchmark}
S.~Tian, L.~Zhang, L.~Morin, O.~D\'{e}forges, A benchmark of {DIBR} synthesized
  view quality assessment metrics on a new database for immersive media
  applications, IEEE Trans. Multimedia 21~(5) (2019) 1235--1247.

\bibitem{video2000final}
VQEG, Final report from the video quality experts group on the validation of
  objective models of video quality assessment, Tech. rep., VQEQ (2000).

\bibitem{hanhart2014calculation}
P.~Hanhart, T.~Ebrahimi, Calculation of average coding efficiency based on
  subjective quality scores, Journal of Visual Communication and Image
  Representation 25 (2014) 555--564.

\end{thebibliography}
\end{document}